\documentclass[11pt,letterpaper]{article}
\usepackage{fullpage}
\usepackage{amsmath}
\usepackage{graphicx}%
\usepackage{amsfonts}%
\usepackage{amssymb}
\usepackage{url}
\usepackage{fancyhdr}
\usepackage{tocbibind}
\usepackage{isomath}
\usepackage{amsmath}
\usepackage{amsbsy}
\usepackage{amssymb}
\usepackage{amscd}
\usepackage{amsfonts}
\usepackage{graphicx}
\usepackage{verbatim}
\usepackage{euscript}
\usepackage{alltt}
\usepackage{stmaryrd}
\usepackage{float}
\usepackage{appendix}
\usepackage{caption}
\usepackage[english]{babel}

\usepackage{subfigure}
\usepackage{color}

\usepackage{amsmath}
\usepackage{amsbsy}
\usepackage{amssymb}
\usepackage{amscd}
\usepackage{amsfonts}

\newcommand{\bfa}{{\mathbold a}}
\newcommand{\bfb}{{\mathbold b}}
\newcommand{\bfc}{{\mathbold c}}
\newcommand{\bfd}{{\mathbold d}}
\newcommand{\bfe}{{\mathbold e}}
\newcommand{\bff}{{\mathbold f}}

\newcommand{\bfn}{{\mathbold n}}

\newcommand{\bfr}{{\mathbold r}}
\newcommand{\bfs}{{\mathbold s}}

\newcommand{\bfu}{{\mathbold u}}
\newcommand{\bfv}{{\mathbold v}}

\newcommand{\bfx}{{\mathbold x}}
\newcommand{\bfy}{{\mathbold y}}
\newcommand{\bfz}{{\mathbold z}}

\newcommand{\bfA}{{\mathbold A}}
\newcommand{\bfB}{{\mathbold B}}

\newcommand{\bfD}{{\mathbold D}}
\newcommand{\bfE}{{\mathbold E}}
\newcommand{\bfF}{{\mathbold F}}

\newcommand{\bfH}{{\mathbold H}}
\newcommand{\bfI}{{\mathbold I}}

\newcommand{\bfS}{{\mathbold S}}
\newcommand{\bfT}{{\mathbold T}}

\newcommand{\bfW}{{\mathbold W}}
\newcommand{\bfX}{{\mathbold X}}
\newcommand{\bfY}{{\mathbold Y}}
\newcommand{\bfZ}{{\mathbold Z}}

\newcommand{\beq}{\begin{equation}}
\newcommand{\eeq}{\end{equation}}
\newcommand{\beqs}{\begin{eqnarray}}
\newcommand{\eeqs}{\end{eqnarray}}
\newcommand{\beql}{\begin{equation} \label}

\newcommand{\bfsigma}{\mathbold{\sigma}}

\newcommand{\bfalpha}{\mathbold{\alpha}}
\newcommand{\bfomega}{\mathbold{\omega}}

\newcommand{\bfPi}{\mathbold{\Pi}}

\newcommand{\bfOmega}{\mathbold{\Omega}}

\newcommand{\bfDelta}{\mathbold{\Delta}}

\newcommand{\grad}{\mathop{\rm grad}\nolimits}
\newcommand{\divergence}{\mathop{\rm div}\nolimits}
\newcommand{\curl}{\mathop{\rm curl}\nolimits}

\usepackage[margin=1in]{geometry}
\newenvironment{rcases}
  {\left.\begin{aligned}}
  {\end{aligned}\right\rbrace}

\begin{document}

\title{On the relevance of generalized disclinations in defect mechanics}
\author{Chiqun Zhang and Amit Acharya\\ \\
Carnegie Mellon University, Pittsburgh, PA 15213, USA \\ }
%%\date{December 23, 2016}
\date{}
\maketitle

\begin{abstract}

\noindent The utility of the notion of generalized disclinations in materials science is discussed within the physical context of modeling interfacial and bulk line defects like defected grain and phase boundaries, dislocations and disclinations. The Burgers vector of a disclination dipole in linear elasticity is derived, clearly demonstrating the equivalence of its stress field to that of an edge dislocation. We also prove that the inverse deformation/displacement jump of a defect line is independent of the cut-surface when its g.disclination strength vanishes. An explicit formula for the displacement jump of a single localized composite defect line in terms of given g.disclination and dislocation strengths is deduced based on the Weingarten theorem for g.disclination theory (Weingarten-gd theorem) at finite deformation. The Burgers vector of a g.disclination dipole at finite deformation is also derived.

\end{abstract}

\section{Introduction}

While the mechanics of disclinations has been studied \cite{wit1970linear, dewit1971relation, dewit1973theory, de1973theory, Nabarro1987, hirth2006disconnections, zubov1997nonlinear, romanov2009application, fressengeas2011elasto}, there appears to be a significant barrier to the adoption of disclination concepts in the practical modeling of physical problems in the mechanics of materials, perhaps due to the strong similarities between the fields of a disclination dipole and a dislocation \cite{romanov2009application}. Furthermore, the finite deformation version of disclination theory is mathematically intricate \cite{zubov1997nonlinear,derezin2011disclinations}, and does not lend itself in a natural way to the definition of the strength of a disclination purely in terms of any candidate field that may be defined to be a disclination density. This has prevented the introduction of a useful notion of a disclination density field \cite{zubov1997nonlinear,derezin2011disclinations}, thereby hindering the development of a finite deformation theory of disclination fields and its computational implementation to generate approximate solutions for addressing practical problems in the mechanics of materials and materials science. 

A recent development in this regard is the development of g.disclination theory (generalized disclination theory) \cite{acharya2012coupled, acharya2015continuum}, that alleviates the significant road-block in the finite deformation setting mentioned above. It does so by adopting a different conceptual standpoint in defining the notion of g.disclinations than what arose in the works of Weingarten and Volterra (as described by Nabarro \cite{Nabarro1987}). This new standpoint also allows the consideration of phase and grain-boundaries and their terminating line defects within a common framework. Briefly, Weingarten asked a question adapted to the theory of linear elasticity which requires the construction of a displacement field on a multiply-connected\footnote{We refer to any non simply-connected body as multiply-connected or multi-connected.} body with a single hole, and the characterization of its jumps across any surface that renders the body simply-connected when `cut' by it. An important constraint of the construction is that the strain of the displacement match a given symmetric second-order tensor field on the simply-connected body induced by the cut; the given symmetric second-order tensor field is assumed twice-differentiable on the original multiply-connected domain and to satisfy the St.-Venant compatibility conditions. There is a well defined analogous question at finite deformations  \cite{casey2004volterra}. The constructed displacement field on the simply-connected domain will in general have a jump (i.e. difference) in the values of its rotation field at corresponding points across the surface but the jump in its strain (similarly defined) necessarily vanishes by definition. However, when viewed from this perspective and keeping physically abundant objects like (incoherent) phase boundaries in mind across which strains are discontinuous as well, there seems to be no reason to begin from a starting point involving a continuous strain field; it is just as reasonable to ask that one is given a smooth third-order tensor field that is curl-free on a multiply-connected domain with a hole (this condition replacing the given strain field satisfying the St.-Venant compatibility condition), and then ask for the construction of a displacement field whose second gradient matches the given third-order tensor field on a cut-surface induced simply-connected domain, and the characterization of the jump of the displacement field across the surface. This allows the whole first gradient of the deformation/displacement field (constructed on the simply-connected domain) to exhibit jumps across the surface, instead of only the rotation. Moreover, this whole argument goes through seamlessly in the context of geometrically nonlinear kinematics; the g.disclination strength is defined as a standard contour integral of the given third order tensor field. The framework naturally allows the calculation of fields of a purely rotational disclination specified as a g.disclination density distribution.

The principal objectives of this paper are to
\begin{itemize}
\item  review the physical situations that  may be associated with the mathematical concept of disclinations (considered as a special case of g.disclinations). Much is known in this regard amongst specialists (cf. \cite{romanov2009application}), and we hope to provide complementary, and on occasion new, perspective to what is known to set the stage for solving physical problems related to disclination mechanics in a forthcoming paper \cite{zhang_acharya_puri} within the framework of g.disclinations.
\item Establish the connection between the topological properties of a g.disclination dipole and a dislocation at finite strains by deducing the formula for the Burgers vector of the g.disclination dipole.
\item Interpret the Weingarten theorem for g.disclinations at finite deformation \cite{acharya2015continuum} in terms of g.disclination kinematics.
\end{itemize}

This paper is organized as follows. Section \ref{sec:disclination_intro} provides a review of related prior literature. A brief Section \ref{sec:notation} introduces the notation utilized in the paper. In Section \ref{sec:physical_exp} we discuss various physical descriptions for disclinations, dislocations, and grain boundaries as well as the interrelations between them. In Section \ref{sec:burgers_lin} we derive the Burgers vector of a disclination dipole within classical elasticity theory, making a direct connection with the stress field of an edge dislocation. Section \ref{sec:g_disclination_theory} provides an overview of generalized disclination statics from \cite{acharya2015continuum}. In Section \ref{sec:weingarten}, the Weingarten theorem for generalized disclination theory from \cite{acharya2015continuum} is recalled for completeness and a new result proving that the displacement jump is independent of the cut-surface under appropriate special conditions is deduced. In this paper, we refer the Weingarten theorem for generalized disclination theory from \cite{acharya2015continuum} as the Weingarten-gd theorem. In Section \ref{sec:rel_gdisclin_wein} the Weingarten-gd theorem is interpreted in the context of g.disclination kinematics, providing an explicit formula for the displacement jump of a single g.disclination in terms of data prescribed to define the two defect densities (g.disclination and dislocation densities) in g.disclination theory. Finally, in Section \ref{sec:burgers} we derive the Burgers vector for a g.disclination dipole.

\section{A brief review of prior work} \label{sec:disclination_intro}
In this section we briefly review some of the vast literature on the mathematical modeling of disclinations and grain boundaries. An exhaustive review of the subject is beyond the scope of this paper.

The definition of the dislocation and the disclination in \emph{solids}\footnote{
There is a difference in meaning between disclinations in solids and in nematic liquid crystals as explained in \cite{romanov2009application, kleman2008disclinations, pourmatin2012fundamental}.
} 
was first introduced by Volterra (as described by Nabarro \cite{Nabarro1987}). Nabarro \cite{Nabarro1987} studied geometrical aspects of disclinations and Li \cite{li1972disclination} presented microscopic interpretations of a grain boundary in terms of a dislocation and a disclination model. The static fields of dislocations and disclinations along with applications have been studied extensively within linear elasticity by DeWit \cite{wit1970linear, dewit1971relation, de1973theory, de1972partial}, as well as in 2-d nonlinear elasticity by Zubov \cite{zubov1997nonlinear} and the school of study led by Romanov \cite{romanov2009application}. In \cite{romanov2009application}, the elastic fields and energies of the disclination are reviewed and the disclination concept is applied to explaining several observed microstructures in crystalline materials. In \cite{romanov1992disclination}, the expression for the Burgers vector for a single-line, two-rotation-axes disclination dipole appropriate for geometrically linear kinematics is motivated from a physical perspective without dealing with questions of invariance of the physical argument w.r.t different cut-surfaces or the topological nature of the displacement jump of a disclination-dipole in contrast to that of a single disclination.

In Fressengeas et al. \cite{fressengeas2011elasto}, the elasto-plastic theory of dislocation fields \cite{acharya2001model} is non-trivially extended to formulate and study time-dependent problems of defect dynamics including both disclination and dislocation fields. Nonlinear elasticity of disclinations and dislocations in 2-d elastic bodies is discussed in \cite{zubov1997nonlinear,derezin2011disclinations}. Dislocations and disclinations are studied within Riemannian geometry in \cite{kupferman2015metric}; Cartan's geometric method to study Riemannian geometry is deployed in \cite{yavari2012riemann} to determine the nonlinear residual stress for 2-d disclination distributions. Similar ideas are also reviewed in \cite{clayton2006modeling}, in a different degree of mathematical detail and without any explicit calculations, in an effort to develop a time-dependent model of mesoscale plasticity based on disclination-dislocation concepts. Another interesting recent work along these lines is the one in \cite{roychowdhury_gupta} that discusses `metrical disclinations' (among other things) that are related to our g.disclinations. The concerns of classical, nonlinear disclination theory related to defining the strength of a single disclination in a practically applicable manner, and therefore studying the mechanics of interactions of collections of individual such defects, remains in this theory and the authors promote the viewpoint of avoiding any type of curvature line-defects altogether.

From the materials science perspective, extensive studies have been conducted on grain boundary structure, kinetics, and  mechanics from the atomistic \cite{sutton1983structure1, sutton1983structure2} as well as from more macroscopic points of view \cite{mullins1956two,cahn2006coupling,cahn1982transitions}. In \cite{kim2006grain,saylor2004distribution, rohrer2010introduction, rohrer2011grain}, the grain boundary character distribution is studied from the point of view of grain boundary microstructure evolution. In \cite{kinderlehrer2006variational, elsey2009diffusion}, a widely used framework for grain boundary network evolution, which involves the variation of the boundary energy density based on misorientation, is proposed. In most cases, these approaches do not establish an explicit connection with the stress and elastic deformation fields caused by the grain boundary \cite{holm2001misorientation}. Phase boundary mechanics considering effects of stress is considered in \cite{porta2013heterogeneity,  agrawal2015dynamic}, strictly within the confines of compatible elastic deformations.

One approach to study a low angle grain boundary is to model it as a series of dislocations along the boundary \cite{read1950dislocation, sutton1983structure1, sutton1983structure2}. In \cite{dai2013structure}, a systematic numerical study is conducted of the structure and energy of low angle twist boundaries based on a generalized Peierls-Nabarro model. The dislocation model has also been applied to study grain boundaries with disconnections \cite{hirth1996steps, hirth1998extended, hirth2006disconnections, hirth2007spacing, howe2009role, hirth2011compatibility, hirth2013interface}. In these work, disconnections are modeled as dislocations at a step and the grain boundaries are represented as a series of coherency dislocations.  Long-range stress fields for disconnections in tilt walls are discussed from both the discrete dislocations and the disclination dipole perspectives in \cite{akarapu2008modeling}. In \cite{vattre2013determining, vattre2014computational, vattre2015partitioning} a combination of the Frank-Bilby equation \cite{frank1950resultant, FRANK195315, Bilby263,sutton1995interfaces,bullough1956continuous} and anisotropic elasticity theory is employed to formulate a computational method for describing interface dislocations. In \cite{dong1998stress} atomic-level mechanisms of dislocation nucleation is examined by dynamic simulations of the growth of misfitting films. 

Although low angle grain boundaries can be modeled by dislocations, the dislocation model is no longer satisfactory for describing high-angle boundaries because the larger misorientations require introducing more dislocations along the boundary interface which shortens the distance between dislocations \cite{balluffi2005kinetics}. Thus, it is difficult to identify the Burgers vectors of grain boundary dislocations in high angle grain boundaries. Alternatively, a grain boundary can also be modeled as an array of disclination dipoles \cite{romanov2009application, nazarov2000disclination}.  In \cite{fressengeas2014continuous}, the crossover between the atomistic description and the continuous representation is demonstrated for a tilt grain boundary by designing a specific array of disclination dipoles. Unlike the dislocation model for a grain boundary, the disclination model is applicable to the modeling of both low and high angle grain boundaries.
 
\section{Notation and terminology} \label{sec:notation}

The condition that $a$ is defined to be $b$ is indicated by the statement $a := b$. 
The Einstein summation convention is implied unless otherwise specified. $\bfA \bfb$ is denoted as the action of a tensor $\bfA$ on
a vector $\bfb$, producing a vector. A $\cdot$
represents the inner product of two vectors; the symbol $\bfA\bfD$
represents tensor multiplication of the second-order tensors
$\bfA$ and $\bfD$. A third-order tensor is treated as a linear
transformation on vectors to a second-order tensors. 

 We employ rectangular Cartesian coordinates and components in this paper; all tensor and vector components are written with respect to a rectangular Cartesian basis  
$(\bfe_i)$, $i$=1 to 3. The symbol $div$ represents the divergence, $grad$ represents the
gradient on the body (assumed to be a domain in ambient space). In component form, 
\begin{equation*}
 \begin{split}
    \left(\bfA\times\bfv\right)_{im} &= e_{mjk} A_{ij} v_{k}    \\
    \left(\bfB\times\bfv\right)_{irm} &= e_{mjk} B_{irj} v_{k}  \\
    \left(\divergence\bfA\right)_{i} &= A_{ij,j}                \\
    \left(\divergence\bfB\right)_{ij} &= B_{ijk,k}              \\
    \left(\curl\bfA\right)_{im} &= e_{mjk} A_{ik,j}             \\
    \left(\curl\bfB\right)_{irm} &= e_{mjk} B_{irk,j},             \\
 \end{split}
\end{equation*}
where $e_{mjk}$ is a component of the alternating tensor $\bfX$.

$\bfF^e$ is the elastic distortion tensor; $\bfW := \left(\bfF^e\right)^{-1}$ is the inverse-elastic 1-distortion tensor; $\bfS$ is the eigenwall tensor ($3^{rd}$-order); $\bfY$ is the inverse-elastic 2-distortion tensor ($3^{rd}$-order); $\bfalpha$ is the dislocation density tensor ($2^{nd}$-order) and $\bfPi$ is the generalized disclination density tensor ($3^{rd}$-order). The physical and mathematical meanings of these symbols will be discussed subsequently in Section \ref{sec:g_disclination_theory}.

In dealing with questions related to the Weingarten-gd theorem, we will often have to talk about a vector field $\bfy$  which will generically define an inverse elastic deformation from the current deformed configuration of the body.

\section{Basic ideas for the description of g.disclinations, dislocations, and grain boundaries} \label{sec:physical_exp}

\begin {figure}
\centering
\subfigure[A compatible/coherent phase boundary, where all atomic planes from either side match along the interface.]{
\includegraphics[width=0.35\textwidth]{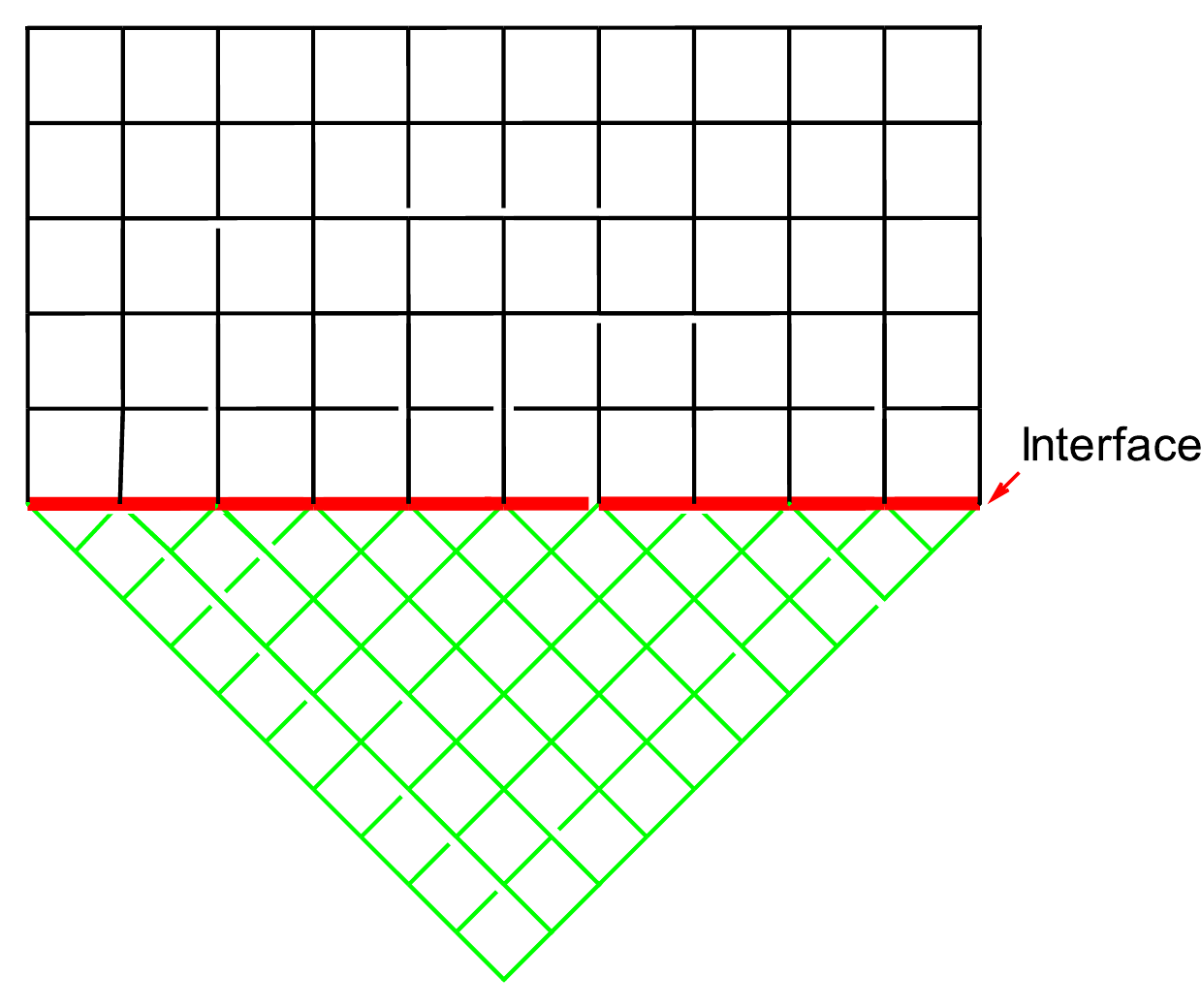}
\label{fig:comp_bound}}\qquad
\subfigure[An incompatible/incoherent phase boundary. There are some mismatches of the atomic planes along the interfaces.]{
\includegraphics[width=0.35\textwidth]{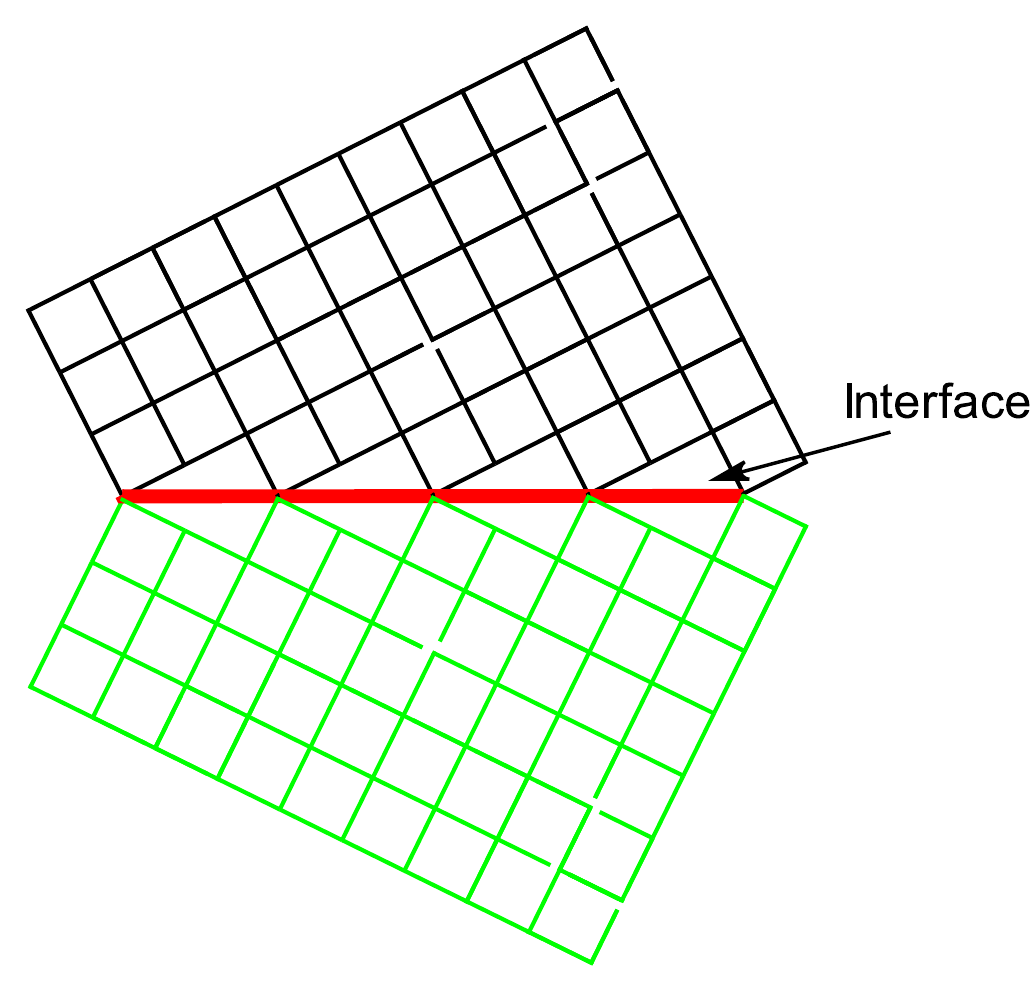}
\label{fig:incomp_bound}}
\caption{Illustration of a compatible/coherent and an incompatible/incoherent phase boundary.}\label{fig:compatible}
\end {figure}

In this section we will discuss various aspects of modeling g.disclinations and their relationship to dislocations, mostly from a physical perspective and through examples. Beginning from a geometric visualization of single disclinations, we will motivate the physical interpretation of such in lattice structures. The formation and movement of a disclination dipole through a lattice will be motivated.  Descriptions of a dislocation in terms of a disclination dipole will be discussed. Finally, we will demonstrate how the description of a low-angle boundary in terms of disclination dipoles may be understood as a dislocated grain-boundary in a qualitative manner.

In many situations in solid mechanics it is necessary to consider a 2-D surface where a distortion measure is discontinuous. In elasticity a distortion corresponds to the deformation gradient; in linear elasticity the distortion will be $\grad \bfu$, where $\bfu$ is the displacement. However, there are many cases where the distortion field cannot be interpreted as a gradient of a vector field on the whole body. In such cases, the distortion will have an incompatible part that is not curl-free. One familiar situation  is to consider the presence of dislocations modeled by the elastic theory of dislocations \cite{kroner1981continuum, willis1967second}.  A 2-d surface of discontinuity of the elastic distortion is referred to as a phase boundary, of which the grain boundary is a particular case.  Based on whether atomic planes from either side of the interface can match with each other at the interface or not, a phase boundary is categorized into a compatible/coherent or an incompatible/incoherent boundary, as shown in Figure~\ref{fig:compatible}. A special compatible phase boundary is called a twin boundary with a highly symmetrical interface, where one crystal is the mirror image of the other, also obtained by a combination of shearing and rotation of one side of the interface with respect to the other. A grain boundary is an interface between two grains with different orientations. The orientation difference between the two grains comprising a grain boundary is called the misorientation at the interface, and it is conventional to categorize grain boundaries based on the misorientation angle. Low angle grain boundaries (LAGBs) are defined as those whose misorientations are less than $11$ degrees and high angle grain boundaries (HAGBs) are those with greater misorientations. In the situation that the phase boundary discontinuity shows gradients along the surface, we will consider the presence of line defects. Following \cite{acharya2012coupled, acharya2015continuum},  the terminating tip-curves of phase boundary discontinuities are called generalized disclinations or g.disclinations.

The classical singular solutions for defect fields contain interesting subtleties. For instance, the normal strain $e_{11}$ in dimension two for a straight dislocation and of a straight disclination in linear elasticity are \cite{dewit1971relation, de1973theory}:
\begin{eqnarray*}
\begin{aligned}
&\text{Straight Dislocation} \\
&\quad e_{11} = -\frac{b_1}{4\pi\left(1-\nu\right)}[\left(1-2\nu\right)\frac{x_2}{\rho^2}+2\frac{x_1^2x_2}{\rho^4}]+\frac{b_2}{4\pi\left(1-\nu\right)}[\left(1-2\nu\right)\frac{x_1}{\rho^2}-2\frac{x_1x_2^2}{\rho^4}], \\ 
&\text{Straight Disclination} \\
&\quad e_{11} = \frac{\Omega_3}{4\pi(1-\nu)}[(1-2\nu)\ln\rho+\frac{x_2^2}{\rho^2}].
\end{aligned}
\end{eqnarray*}
The strain fields blow up in both the dislocation and the disclination cases. In addition, on approaching the core (i.e. the coordinate origin in the above expressions) in dislocation solutions, the elastic strain blows up as $\frac{1}{\rho}$, $\rho$ being the distance from the dislocation core. Thus the linear elastic energy density diverges as $\frac{1}{\rho^2}$, causing unbounded total energy for a finite body for the dislocation whereas the total energy of a disclination is bounded. The disclination, however, has more energy stored in the far-field (w.r.t the core) than the dislocation, and this is believed to be the reason for a single disclination being rarely observed as opposed to a dislocation. Our modeling philosophy and approach enables defects to be represented as non-singular defect lines and surfaces, always with bounded total energy (and even local stress fields).

\subsection{Disclinations}

\begin {figure}%%[H]
\centering
\includegraphics[width=0.5\textwidth]{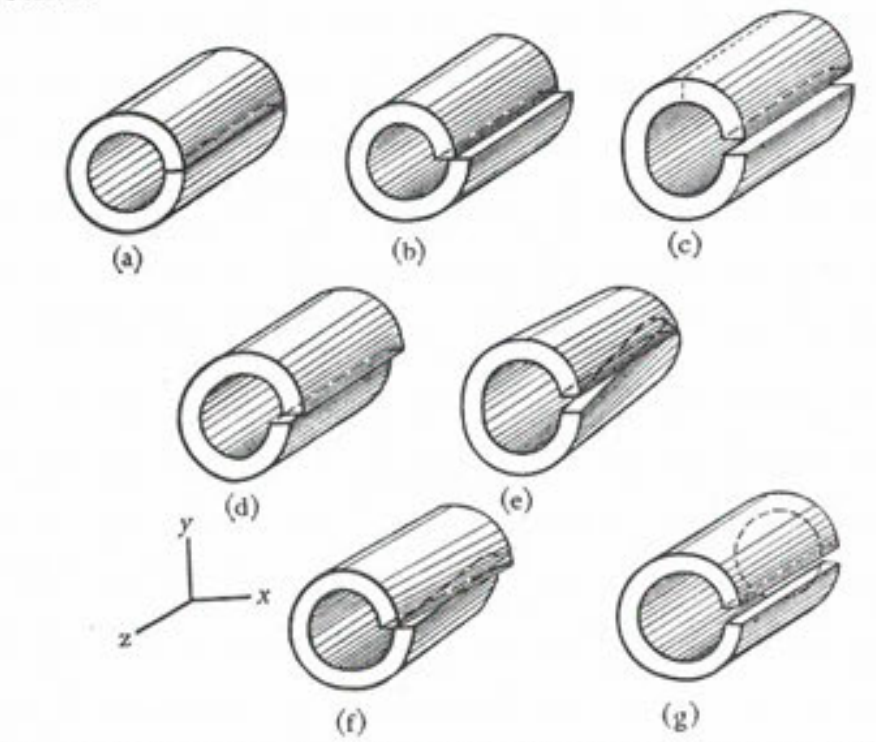}
\caption{Descriptions of Volterra dislocations and disclinations . Figure (a) is a cylinder with an inner hole along the axis. Figure (b) and (c) are the edge dislocations. Figure (d) is the screw dislocation. Figure (e) and (f) are twist disclinations and Figure (g) is the wedge disclination. (Figure reprinted from \cite{Nabarro1987} with permission from Dover Publications).}
\label{fig:cylinder}
\end {figure}

Volterra \cite{nabarro1985development} described dislocations and disclinations by considering a cylinder with a small inner hole along the axis, as shown in Figure \ref{fig:cylinder} (the hole is exaggerated in the figure). Figure \ref{fig:cylinder}(e)(f)(g) show configurations of disclinations. Imagine cutting the cylinder with a half plane, rotating the cut surfaces by a vector $\bfomega$, welding the cut surfaces together and relaxing (i.e. letting the body attain force equilibrium). Then a rotation discontinuity occurs on the cut surface and the vector $\bfomega$ is called the Frank vector. If the Frank vector is parallel to the cylinder's axis, the disclination is called a wedge disclination; if the Frank vector is normal to the cylinder's axis, the line defect is called a twist disclination. In the following, we will mostly focus on wedge disclinations.

\begin {figure}%%[H]
\centering
\includegraphics[width=0.8\textwidth]{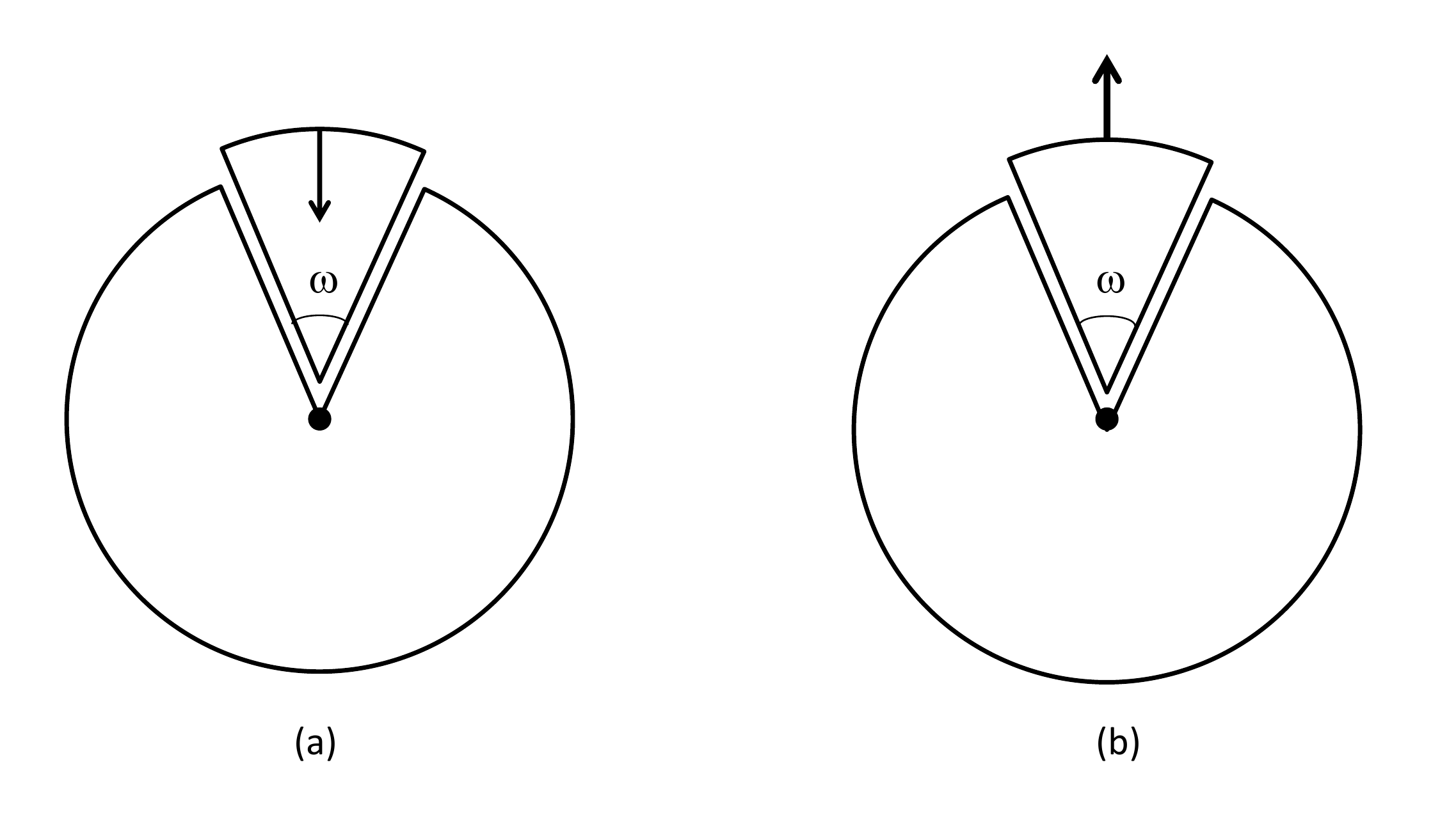}
\caption{A planar illustration for wedge disclinations. Figure (a) is a negative wedge disclination, where a wedge is inserted into a vertical cut causing compressive circumferential stress after `welding' the wedge to the body. Figure (b) is a positive wedge disclination, where the wedge is taken out of the original structure and the exposed faces welded together. $\omega$ is the wedge angle as well as the magnitude of the Frank vector. (Figure reproduced from \cite{nazarov2013disclinations} with permission from publisher of article under an open-access Creative Commons license).}\label{fig:pizza}
\end {figure}

A wedge disclination can be visualized easily \cite{nazarov2013disclinations}, as shown in Figure \ref{fig:pizza}. By taking away or inserting a wedge of an angle $\omega$, a positive or negative wedge disclination is formed. In Figure \ref{fig:pizza}(a) is a negative wedge disclination and (b) is a positive wedge disclination in a cylindrical body. After eliminating the overlap/gap-wedge and welding and letting the body relax, the body is in a state of internal stress corresponding to that of the wedge disclination (of corresponding sign). 

\begin {figure}%%[H]
\centering
\subfigure[A positive wedge disclination with a gap-wedge between two orientations.  The red dot is the positive wedge disclination core where the interface of the orientation-discontinuity terminates.]{
\includegraphics[width=0.45\textwidth]{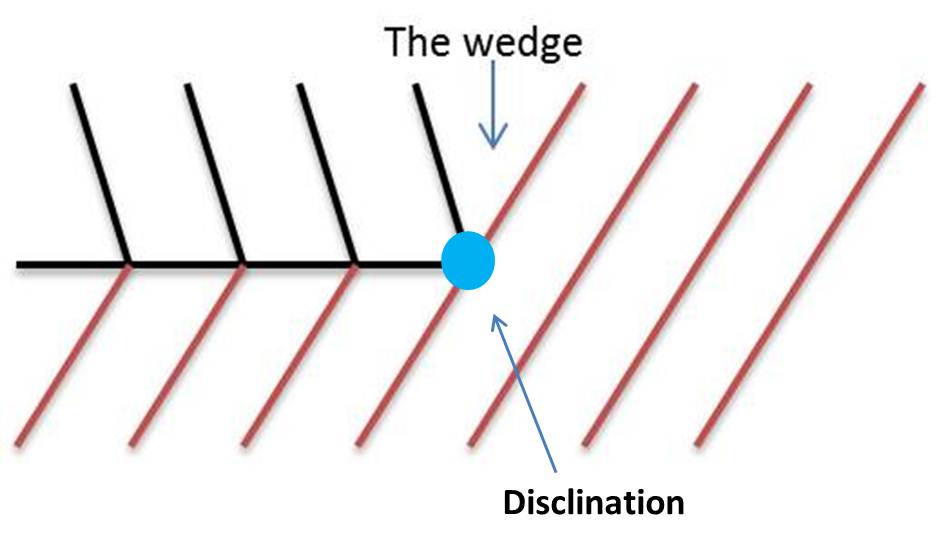}
\label{fig:single_disclination_1}
}\qquad
\subfigure[A negative wedge disclination with an overlap-wedge between two orientations.  The green dot is the negative wedge disclination core where the interface of the orientation-discontinuity terminates.]{
\includegraphics[width=0.45\textwidth]{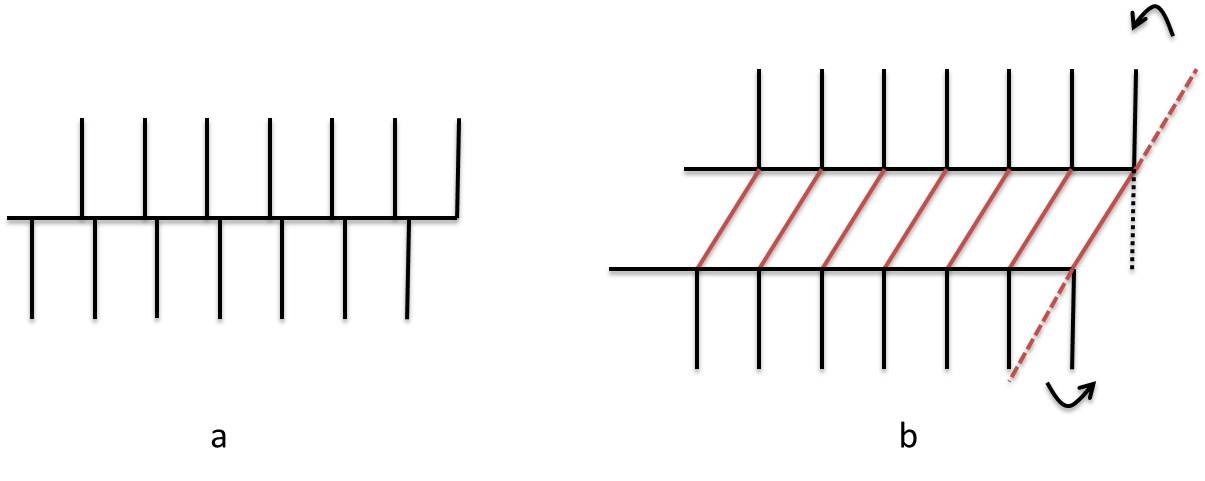}
\label{fig:single_disclination_2}
}
\caption{An elastic distortion based description of wedge disclinations.}\label{fig:single_disclination}
\end {figure}

In this work, we introduce a description for the disclination configuration based on the elastic distortion field, as shown in Figure \ref{fig:single_disclination}. In Figure \ref{fig:single_disclination}, red lines represent one elastic  distortion field (possibly represented by the Identity tensor); black lines represent another distortion field. Thus, there is a surface of discontinuity between these two distortion fields and a terminating line (which is a point on the 2-d plane) on the interface is called a disclination. Also, there is a gap-wedge between the red part and the black part as shown in Figure \ref{fig:single_disclination_1}, indicating it as a positive disclination; an overlap-wedge in Figure \ref{fig:single_disclination_2} corresponds to a negative disclination. Since a gap-wedge is eliminated for a positive disclination, there is circumferential tension around the core. Similarly, there is circumferential compression around the core for the negative disclination because of the inserted wedge. These physical arguments allow the inference of some features of the internal state of stress around disclination defects without further calculation.

\begin {figure}
\centering
\includegraphics[width=0.6\textwidth]{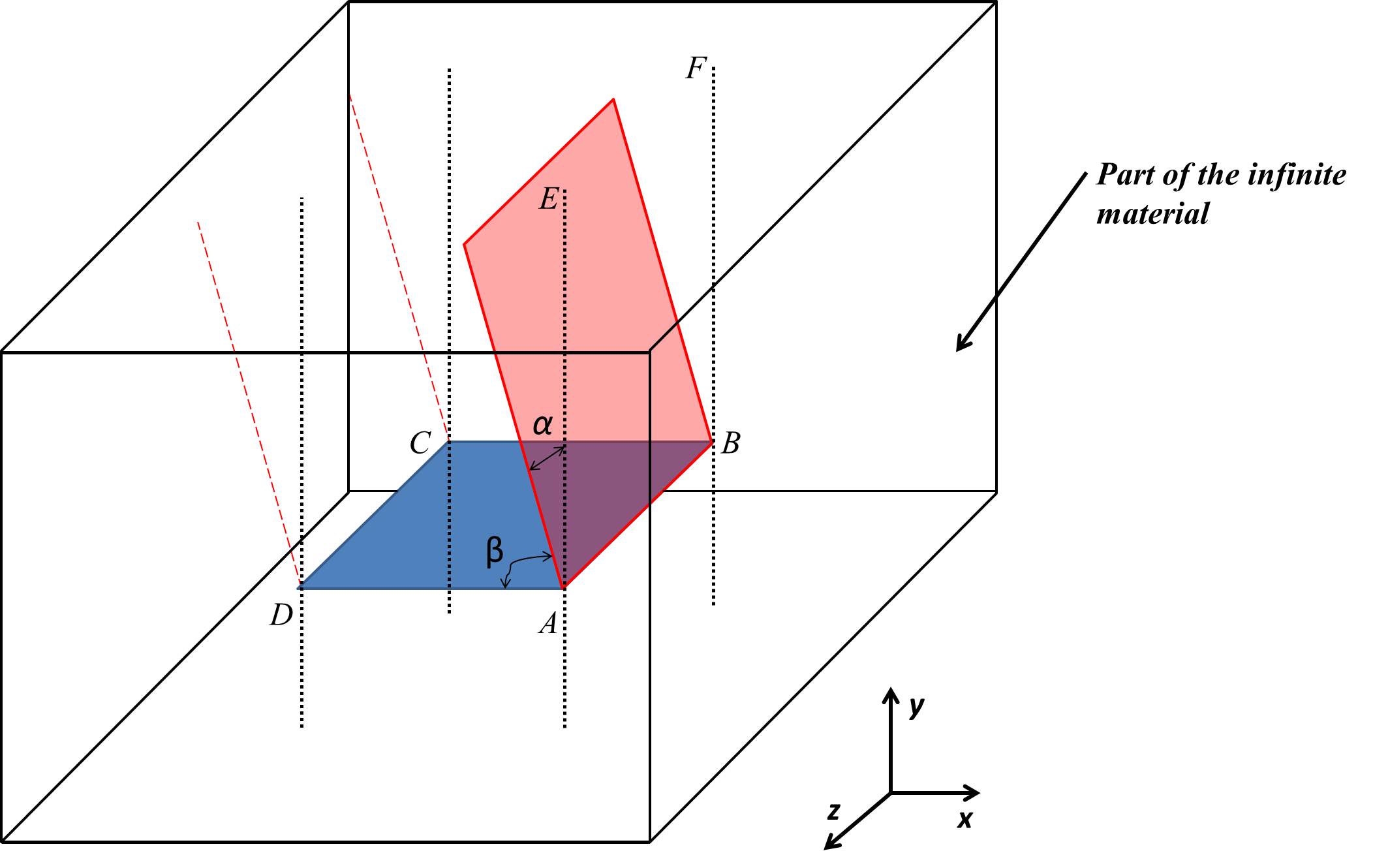}
\caption{A 3D description for a disclination loop in an infinite block. ABCD is the disclination loop in a parallelepiped. Wedge disclinations exist along AB and CD while twist disclinations exist along AD and BC.}\label{fig:loop_disclination}
\end {figure}

A disclination loop is formed if an inclusion of a crystal with one orientation, and in the shape of a parallelepiped with infinite length, is inserted in another infinite crystal of a different orientation, as shown in Figure \ref{fig:loop_disclination}. Focusing on the bottom surface of the parallelepiped, we consider the `exterior' crystal as having one set of atomic planes parallel to the $y-z$ plane bounded by unbounded black rectangles in Figure \ref{fig:loop_disclination}. The interior crystal has one set of planes at an angle of $\alpha$  to the $y-z$ plane. The line of intersection $AB$ represents a termination of a gap-wedge formed by the red plane of the interior crystal and the plane $ABFE$ of the exterior crystal. Because the misorientation vector is directed along line $AB$ ($z$ axis), the latter serves as a wedge disclination. Similarly, there is a wedge disclination along intersection line $CD$ of opposite sign to $AB$. For intersection lines $BC$ and $DA$, the misorientation vector is perpendicular to the direction of intersection lines and twist disclinations of opposite signs are formed along $BC$ and $DA$. The curve $ABCD$ forms a disclination loop in the body (on elimination of the gap and overlap wedges). 

\subsection{Disclination dipole formation and movement in a lattice}

\begin{figure}
\centering
\subfigure[Structure of a hexagonal lattice.]{
\includegraphics[width=0.4\textwidth]{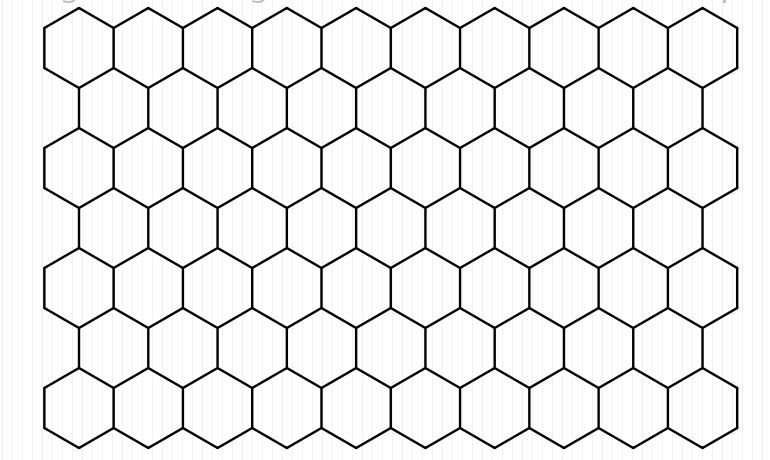}
\label{fig:disclination_dipole_1}
}\qquad
\subfigure[Break and rebuild atomic bonds to form a disclination dipole (pentagon-heptagon pair).]{
\includegraphics[width=0.4\textwidth]{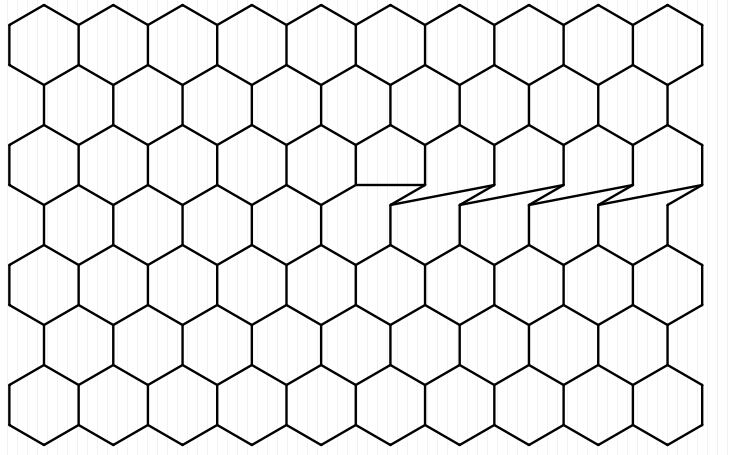}
\label{fig:disclination_dipole_2}
}
\subfigure[Relaxed configuration with a disclination dipole.]{
\includegraphics[width=0.4\textwidth]{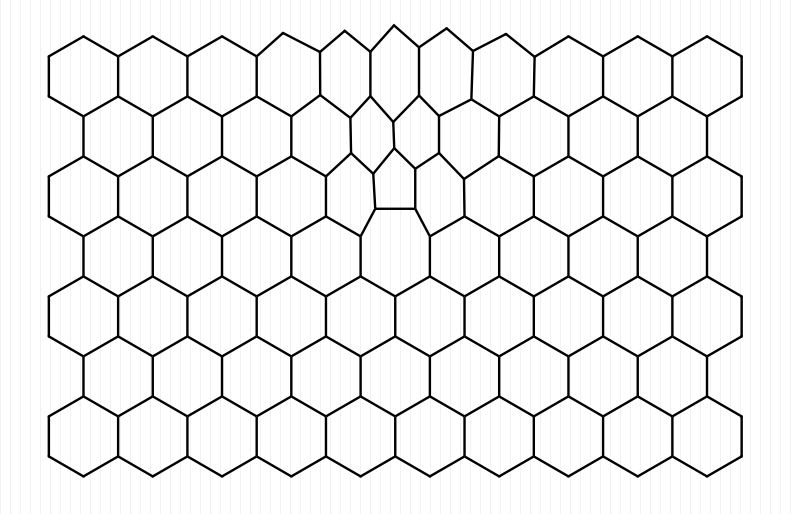}
\label{fig:disclination_dipole_3}
}
\caption{Kinematics of formation of a disclination-dipole in a hexagonal lattice. Figures constructed with \textit{Chemdoodle}\cite{chem}.}
\label{fig:disclination_dipole}
\end {figure}

Due to the addition and subtraction of matter over large distances involved in the definition of a disclination in the interior of a body, it is intuitively clear that a single disclination should cause long-range elastic stresses, which can also be seen from the analytical solution given in Section \ref{sec:disclination_intro}. Thus, a disclination rarely exists alone. Instead, usually, disclinations appear in pairs in the form of dipoles, namely a pair(s) of disclinations with opposite signs. Figure \ref{fig:disclination_dipole} shows a schematic of how a disclination dipole can form in a hexagonally coordinated structure. Figure \ref{fig:disclination_dipole_1} is the original structure with a hexagonal lattice; Figure \ref{fig:disclination_dipole_2} shows how bonds can be broken and rebuilt to transform a hexagon pair to a pentagon-heptagon pair in a topological sense (this may be thought of as a situation before relaxation); Figure \ref{fig:disclination_dipole_3} presents the relaxed configuration with a disclination dipole (the penta-hepta pair) after the transformation. 

In a stress-free hexagonal lattice, removing an edge of a regular hexagon to form a pentagon can be associated with forming a positive wedge disclination at the center of the regular polygon (due to the tensile stress created in the circumferential direction); similarly, adding an edge to form a heptagon may be considered the equivalent of forming a negative wedge disclination. Hence, a heptagon-pentagon pair in a nominally hexagonal lattice is associated with a disclination dipole. It should be clear by the same logic that in a lattice with regular $n$-sided repeat units, an $(n-1)-(n+1)$ polygon pair may be viewed as a disclination dipole.

\begin{figure}
\centering
\subfigure[Structure of a hexagonal lattice with a disclination dipole.]{
\includegraphics[width=0.4\textwidth]{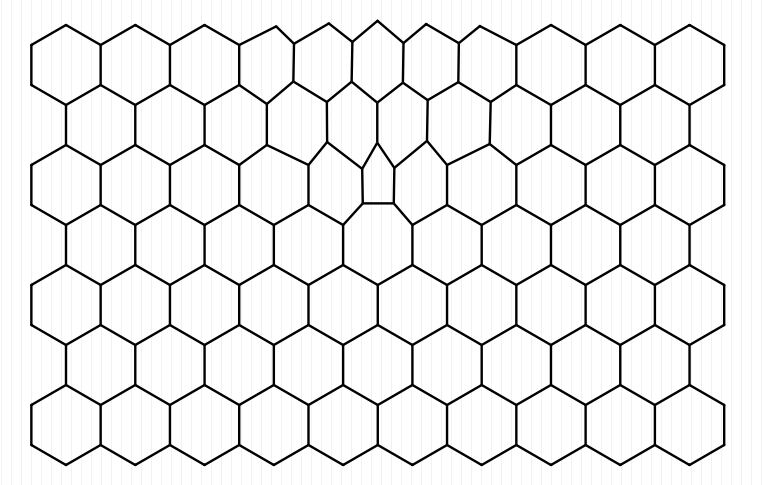}
\label{fig:dipole_movement_1}
}\qquad
\subfigure[Break and rebuild atomic bonds to move a disclination dipole.]{
\includegraphics[width=0.4\textwidth]{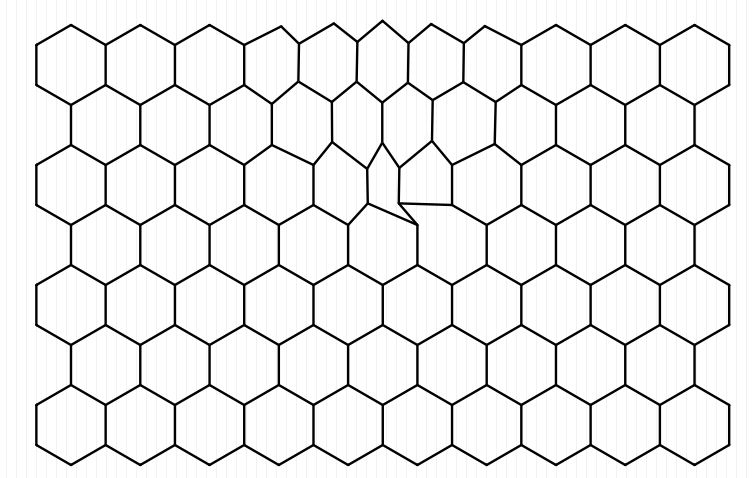}
\label{fig:dipole_movement_2}
}
\subfigure[Relaxed configuration with  disclination dipole having moved through the material to the right.]{
\includegraphics[width=0.4\textwidth]{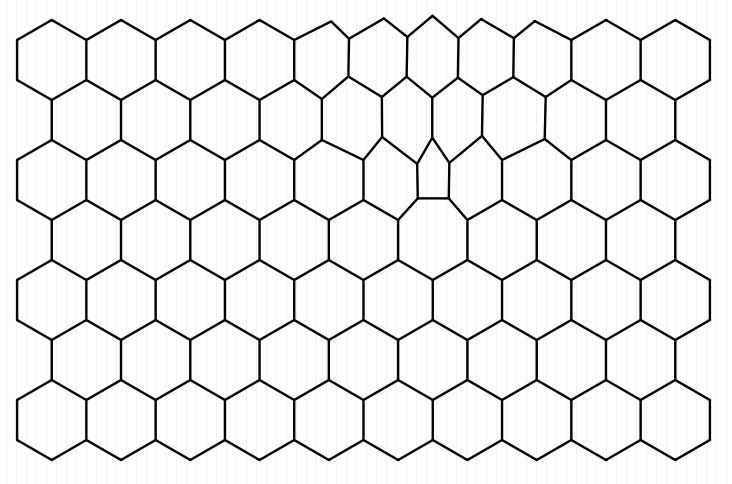}
\label{fig:dipole_movement_3}
}
\caption{Kinematics of motion of a disclination dipole. Figures constructed with \textit{Chemdoodle}\cite{chem}.}
\label{fig:dipole_movement}
\end {figure}

Figure \ref{fig:dipole_movement} shows how a disclination dipole moves by local crystal rearrangement under some external force. Figure \ref{fig:dipole_movement_1} shows the configuration for a hexagonal lattice with a disclination dipole; then some atomic bonds nearby are broken and rebuilt in Figure \ref{fig:dipole_movement_2}; Figure \ref{fig:dipole_movement_3} shows the relaxed configuration, where the disclination dipole has moved the right. The movement of a disclination dipole is a local rearrangement instead of a global rearrangement required to move a single disclination.

\subsection{Descriptions of a dislocation by a (g.)disclination dipole}\label{sec:dislocation_des}
\begin {figure}
\centering
\subfigure[A perfect crystal structure, where the black lines represent atomic planes.]{
\includegraphics[width=0.3\textwidth]{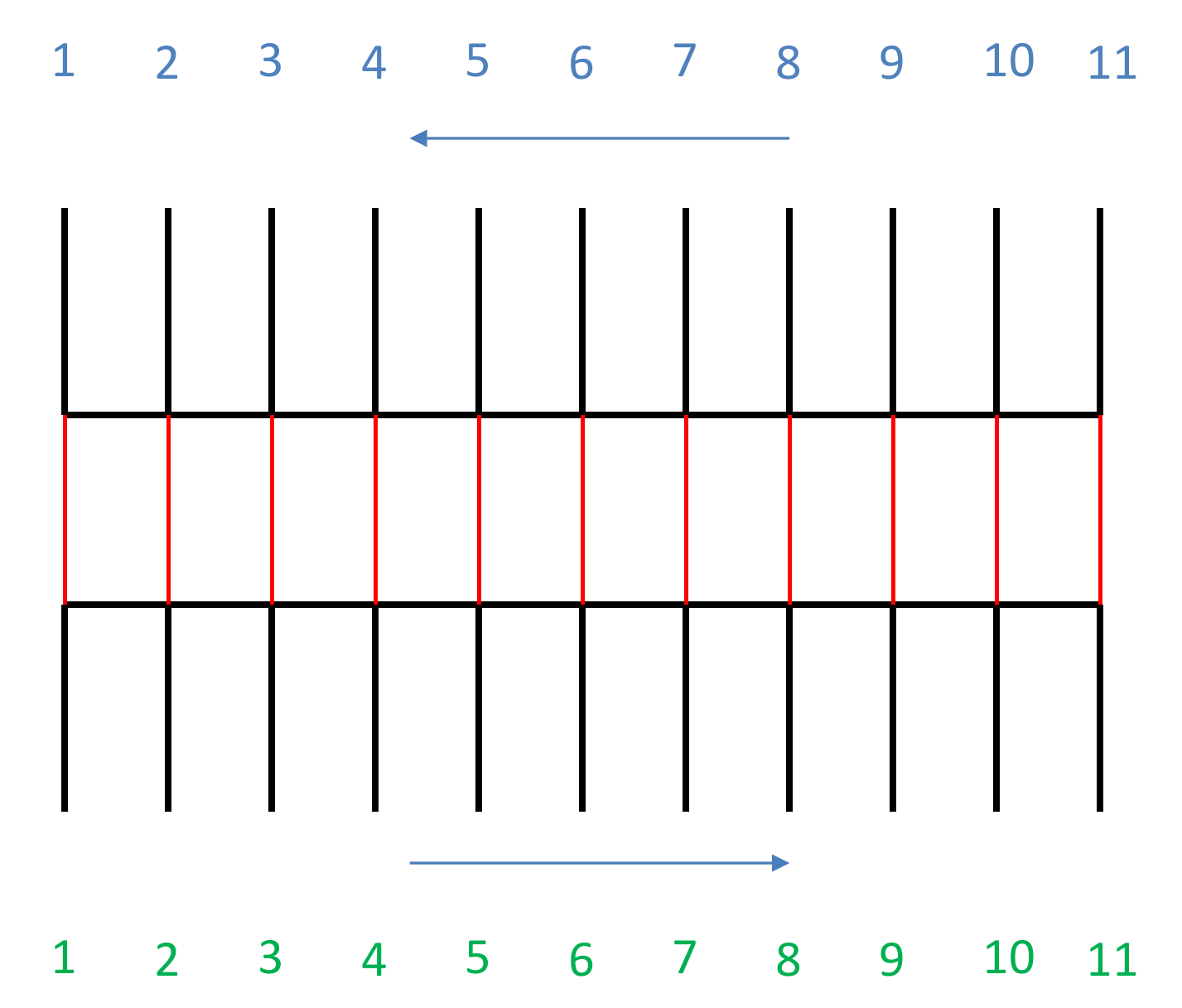}
\label{fig:dislocation_1}
}\qquad
\subfigure[Half-planes of atoms ($1 - 7$) in the top-block change topological connections to their counterparts in the bottom block on shearing, resulting in the appearance of an `extra' half-plane in the bottom block. No extra atoms are introduced in the structure.]{
\includegraphics[width=0.3\textwidth]{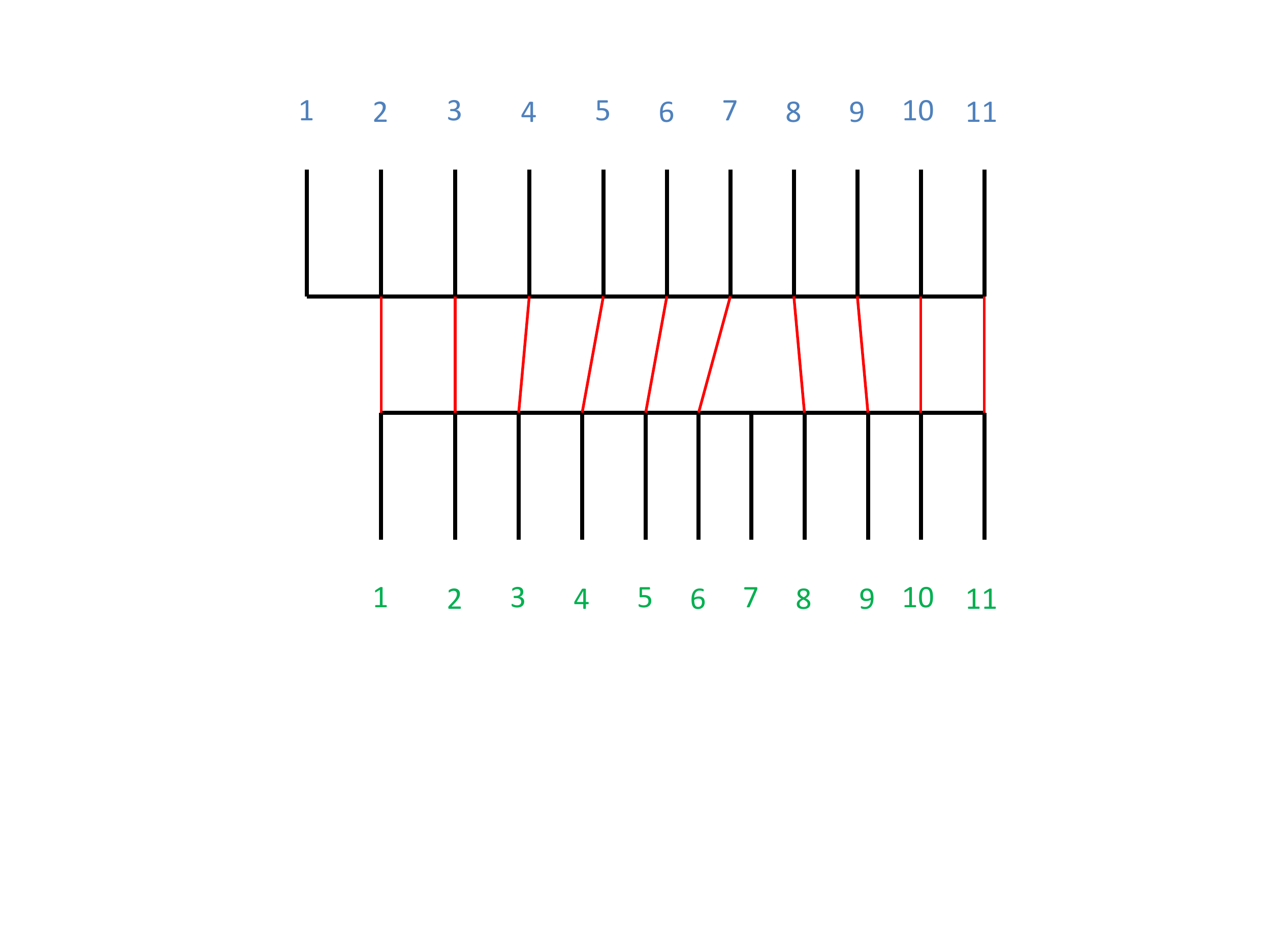}
\label{fig:dislocation_2}
}\\
\subfigure[Interpretation of defected structure as a disclination dipole. The red and green dots represent positive and negative wedge disclinations, respectively.]{
\includegraphics[width=0.3\textwidth]{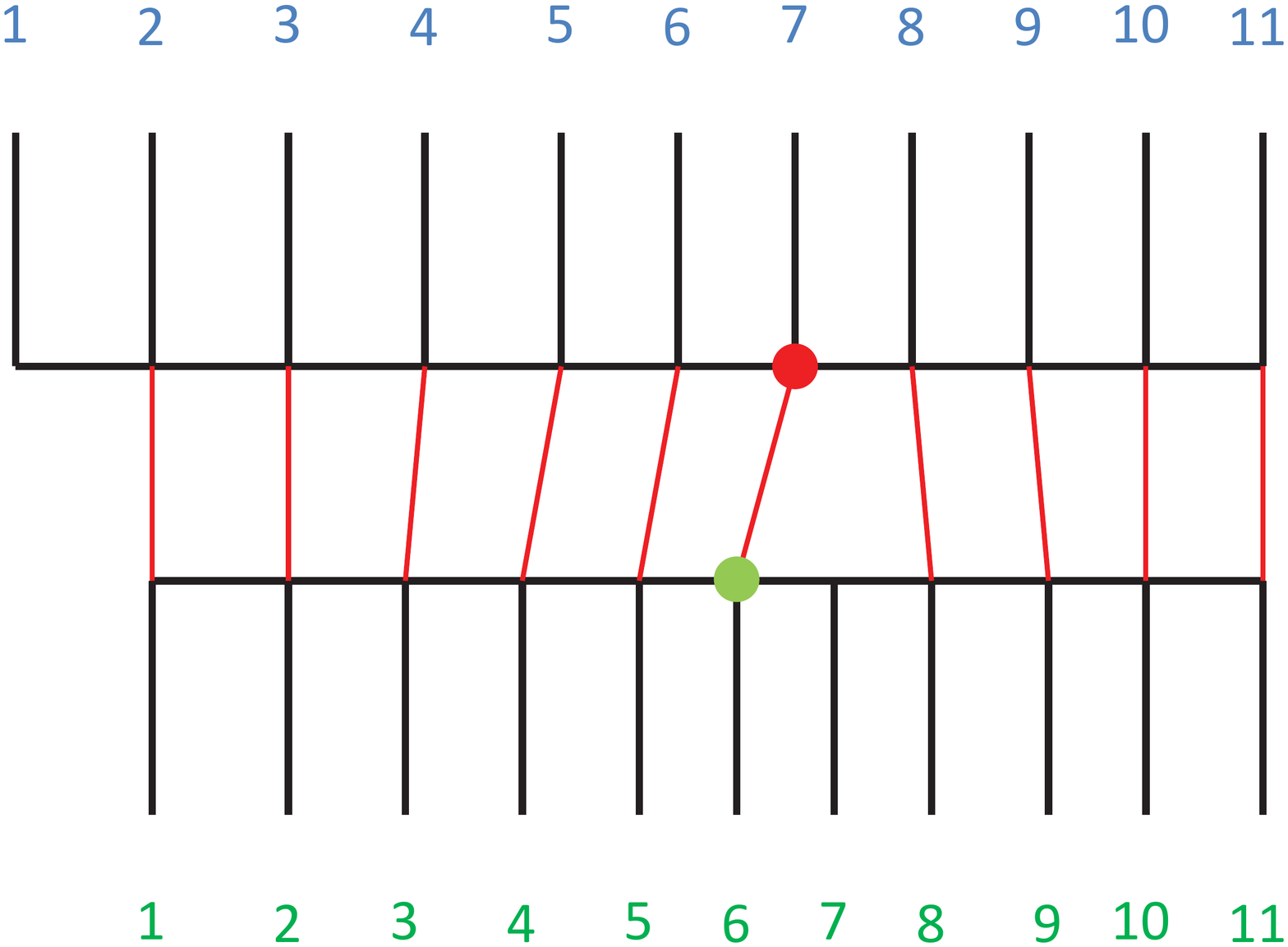}
\label{fig:dislocation_3}
}\qquad
\subfigure[Disclination dipole in (c) viewed at a larger length scale (weaker resolution). The disclination dipole appears as an edge dislocation.]{
\includegraphics[width=0.3\textwidth]{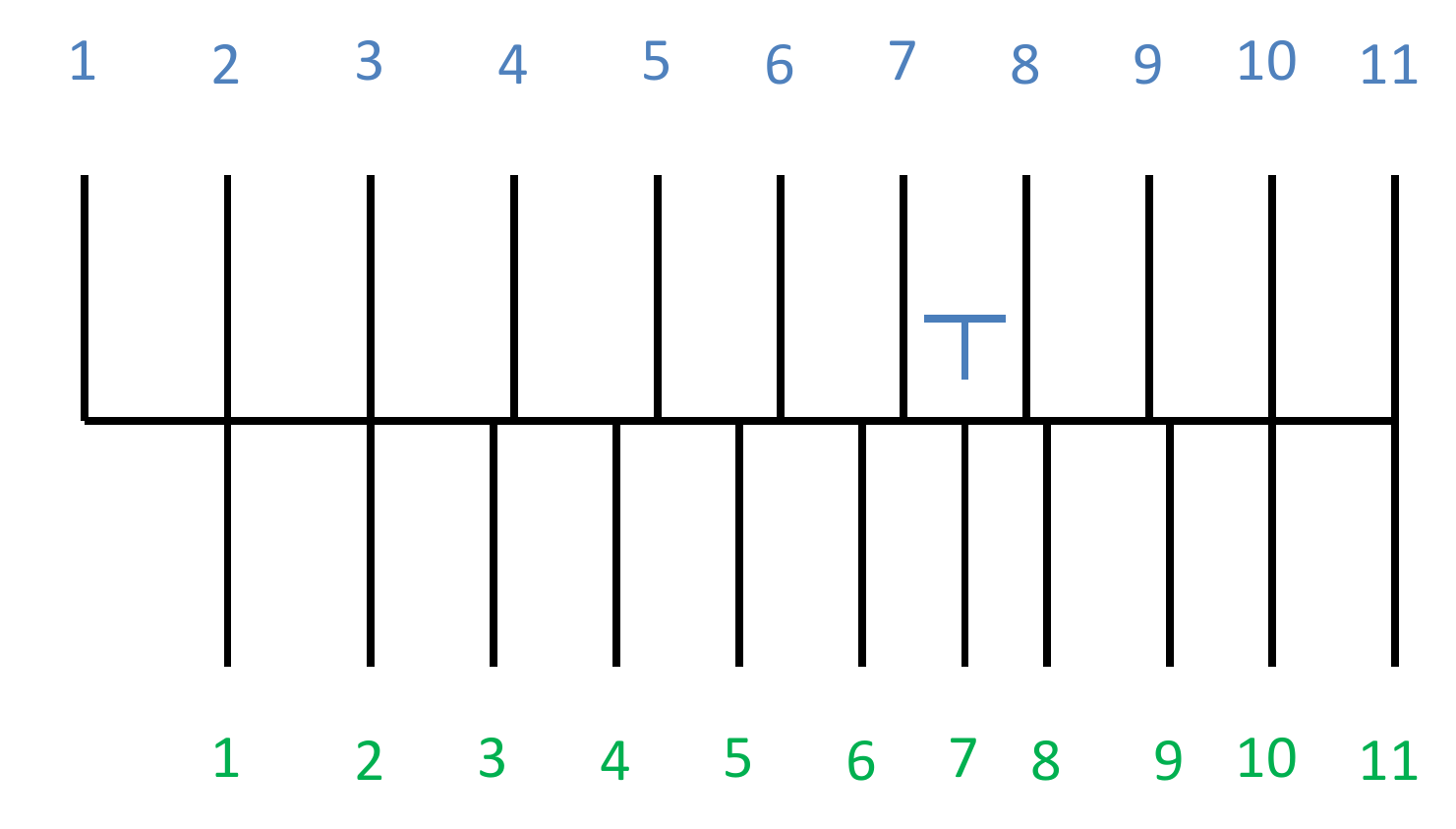}
\label{fig:dislocation_4}
}
\caption{Interpretation of a wedge disclination dipole as an edge dislocation.}
\label{fig:dislocation}
\end {figure}

In this section we consider two physically distinct constructions that motivate why a straight edge dislocation may be thought of as being closely related to a (g.)disclination dipole.  Figure \ref{fig:dislocation_1} is a perfect crystal structure. Black lines represent atomic planes and red lines are the atomic bonds between two horizontal atomic planes. We apply a shear on the top and the bottom of this body along the blue arrows shown in Figure \ref{fig:dislocation_1}. After shearing, an extra half plane is introduced in the bottom part, as shown in Figure \ref{fig:dislocation_2}. This dislocation can as well be interpreted as a disclination dipole; a positive disclination (the red dot in Figure \ref{fig:dislocation_3}) exists in the top part and a negative disclination (the green dot in Figure \ref{fig:dislocation_3}) exists in the bottom part. Figure \ref{fig:dislocation_4} shows a zoomed-out macroscopic view of the final configuration with an extra half-plane (of course obtained by a  process where no new atoms have been introduced). Thus, a dislocation can be represented as a disclination dipole with very small separation distance. The Burgers vector of the dislocation is determined by the misorientation of the disclinations as well as the interval distance, as discussed in detail in Section \ref{sec:burgers}. The upper disclination has a gap-wedge, namely a positive disclination, while the lower disclination has an overlap-wedge which is a negative disclination. Thus, the upper part is under tension and the bottom part under compression, consistent with the dislocation description with an `extra' half-plane in the bottom part. Our rendition here is a way of understanding how a two-line, two-rotation axes disclination dipole \cite{romanov2009application} results in an edge dislocation in the limit of the distance between the two planes vanishing.

\begin {figure}
\centering
\subfigure[A defected grain boundary.]{
\includegraphics[width=0.35\textwidth]{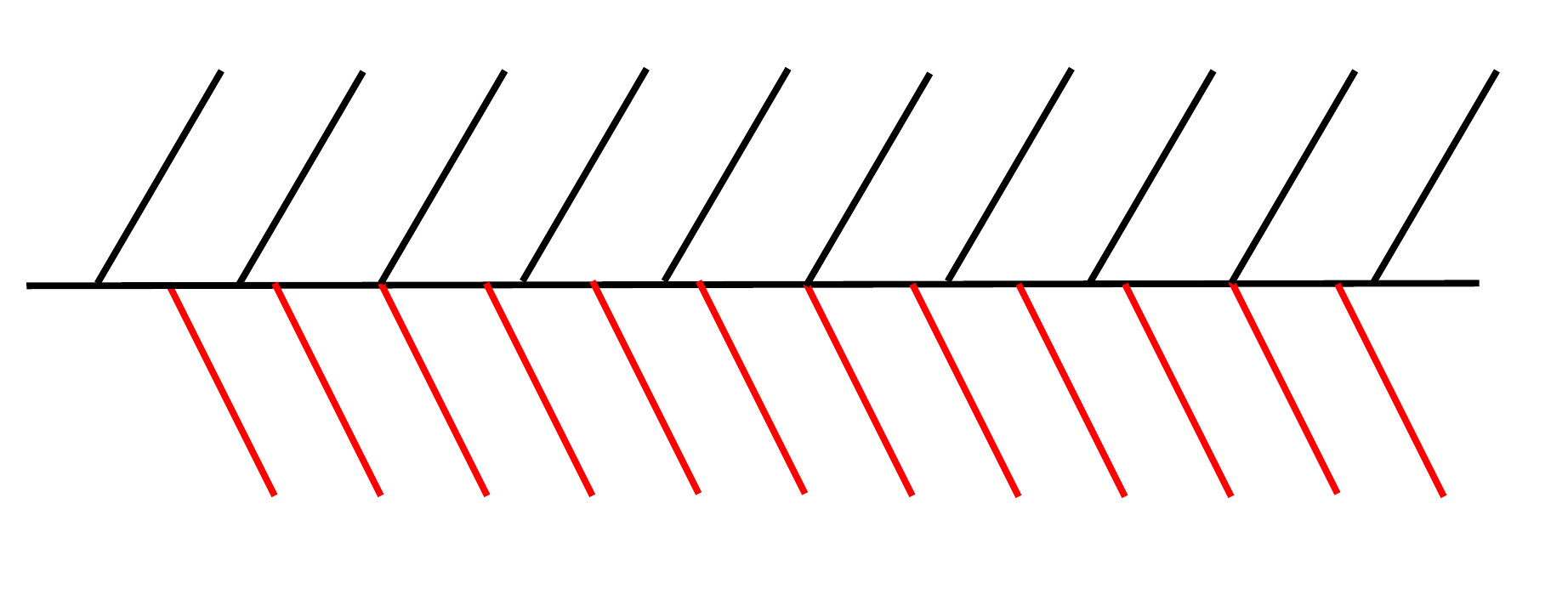}
\label{fig:dipole_grain_1}
} \qquad
\subfigure[A disclination dipole representing one defect of the grain boundary.]{
\includegraphics[width=0.4\textwidth]{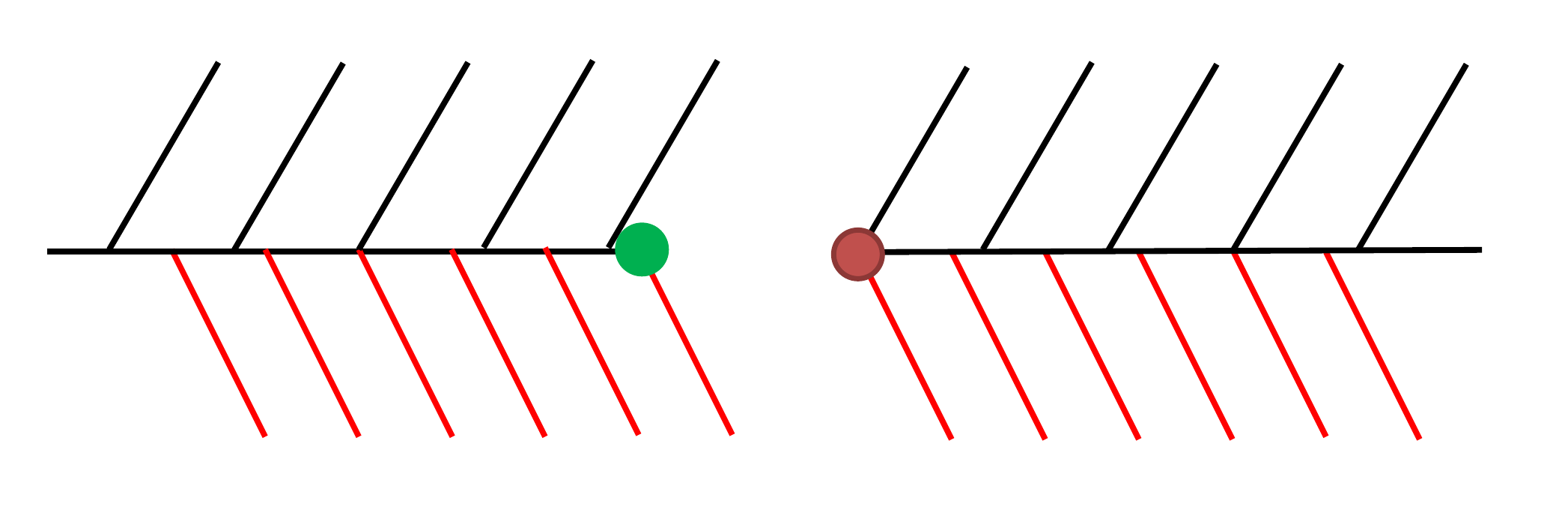}
\label{fig:dipole_grain_2}
}
\subfigure[Stress $\sigma_{yy}$ around a single defect in the grain boundary, calculated from the (g.)disclination dipole model.]{
\includegraphics[width=0.4\textwidth]{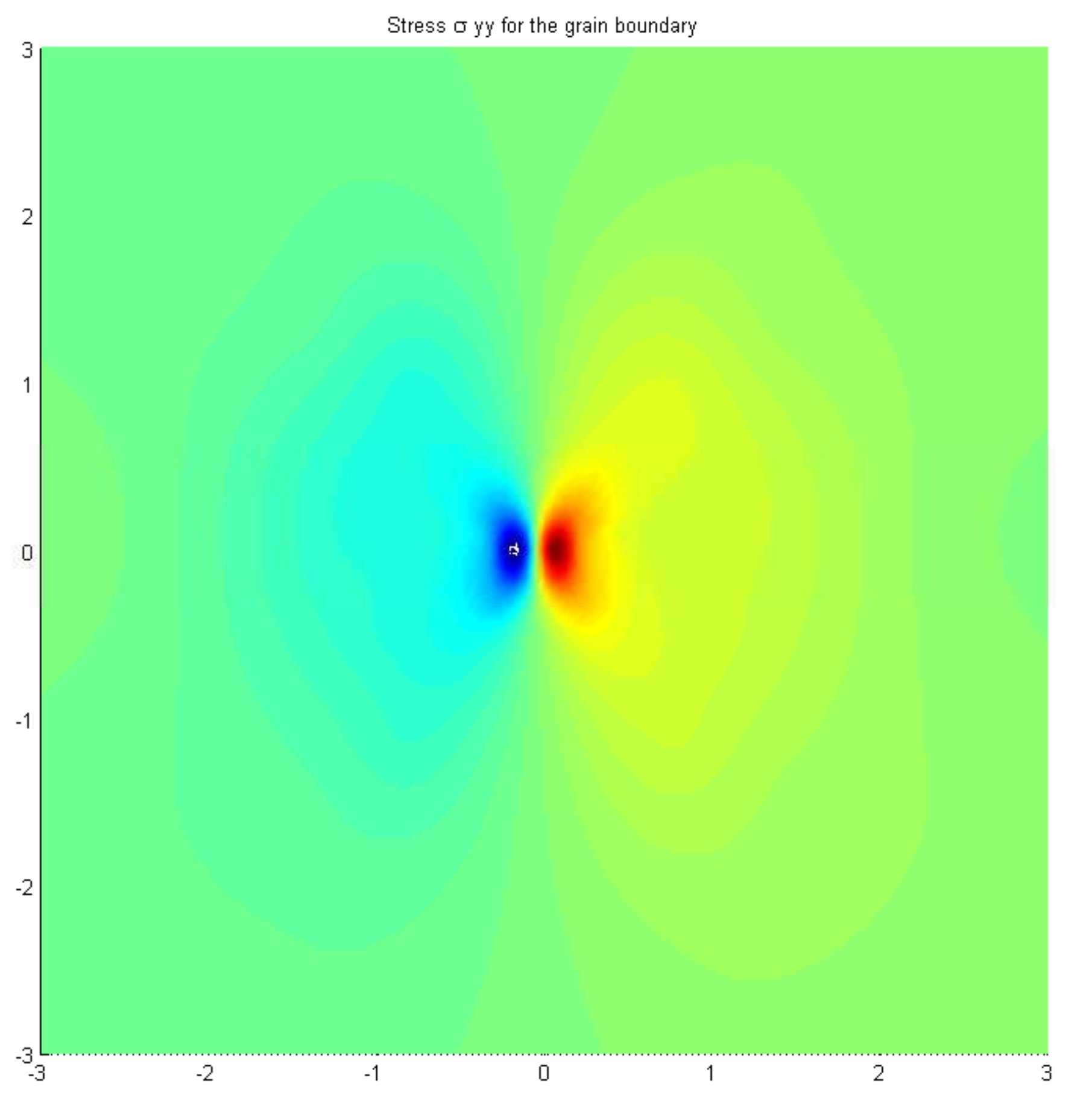}
\label{fig:dipole_grain_3}
}
\caption{The disclination dipole description of a defect in the grain boundary.}
\label{fig:grain_boundary}
\end {figure}

Another way in which a dislocation can be associated with a disclination dipole is one that is related to the description of incoherent grain boundaries. Figure \ref{fig:dipole_grain_1} is an incompatible grain boundary represented by orientation fields, where black and red lines represent two different orientations. In Figure \ref{fig:dipole_grain_2},  the grain boundary interface is cut in two parts and the cut points are treated as a disclination dipole; the red dot is the positive disclination and the green dot is the negative disclination. In contrast to the description in Fig. \ref{fig:dislocation}, here the discontinuity surfaces being terminated by the disclinations are coplanar. In Figure \ref{fig:dipole_grain_2}, the disclination on the left is of negative strength while the disclination on the right is positive. It is to be physically expected  that the disclination on the left of the dipole produces a compressive stress field in the region to the left of the dipole. Similarly, the disclination on the right of the pair should produce a tensile stress field to the right of the dipole. Figure \ref{fig:dipole_grain_3} is the stress field for the grain boundary in Figure \ref{fig:grain_boundary}(a) modeled by a single disclination dipole through a numerical approximation of a theory to be described in Section \ref{sec:g_disclination_theory}. Indeed, the calculation bears out the physical expectation - the blue part represents a region with compressive stress and the red part a region with tensile stress. The stress field may be associated with that of an edge dislocation with Burgers vector in the vertical direction with an extra half plane of atoms in the right-half plane of the figure.  This description of a dislocation by a disclination dipole is a way of understanding a single-line, two-rotation-axes dipole \cite{romanov2009application}.

%%%%
%%%%

\subsection{Grain boundaries via (g.)disclinations} \label{sec:disclination_grain_des}
\begin {figure}
\centering
\subfigure[Two grains with different crystal structures.]{
\includegraphics[width=0.4\textwidth]{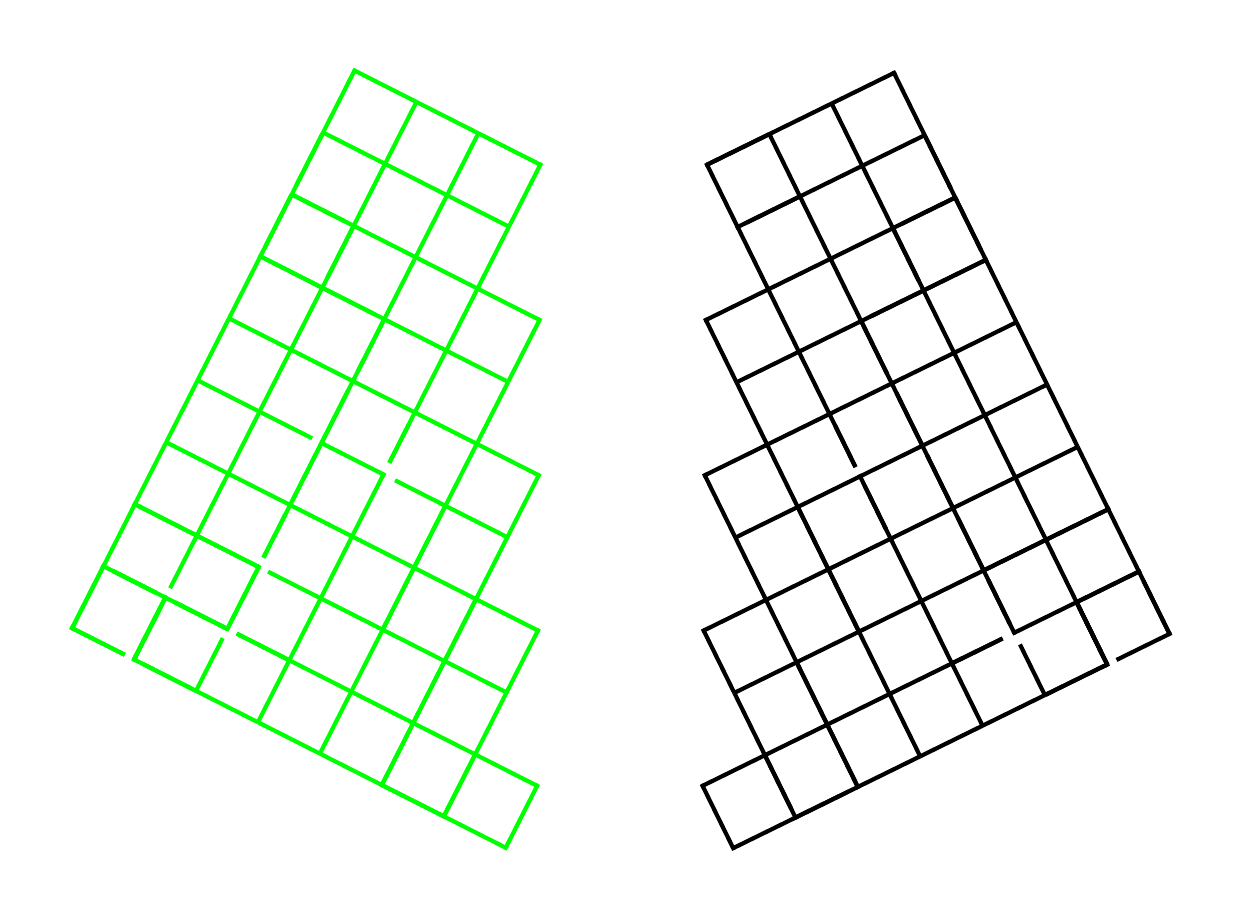}
}\qquad
\subfigure[Bi-crystal after merging two grains together.]{
\includegraphics[width=0.35\textwidth]{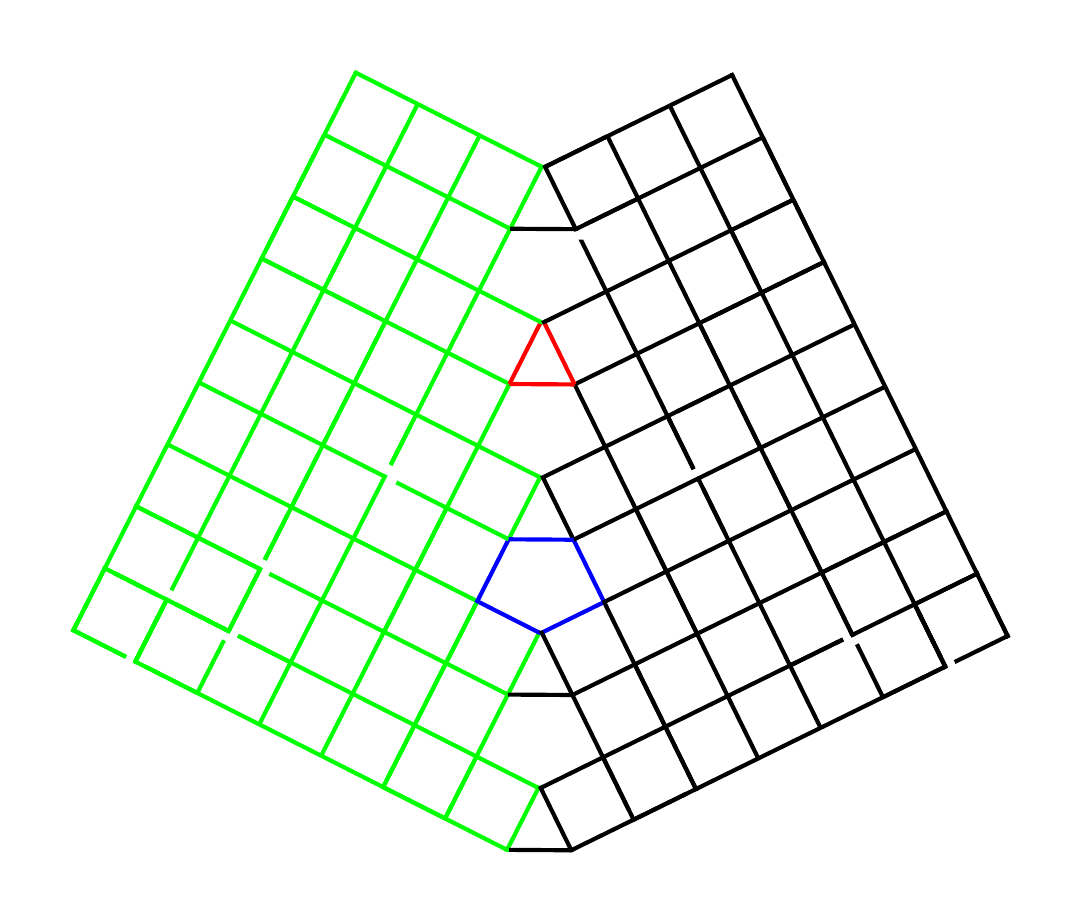}
}
\caption{Schematic of forming a series of disclination dipoles in a grain boundary. The red triangle is a positive disclination while the blue pentagon is a negative disclination. The pentagon-triangle disclination dipoles (in a 4-coordinated medium) exist along the interface.}\label{fig:grain_boundary2}
\end {figure}

We have already seen in the last section that a disclination-dipole model can be relevant to modeling the geometry and mechanics of grain boundaries.  Figure \ref{fig:grain_boundary2} motivates how disclination dipoles arise naturally in the idealized description  of a grain boundary from a microscopic view. In Figure \ref{fig:grain_boundary2}(a), there are two grains with different orientations. After putting these two grains together and connecting the adjacent atomic bonds, we form a grain boundary, as shown in Figure \ref{fig:grain_boundary2}(b). There exists a series of disclination dipoles along the boundary, as shown in Figure \ref{fig:grain_boundary2}(b) where a blue pentagon is a negative disclination while a red triangle is a positive disclination. 

\begin {figure}
\centering
\includegraphics[width=0.5\textwidth]{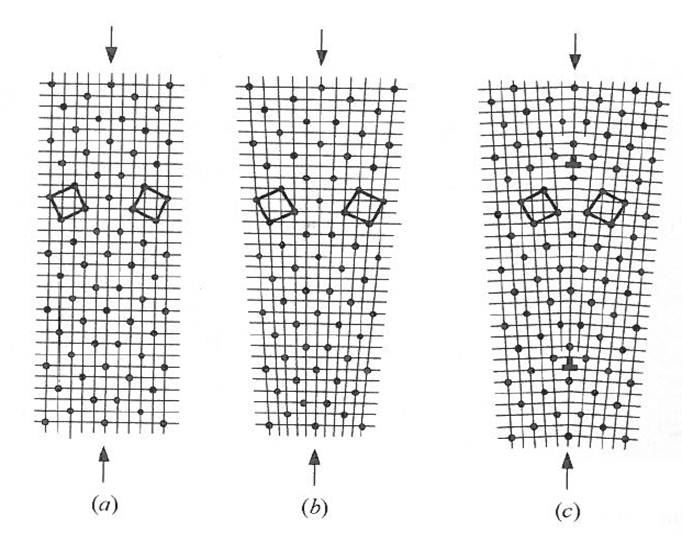}
\caption{(a) A common tilt grain boundary with 53.1 degree misorientation. (b) The configuration after applying additional tilt angle without any rearrangement. (c) The configuration with some dislocations introduced along the interface to eliminate far field deformation. (Figures reproduced from \cite{balluffi2005kinetics} with permission from John Wiley and Sons).}
\label{fig:grain_boundary3}
\end {figure}

In some cases, grain boundaries involve other types of defects beyond disclination dipoles, such as dislocations. Figure \ref{fig:grain_boundary3} is an example of a vicinal crystal interface \cite{balluffi2005kinetics}, which consists of a combination of dislocations  and disclinations along the interface. In Figure \ref{fig:grain_boundary3}(a) a high-angle tilt boundary with a tilt of $53.1^\circ$  is viewed along the $<100>$ tilt axis.  If we slightly increase the tilt angle while keeping the topology of bond connections near the boundary fixed,  high elastic deformations are generated, as shown in the Figure \ref{fig:grain_boundary3}(b). Instead, an array of dislocations is often observed along the boundary as shown in the configuration Figure \ref{fig:grain_boundary3}(c), presumably to eliminate long-range elastic deformations. In \cite{zhang_acharya_puri}, we calculate the elastic fields of such boundaries utilizing both g.disclinations and dislocations.

\subsubsection{Relationship between the disclination and dislocation models of a low-angle grain boundary} \label{sec:relation_disclination_dislocation}

Normally, a grain boundary is modeled by an array of dislocations. As discussed in Section \ref{sec:disclination_intro}, a dislocation model cannot deal with a high-angle grain boundary. An alternative is to interpret the grain boundary through a disclination model as we discussed above. In this section, the relation between the disclination and dislocation models for a low-angle grain boundary is explained. 

\begin{figure}
  \centering
  \subfigure[The perfect crystal configuration.]{
    \includegraphics[height=0.35\textheight]{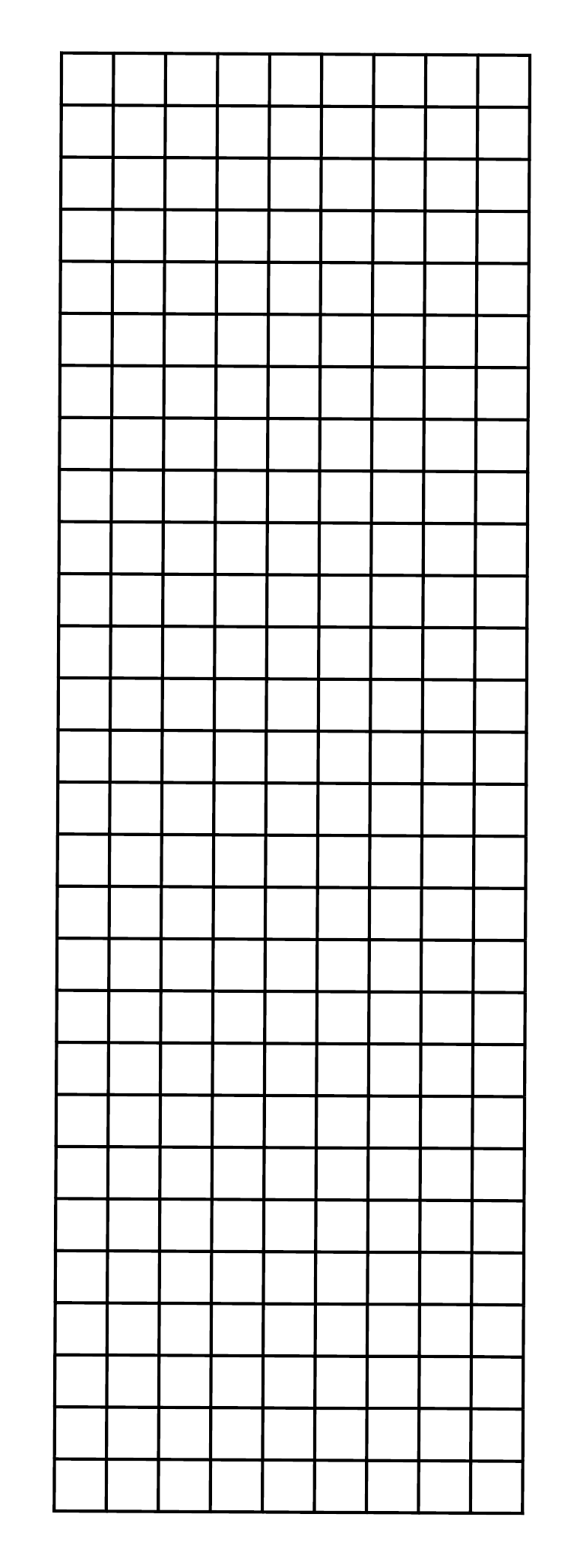}
    \label{fig:relation_1} 
  }\qquad
  \subfigure[Cut the material into four parts; introduce a positive disclination at the top and a negative disclination at the bottom.]{
    \includegraphics[height=0.35\textheight]{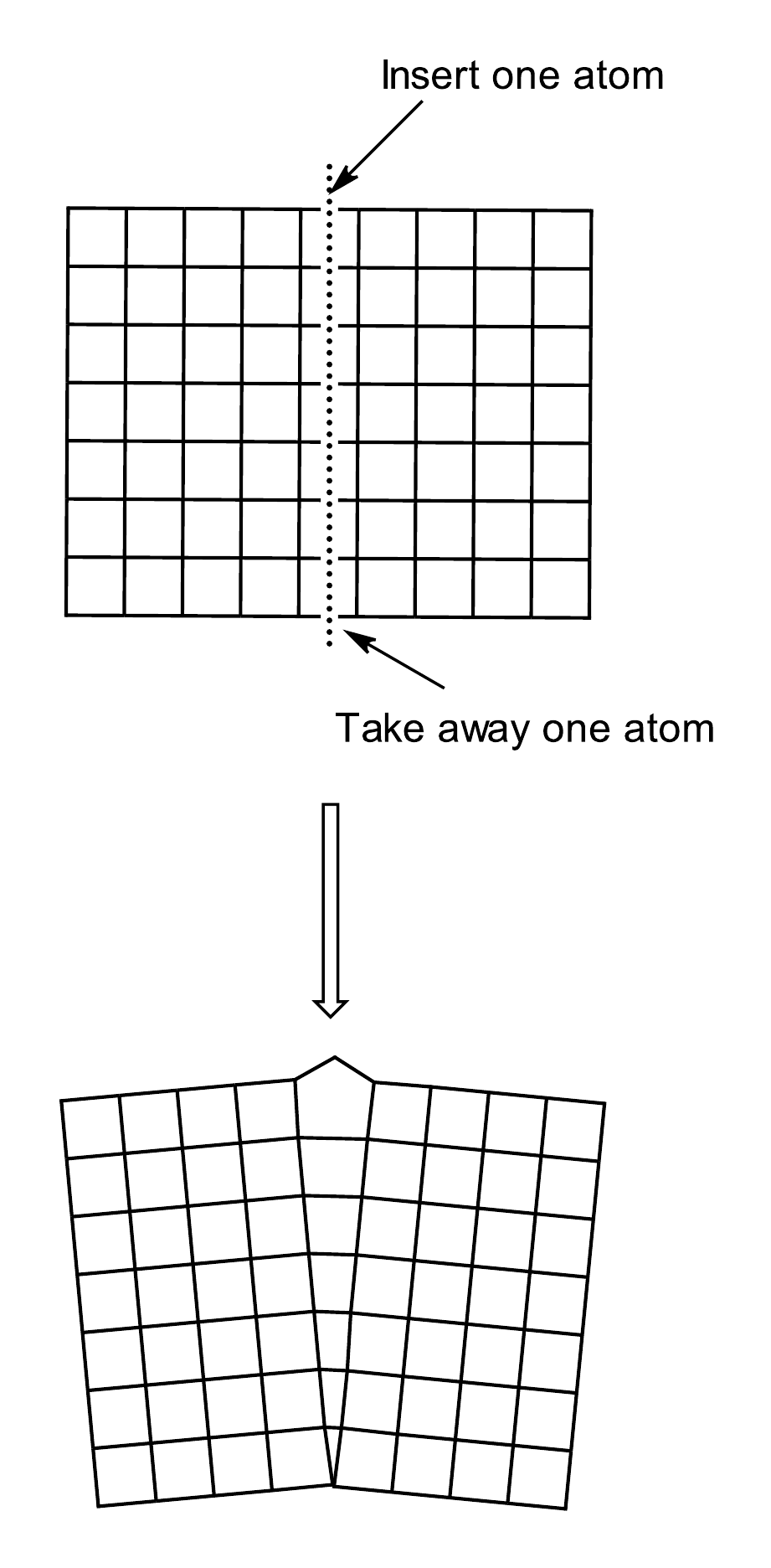}
    \label{fig:relation_2} 
 } \\
  \subfigure[Repeat the same procedure for all four parts.]{
    \includegraphics[height=0.35\textheight]{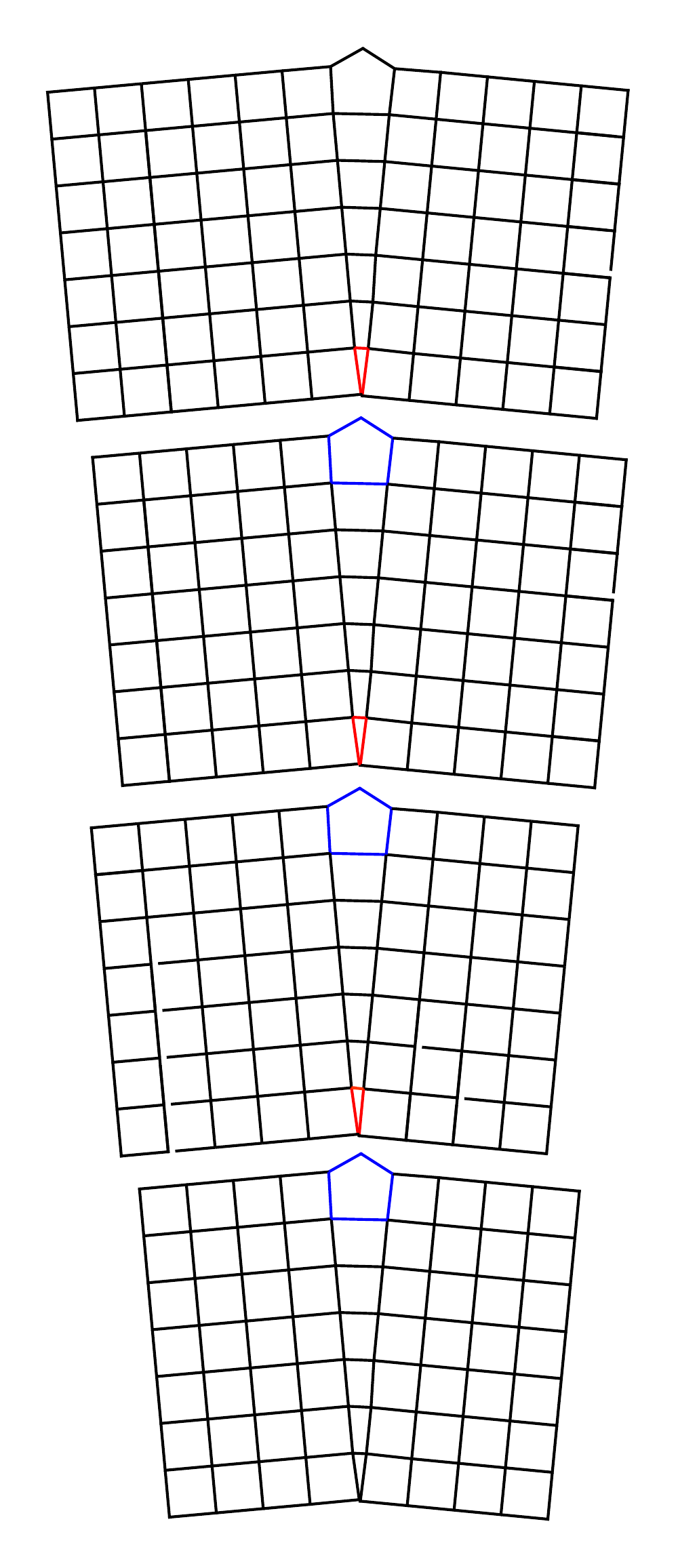}
    \label{fig:relation_3} 
 }\qquad
  \subfigure[Weld four parts together and form a grain boundary whose interface is shown as the dotted blue line.]{
    \includegraphics[height=0.35\textheight]{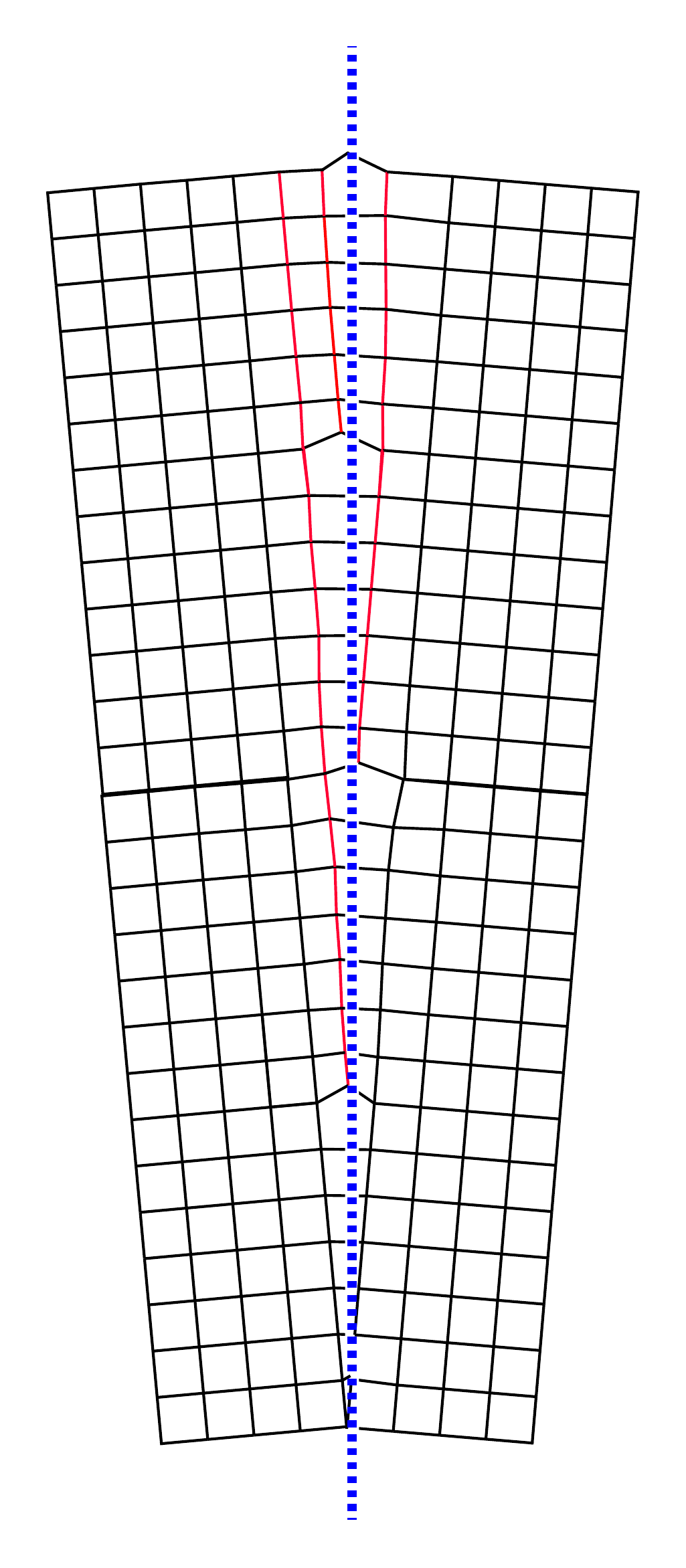}
    \label{fig:relation_4} 
 }
  \caption{Schematic of how disclination dipoles may fade into dislocations along a grain boundary.}
  \label{fig:relation} 
\end{figure}

Consider a defect-free crystal as shown in Figure \ref{fig:relation_1}. First, we horizontally cut the material into four parts, as shown in the upper configuration in Figure \ref{fig:relation_2}. Now, for every part, we cut the material along its center surface (the dashed line in Figure \ref{fig:relation_2}), insert one atom at the top and take away one atom from the bottom, weld the two half parts together again and relax the material. Then the configuration for every part will become the configuration at the bottom in Figure \ref{fig:relation_2}. By inserting and taking away atoms, we generate a negative disclination at the top and a positive disclination at the bottom. Repeating the same procedure for the remaining three parts, we finally obtain four parts with configurations as in Figure \ref{fig:relation_3}. The blue pentagons are negative disclinations while the red triangles are positive disclinations; a pair of a blue pentagon and an adjacent red triangle forms a disclination dipole. Next, we weld back these four parts and relax the whole material. Finally, a crystal configuration as in Figure \ref{fig:relation_4} is generated, which is a grain boundary with the boundary interface shown as the blue dashed line. Along the grain boundary, dislocations exist along the interface with extra atomic planes shown as red lines in Figure \ref{fig:relation_4}. When the pentagon-triangle (5-3) disclination dipole is brought together to form a dislocation, the pentagon-triangle structure actually disintegrates and becomes a pentagon-square-square (5-4-4) object. Thus, a grain boundary can be constructed from a series of disclination dipoles; at the same time, we can see dislocation structures at the grain boundary interface. It is as if the 5-3 disclination dipole structure fades into the dislocation structure on coalescing the two disclinations in a dipole.

A comparison between the dislocation model and the disclination model has also been discussed in \cite{li1972disclination} where the possibility of modeling a dislocation by a disclination dipole is proposed within the context of the theory of linear elasticity. In this paper, we have elucidated the physical picture of forming a dislocation from a disclination dipole and, in subsequent sections, we also derive the general relationship between the Burgers vector of a dislocation  and the disclination dipole for  both the 
small and finite deformation cases, capitalizing crucially on a g.disclination formulation of a disclination dipole.

\section{The Burgers vector of a disclination dipole in linear elasticity}\label{sec:burgers_lin}

\begin{figure}
\centering
\subfigure[A positive disclination located at the coordinate origin with a field point located at $\bfr$.]{
\includegraphics[width=0.45\textwidth]{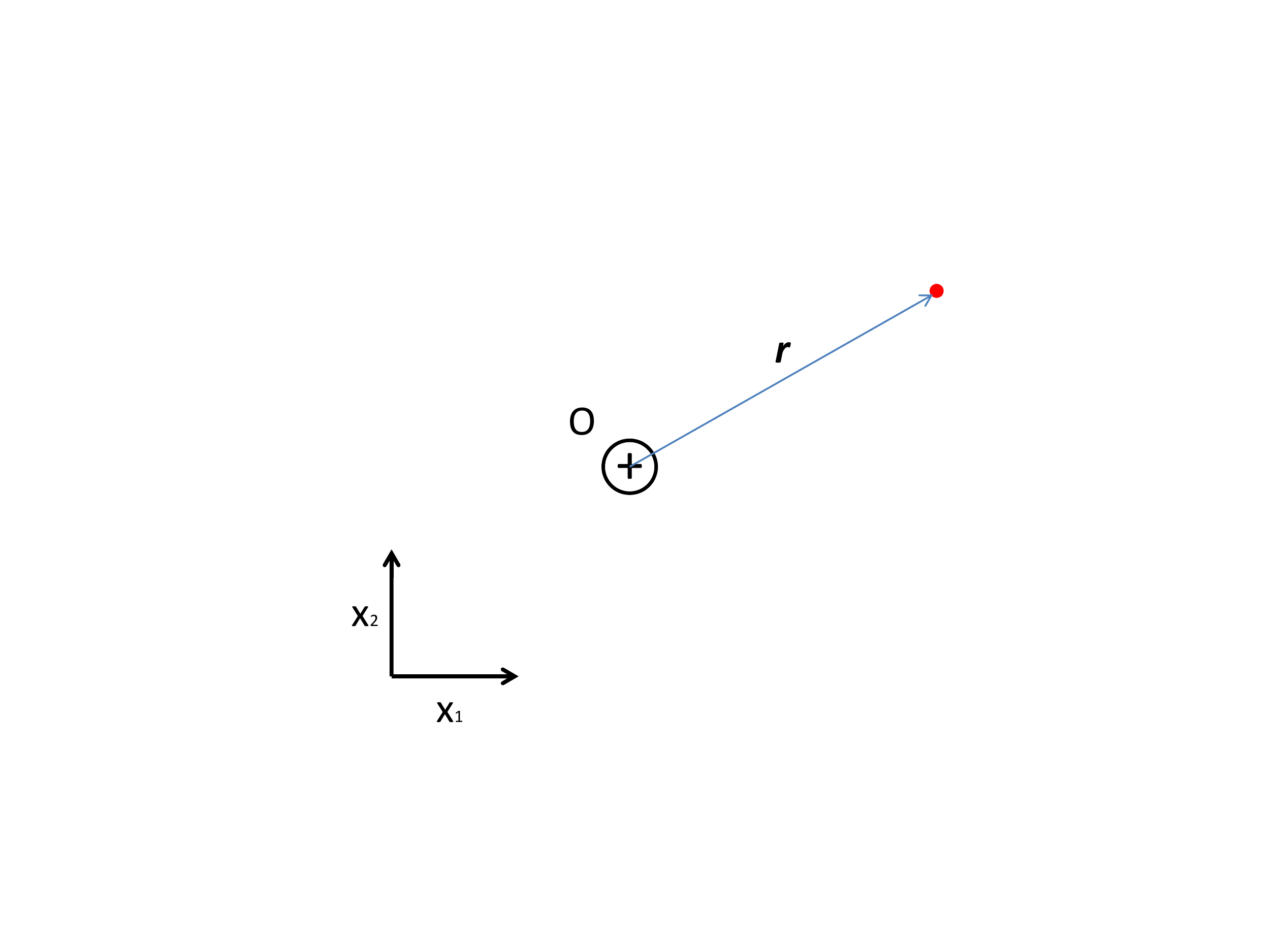}
\label{fig:burgers_1}
}\qquad
\subfigure[Move the field point to $\bfr + \delta \bfr$.]{
\includegraphics[width=0.45\textwidth]{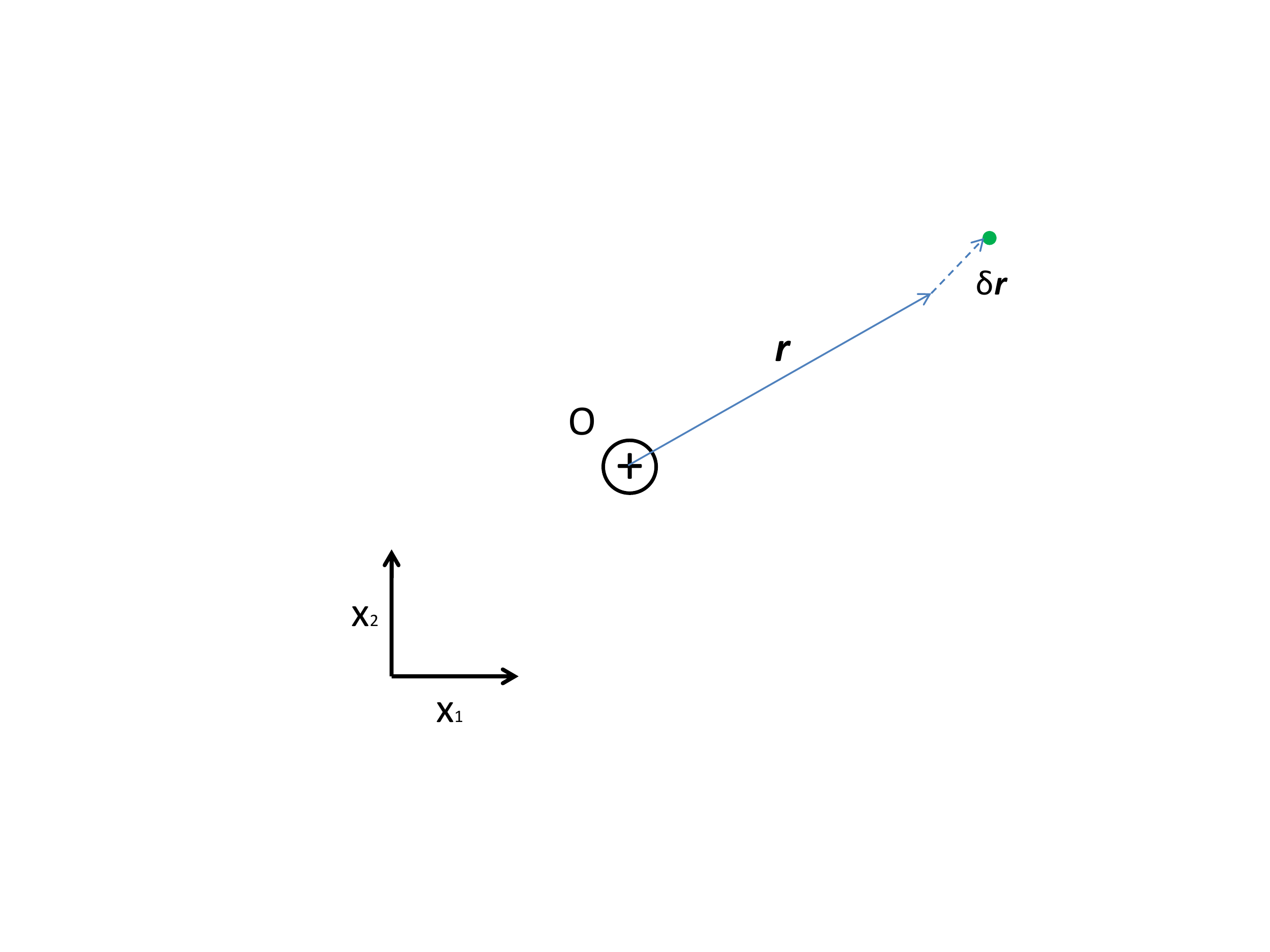}
\label{fig:burgers_2}
}
\subfigure[The configuration with disclination source moved to $- \delta \bfr$.]{
\includegraphics[width=0.45\textwidth]{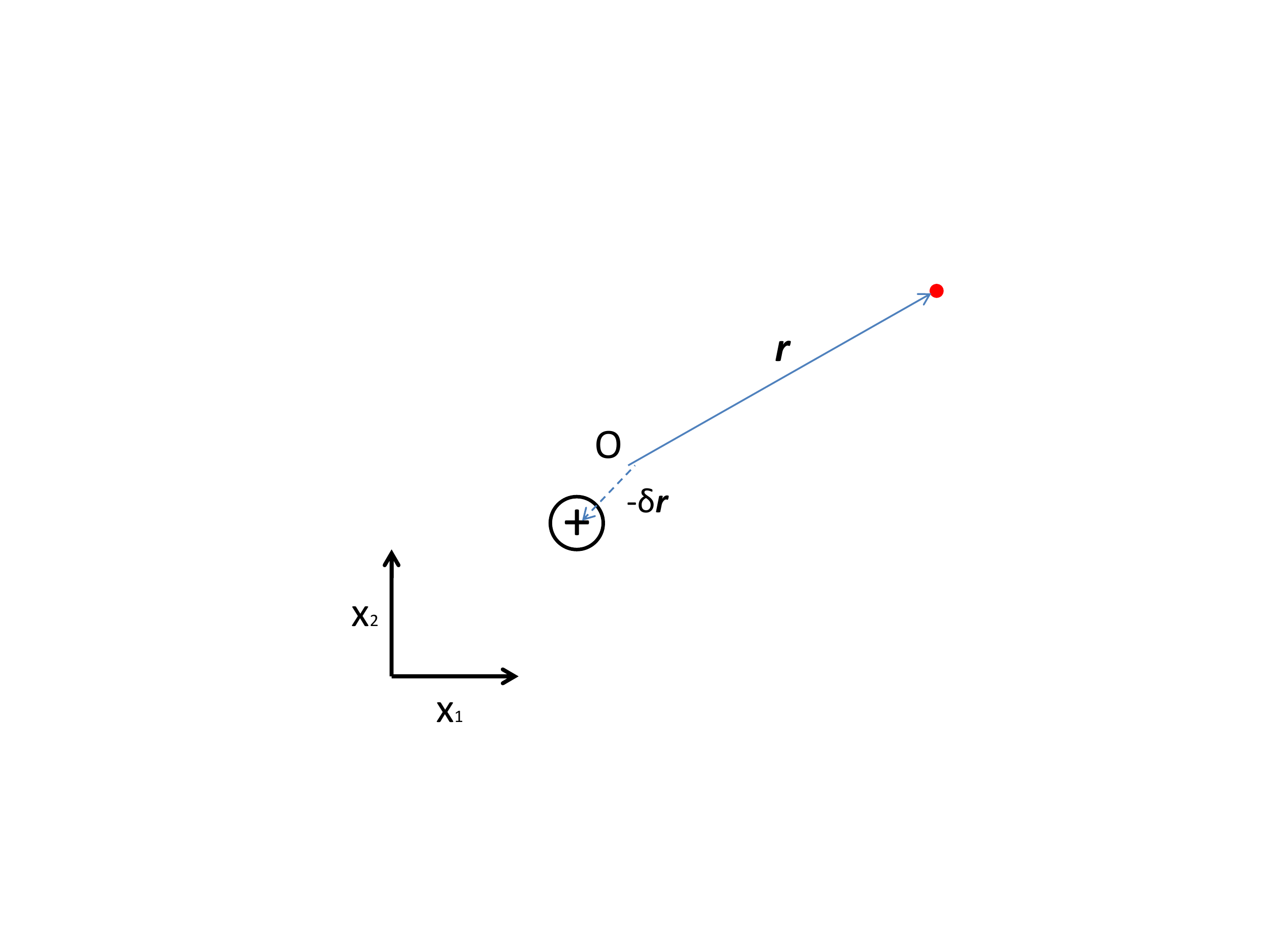}
\label{fig:burgers_3}
}\qquad
\subfigure[Place a disclination dipole with the separation vector $\delta \bfr$ and keep the field point at $\bfr$.]{
\includegraphics[width=0.45\textwidth]{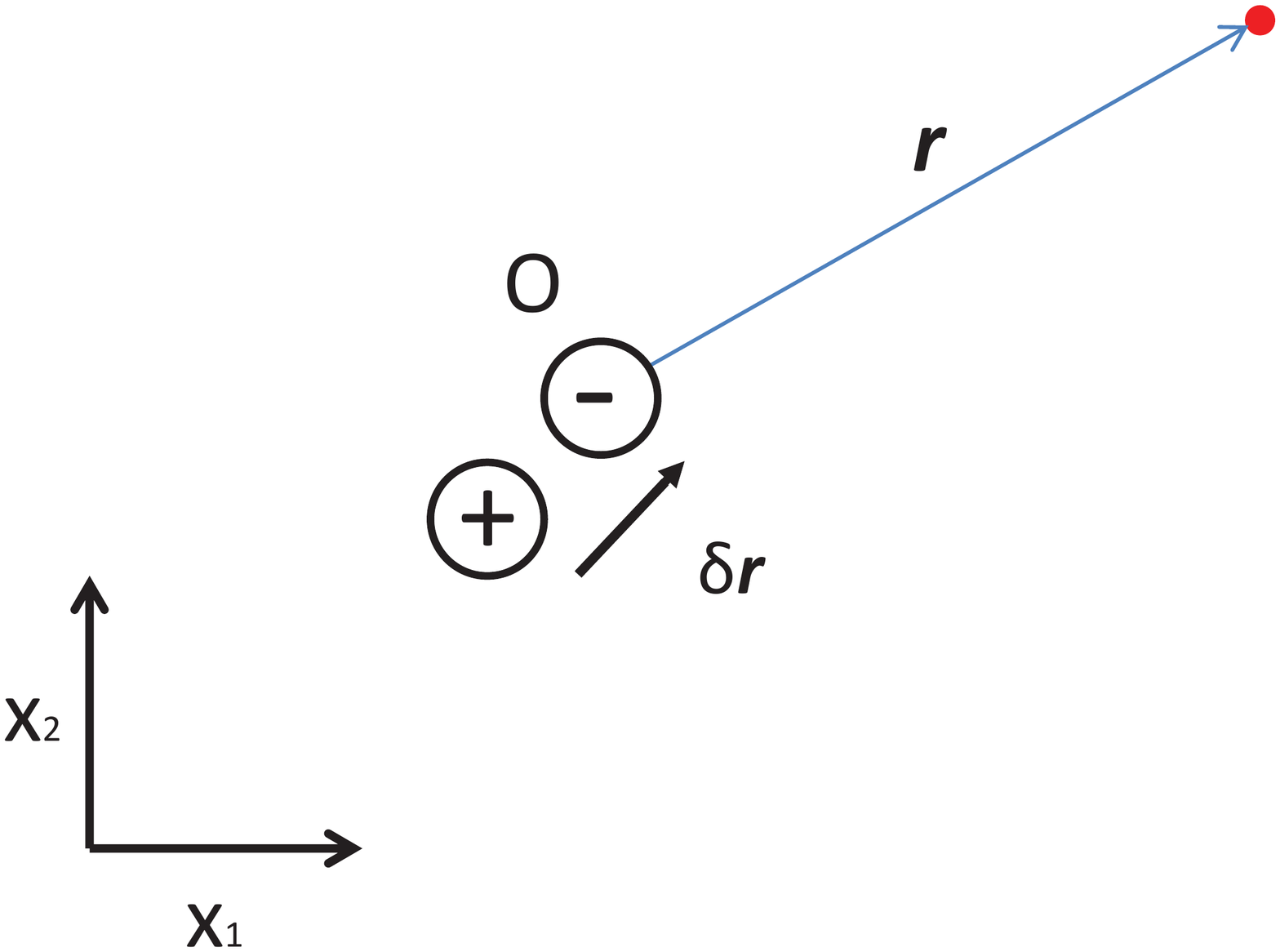}
\label{fig:burgers_4}
}
\caption{Schematic in support of calculation of stress field of a wedge-disclination dipole in linear, plane, isotropic elasticity. }
\label{fig:burgers}
\end{figure}

We derive a formula for the Burgers vector of a wedge disclination dipole utilizing the linear theory of plane isotropic elasticity. Consider a positive disclination located at the origin $O$, as shown in Figure \ref{fig:burgers_1}. We denote the stress at point $\bfc$ of a single disclination located at $\bfa$ with Frank vector $\bfOmega$  as $\bfsigma(\bfc; \bfa, \bfOmega)$. Thus, the stress field at $\bfr$ in Figure \ref{fig:burgers_1} is $\bfsigma(\bfr;\bf0,\bfOmega)$. Next we consider the field point $\bfr+\delta\bfr$ marked by the green point as in Figure \ref{fig:burgers_2}, with the disclination kept at the origin $O$. The stress tensor at this point is given by $\bfsigma(\bfr+\delta\bfr;\bf0,\bfOmega)$.

Instead of moving the field point in Figure \ref{fig:burgers_2}, we next consider the field point as fixed at $\bfr$ with the disclination moved from $\bf0$ to $-\delta\bfr$ as shown in Figure \ref{fig:burgers_3}. The value of the stress at $\bfr$ now is $\bfsigma(\bfr;-\delta\bfr,\bfOmega)$.  Utilizing the results in \cite{dewit1973theory}, the stress  of a disclination, of fixed strength $\bfOmega$ and located at $\bfa=a_1\bfe_1+a_2\bfe_2+a_3\bfe_3$, at the field point $\bfc= c_1\bfe_1 + c_2\bfe_2 + c_3\bfe_3$ is given by
\begin{equation}\label{eqn:classical_stress}
\begin{aligned}
\sigma_{11}(\bfc; \bfa, \bfOmega) =& -\frac{G\Omega_1(c_3-a_3)}{2\pi(1-\nu)}\left[ \frac{c_1-a_1}{\rho^2}- 2 \frac{(c_1-a_1)(c_2-a_2)^2}{\rho^4}\right] \\
& - \frac{G\Omega_2(c_3-a_3)}{2\pi(1-\nu)}\left[\frac{c_2-a_2}{\rho^2}+ 2\frac{(c_1-a_1)^2(c_2-a_2)}{\rho^4}\right]\\
& + \frac{G\Omega_3}{2\pi(1-\nu)}\left[ \ln \rho + \frac{(c_2-a_2)^2}{\rho^2} + \frac{\nu}{1-2\nu}\right]\\
\sigma_{22}(\bfc; \bfa, \bfOmega) =& -\frac{G\Omega_2(c_3-a_3)}{2\pi(1-\nu)}\left[ \frac{c_1-a_1}{\rho^2}+ 2 \frac{(c_1-a_1)(c_2-a_2)^2}{\rho^4}\right] \\
&- \frac{G\Omega_2(c_3-a_3)}{2\pi(1-\nu)}\left[\frac{c_2-a_2}{\rho^2}- 2\frac{(c_1-a_1)^2(c_2-a_2)}{\rho^4}\right] \\
& + \frac{G\Omega_3}{2\pi(1-\nu)}\left[ \ln \rho + \frac{(c_1-a_1)^2}{\rho^2} + \frac{\nu}{1-2\nu}\right]\\
\sigma_{33}(\bfc; \bfa, \bfOmega) =& -\frac{G\nu (c_3-a_3)}{\pi(1-\nu)\rho^2}(\Omega_1(c_1-a_1)+\Omega_2(c_2-a_2))+\frac{G\Omega_3}{2\pi(1-\nu)}\left[2\nu \ln \rho + \frac{\nu}{1-2\nu} \right] \\
\sigma_{12}(\bfc; \bfa, \bfOmega) =& \frac{G\Omega_1(c_3-a_3)}{2\pi(1-\nu)}\left[ \frac{c_2-a_2}{\rho^2}- 2 \frac{(c_1-a_1)^2(c_2-a_2)}{\rho^4}\right] \\
& + \frac{G\Omega_2(c_3-a_3)}{2\pi(1-\nu)}\left[\frac{c_1-a_1}{\rho^2}- 2\frac{(c_1-a_1)^2(c_2-a_2)}{\rho^4}\right] - \frac{G\Omega_3(c_1-a_1)(c_2-a_2)}{2\pi(1-\nu)\rho^2}\\
\sigma_{23}(\bfc; \bfa, \bfOmega) =& \frac{G\Omega_1(c_1-a_1)(c_2-a_2)}{2\pi(1-\nu)\rho^2} - \frac{G\Omega_2}{2\pi(1-\nu)} \left[ (1-2\nu) \ln \rho + \frac{(c_1-a_1)^2}{\rho^2}\right] \\
\sigma_{13}(\bfc; \bfa, \bfOmega) =& -\frac{G\Omega_1}{2\pi(1-\nu)}\left[ (1-2\nu) \ln \rho +\frac{(c_2-a_2)^2}{\rho^2} \right] + \frac{G\Omega_2(c_1-a_1)(c_2-a_2)}{2\pi (1-\nu)\rho^2},
\end{aligned}
\end{equation}
where $\rho$ is the distance between the field point $\bfc$ and the source point $\bfa$, $\rho=|\bfc-\bfa|$. Equation (\ref{eqn:classical_stress}) shows that the stress fields only depend on the relative location of the field and disclination source points. In other words, given a disclination at $\bfa$ with Frank vector $\bfOmega$, the stress field at point $\bfc$ can be expressed as
\begin{equation*}
\bfsigma(\bfc; \bfa, \bfOmega) = \bff(\bfc-\bfa; \bfOmega),
\end{equation*}
where $\bff$ is the formula for the stress field of the wedge disclination in linear isotropic elasticity whose explicit expression in Cartesian coordinates is given in (\ref{eqn:classical_stress}). From (\ref{eqn:classical_stress}), we have
\begin{equation}\label{eqn:f_property}
\bff(\bfx;\bfOmega) = -\bff(\bfx;-\bfOmega),
\end{equation}
for any given $\bfx$. Thus, for the stress fields, we have
\begin{equation*}
\bfsigma(\bfc;\bfa,\bfOmega) = -\bfsigma(\bfc;\bfa,-\bfOmega).
\end{equation*}
Hence, the stress field in Figure \ref{fig:burgers_1} can be written as 
\begin{equation*}
\bfsigma(\bfr;\bf0,\bfOmega) = \bff(\bfr; \bfOmega).
\end{equation*}
The stress field corresponding to Figure \ref{fig:burgers_2} is 
\begin{equation*}
\bfsigma(\bfr+\delta\bfr;\textbf{0},\bfOmega) = \bff(\bfr+\delta\bfr;\bfOmega).
\end{equation*}
Also, the stress field in Figure \ref{fig:burgers_3} is 
\begin{equation*}
\bfsigma(\bfr;-\delta\bfr, \bfOmega) = \bff(\bfr- (-\delta\bfr); \bfOmega) = \bff(\bfr+\delta\bfr; \bfOmega).
\end{equation*}

In Figure \ref{fig:burgers_4}, a disclination dipole is introduced. A negative disclination with Frank vector $-\bfOmega$ is at $\bf0$ and the positive disclination with Frank vector $\bfOmega$ is at $-\delta \bfr$. Thus, $\delta\bfr$ is the separation vector of the dipole, pointing from the positive disclination to the negative disclination and we are interested in calculating the stress at $\bfr$, represented as the red dot. Let the stress field for the disclination configuration in Figure \ref{fig:burgers_4} be denoted as $\hat{\bfsigma}$. Due to superposition in linear elasticity, the stress field of Figure \ref{fig:burgers_4} can be written as
\begin{eqnarray*}
\hat{\bfsigma}(\bfr;\delta\bfr, \bfOmega) := \bfsigma(\bfr;-\delta\bfr,\bfOmega) + \bfsigma(\bfr; \bf0, -\bfOmega) \\
\Rightarrow \hat{\bfsigma}(\bfr;\delta\bfr, \bfOmega) = \bff(\bfr+\delta\bfr;\bfOmega) + \bff(\bfr; -\bfOmega).
\end{eqnarray*}
On applying (\ref{eqn:f_property}), we have
\begin{eqnarray}
\hat{\bfsigma}(\bfr;\delta\bfr, \bfOmega) = \bff(\bfr+\delta\bfr;\bfOmega) - \bff(\bfr; \bfOmega)  \nonumber \\
\Rightarrow \hat{\bfsigma}(\bfr;\delta\bfr, \bfOmega) = \bfsigma(\bfr+\delta\bfr;\bf0, \bfOmega) - \bfsigma(\bfr;\bf0, \bfOmega).  \label{eqn:stress_difference}
\end{eqnarray}
Therefore, we have shown that the stress field in Figure \ref{fig:burgers_4} equals the difference between the stress fields in Figure \ref{fig:burgers_2} and the one in Figure \ref{fig:burgers_1}. 

Specializing to the plane case with $\bfr = x_1\bfe_1+x_2\bfe_2$, the stress field corresponding to Figure \ref{fig:burgers_1} , given the Frank vector $\bfOmega=\Omega_3 \bfe_3$, is  
\begin{eqnarray*}
\sigma_{11}(\bfr;\textbf{0}, \bfOmega)  = f_{11}(x_1,x_2; \Omega_3) = \frac{G\Omega_3}{2\pi(1-\nu)}\left[ \ln r + \frac{x_2^2}{r^2}+\frac{\nu}{1-2\nu}\right] \\
\sigma_{22}(\bfr;\textbf{0}, \bfOmega)  = f_{22}(x_1,x_2; \Omega_3) = \frac{G\Omega_3}{2\pi(1-\nu)}\left[\ln r + \frac{x_1^2}{r^2}+\frac{\nu}{1-2\nu}\right] \\
\sigma_{12}(\bfr;\textbf{0}, \bfOmega)  = f_{12}(x_1,x_2; \Omega_3) = -\frac{G\Omega_3x_1x_2}{2\pi(1-\nu)r^2},
\end{eqnarray*}
where $r$ is the norm of $\bfr$, $G$ is the shear modulus and $\nu$ is the Poisson ratio. Assuming
\begin{equation}\label{sep_vec}
\delta \bfr := \delta x_1 \bfe_1 + \delta x_2 \bfe_2
\end{equation}
to be small, the Taylor expansion of $\bfsigma(\bfr+\delta\bfr; \textbf{0}, \bfOmega)$ is
\begin{equation*}
\bfsigma(\bfr+\delta\bfr; \textbf{0}, \bfOmega) = \bff(\bfr; \bfOmega) + \frac{\partial \bff(\bfr; \bfOmega)}{\partial \bfr} \delta \bfr + O(\delta\bfr^2).
\end{equation*}
After substituting $\bfr= x_1\bfe_1+x_2\bfe_2$ and $\delta \bfr = \delta x_1 \bfe_1 + \delta x_2 \bfe_2$, we have
\begin{eqnarray*}
\sigma_{11}(\bfr+\delta\bfr; \textbf{0}, \bfOmega) = f_{11}(x_1,x_2;\Omega_3) + \frac{\partial f_{11}(x_1,x_2;\Omega_3)}{\partial x_1} \delta x_1 + \frac{\partial  f_{11}(x_1,x_2;\Omega_3)}{\partial x_2} \delta x_2 + O(\delta\bfr^2) \\
\sigma_{22}(\bfr+\delta\bfr; \textbf{0}, \bfOmega) = f_{22}(x_1,x_2;\Omega_3) + \frac{\partial f_{22}(x_1,x_2;\Omega_3)}{\partial x_1} \delta x_1 + \frac{\partial  f_{22}(x_1,x_2;\Omega_3)}{\partial x_2} \delta x_2 + O(\delta\bfr^2) \\
\sigma_{12}(\bfr+\delta\bfr; \textbf{0}, \bfOmega) = f_{12}(x_1,x_2;\Omega_3) + \frac{\partial f_{12}(x_1,x_2;\Omega_3)}{\partial x_1} \delta x_1 + \frac{\partial  f_{12}(x_1,x_2;\Omega_3)}{\partial x_2} \delta x_2 + O(\delta\bfr^2).;
\end{eqnarray*}
After substituting $\bfsigma(\bfr;\textbf{0},\bfOmega)$ and $\bfsigma(\bfr+\delta\bfr;\textbf{0},\bfOmega)$ into  (\ref{eqn:stress_difference}) and omitting the higher order terms, we get
\begin{equation}\label{disloc_from_disclin_1}
\begin{aligned}
& \hat{\sigma}_{11}(x_1,x_2;\delta x_1,\delta x_2, \Omega_3) = \frac{G\Omega_3 \delta x_2}{2\pi(1-\nu)}\left[\frac{x_2}{r^2}+2\frac{x_1^2x_2}{r^4}\right] + \frac{G\Omega_3 \delta x_1}{2\pi(1-\nu)}\left[\frac{x_1}{r^2}-2\frac{x_1x_2^2}{r^4}\right] \\
& \hat{\sigma}_{22}(x_1,x_2;\delta x_1,\delta x_2, \Omega_3) = \frac{G\Omega_3 \delta x_2}{2\pi(1-\nu)}\left[\frac{x_2}{r^2}-2\frac{x_1^2x_2}{r^4}\right] + \frac{G\Omega_3 \delta x_1}{2\pi(1-\nu)}\left[\frac{x_1}{r^2}+2\frac{x_1x_2^2}{r^4}\right] \\
& \hat{\sigma}_{12}(x_1,x_2;\delta x_1,\delta x_2, \Omega_3) = -\frac{G\Omega_3 \delta x_2}{2\pi(1-\nu)}\left[\frac{x_2}{r^2}-2\frac{x_1x_2^2}{r^4}\right] - \frac{G\Omega_3 \delta x_1}{2\pi(1-\nu)}\left[\frac{x_2}{r^2}-2\frac{x_1^2x_2}{r^4}\right].
\end{aligned}
\end{equation}
The stress field of the single edge dislocation in 2-D, isotropic elasticity is \cite{de1973theory} 
\begin{equation}\label{disloc_from_disclin_2}
\begin{aligned}
\sigma_{11}^b(x_1,x_2;b_1, b_2) = -\frac{Gb_1}{2\pi(1-\nu)}\left[\frac{x_2}{r^2}+2\frac{x_1^2x_2}{r^4}\right] + \frac{Gb_2}{2\pi(1-\nu)}\left[\frac{x_1}{r^2}-2\frac{x_1x_2^2}{r^4}\right] \\
\sigma_{22}^b(x_1,x_2;b_1, b_2) = -\frac{Gb_1}{2\pi(1-\nu)}\left[\frac{x_2}{r^2}-2\frac{x_1^2x_2}{r^4}\right] + \frac{Gb_2}{2\pi(1-\nu)}\left[\frac{x_1}{r^2}+2\frac{x_1x_2^2}{r^4}\right] \\
\sigma_{12}^b (x_1,x_2;b_1, b_2)= \frac{Gb_1}{2\pi(1-\nu)}\left[\frac{x_2}{r^2}-2\frac{x_1x_2^2}{r^4}\right] - \frac{Gb_2}{2\pi(1-\nu)}\left[\frac{x_2}{r^2}-2\frac{x_1^2x_2}{r^4}\right].
\end{aligned}
\end{equation}
On defining the Burgers vector of a disclination dipole with separation vector $\delta \bfr$ (\ref{sep_vec})  and strength $\bfOmega$ as
\begin{equation} \label{eqn:dipole_burgers_elasticity}
\bfb :=-\Omega_3 \delta r_2 \bfe_1 + \Omega_3 \delta r_1 \bfe_2 = \bfOmega \times \delta \bfr,
\end{equation}
we see that the stress field of the disclination dipole (\ref{disloc_from_disclin_1}) exactly matches that of the single edge  dislocation (\ref{disloc_from_disclin_2}). 

This establishes the correspondence between the Burgers vector of the wedge disclination dipole and the edge dislocation in 2-d, isotropic, plane, linear elasticity. In Section \ref{sec:burgers} we establish the general form of this geometric relationship in the context of exact kinematics, valid for any type of material (i.e. without reference to material response).

\section{Generalized disclination theory and associated Weingarten's theorem}\label{sec:g_disclination_theory}

The connection between g.disclinations (and dislocations) represented as fields and their more classical representation following Weingarten's pioneering work is established in Sections \ref{sec:rel_gdisclin_wein} and \ref{sec:burgers}.  In this Section we briefly review the defect kinematics of g.disclination theory  and a corresponding Weingarten-gd theorem developed in \cite{acharya2015continuum} that are necessary prerequisites for the arguments in the aforementioned sections. We also develop a new result in Section \ref{sec:cut_independ} related to the Weingarten-gd theorem, proving that the inverse deformation jump across the cut-surface is independent of the surface when the g.disclination density vanishes.

\begin{figure}
\centering
\includegraphics[width=0.4\textwidth]{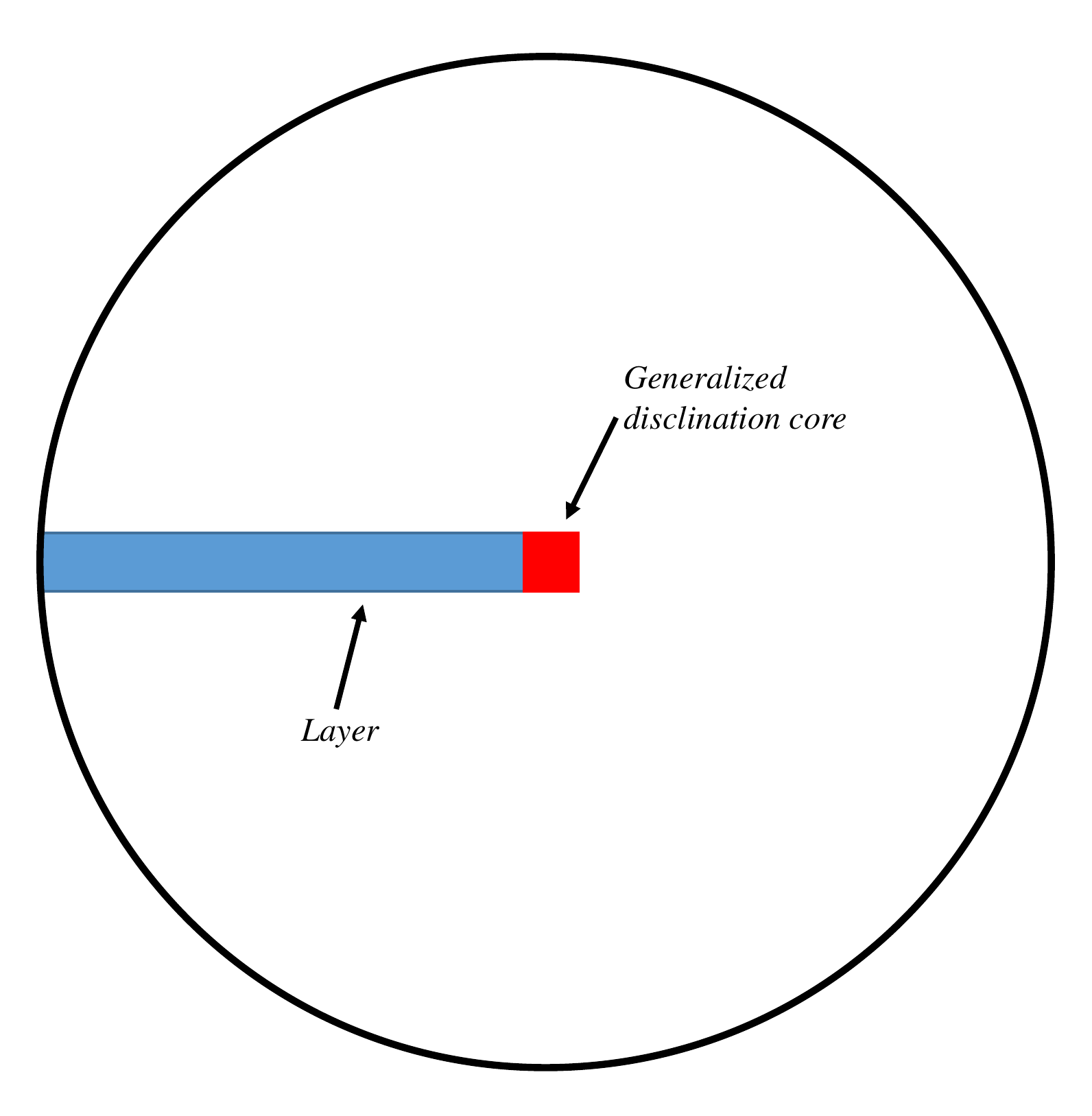}
\caption{Physical regularization of classical terminating discontinuity. Treat the distortion discontinuity as the eigenwall field $\bfS$ with support in a layer. }
\label{fig:g_theory}
\end{figure}

As defined in Section \ref{sec:disclination_intro}, a single g.disclination is a line defect terminating a distortion discontinuity. Developed as a generalization of eigendeformation theory of Kroner, Mura and deWit, the generalized disclination has a core and the discontinuity is modeled by an eigenwall field with support in a layer \cite{acharya2015continuum}, as shown in Figure \ref{fig:g_theory}. The representation of a discrete g.disclination involves a continuous elastic 2-distortion field $\bfY$, assumed to be irrotational outside the generalized disclination core ($ \bfY = \grad \, \grad \left(\bfx^{-1}\right)$ in the case without defects, where $\bfx$ is the deformation map). The \emph{strength of the discrete generalized disclination is given by the second order tensor obtained by integrating the 2-distortion field along any closed curve encircling the core; when defined from a terminating distortion discontinuity, it is simply the difference of the two distortions involved in defining the discontinuity}. One way of setting up the generalized disclination density tensor field, which is a third order tensor, is to assign the tensor product of the strength tensor and the core line direction vector as a uniformly distributed field within the generalized disclination core, and zero outside it. In the case of a disclination, the strength tensor is necessarily the difference of two orthogonal tensors. 

The fundamental kinematic decomposition of generalized disclination theory \cite{acharya2015continuum} is to write
\begin{equation}\label{eqn:Y-decomp}
\bfY = \grad\, \bfW + \bfS,
\end{equation}
where $\bfW$ is the i-elastic 1-distortion ($\bfF^{-1}$ in the defect-free case, where $\bfF$ is the deformation gradient) and $\bfS$ ($3^{rd}$-order tensor) is the eigenwall field.

With this decomposition of $\bfY$, it is natural to measure the generalized disclination density as
\begin{equation}\label{Pi-measure}
\curl\left( \bfY - \grad \bfW \right) = \curl \bfS =: \bfPi.
\end{equation}
It characterizes the closure failure of integrating $\bfY$ on closed contours in the body:
\begin{equation}\label{eqn:pi_characterize}
\int_a \bfPi \bfn da = \int_c \bfY d\bfx
\end{equation}
where $a$ is any area patch with closed boundary contour $c$ in
the body. Physically, it is to be interpreted as a density of
lines (threading areas) in the current configuration, carrying a
tensorial attribute that reflects a jump in $\bfW$.

The dislocation density is defined as \cite{acharya2015continuum}
\begin{equation}
\bfalpha := \bfY : \bfX = \left( \bfS + \grad\bfW \right) : \bfX.
\label{eqn:dislocation}
\end{equation}
In the case that there is no distortion discontinuity, namely $\bfS = \bf0$, (\ref{eqn:dislocation}) becomes $\bfalpha = - \curl\bfW$, since $\curl \bfA = - \grad \bfA :\bfX$ for any smooth tensor field $\bfA$. The definition of the dislocation density (\ref{eqn:dislocation}) is motivated by the displacement-jump formula (\ref{eqn:weingarten}) \cite{acharya2015continuum} corresponding to a single, isolated defect line terminating an i-elastic distortion jump in the body. In this situation, the displacement jump for an isolated
defect line, measured by integrating $\bfW$ along any closed curve encircling the defect core cylinder\footnote{In \cite{acharya2015continuum} a typographical error suggests that the displacement jump is obtained by integrating $\bfalpha$ on area patches; $\bfalpha$ there should have been replaced by $\curl\, \bfW$.}, is no longer a topological object independent of the curve (in the class of curves encircling the core) due to the fact that in the presence of a g.disclination density localized in the core cylinder the field $\bfS$ cannot be localized in the core - it is, at the very least, supported in a layer extending to the boundary from the core, or, when $\divergence \,\bfS = \bf0$, completely delocalized over the entire domain.

Now we apply a Stokes-Helmholtz-like orthogonal decomposition of the field $\bfS$ into a compatible part and an incompatible part:
\begin{eqnarray}\label{eqn:S_decomposition}
\begin{aligned}
\bfS &= \bfS^{\perp} + \grad \, \bfZ^{s} \\ 
\curl \bfS^{\perp} &= \bfPi \\
\divergence\bfS^{\perp} &= \textbf{0} \\  
\mbox{with} \ \ \bfS^{\perp}\bfn &= \textbf{0} \ \ \mbox{on the boundary}.
\end{aligned}
\end{eqnarray}
It is clear that when $\bfPi = \textbf{0}$ then $\bfS^{\perp} = \textbf{0}$. 
%%Furthermore, we can put (\ref{eqn:S_decomposition}) back in to (\ref{eqn:Y-decomp}) to obtain
%%\begin{equation} \label{eqn:Y_S}
%%\bfY = grad\bfW + grad\bfW^s + \bfS^{\perp}.
%%\end{equation}
%%The orientation field is a part of $\bfW$ and its gradient is a component of the compatible part of $\bfY$. For example, in a grain boundary case, the compatible part of $\bfY$ represents a smoothed grain boundary without kinks or corners. On the other hand, the incompatible part $\bfS^{\perp}$, characterizes grain boundaries with kinks or corners. 

In summary, the governing equations for computing the elastic fields for static generalized disclination theory (i.e. when the disclination and dislocation fields are specified) are 
\begin{eqnarray}\label{eqn:summary}
\begin{aligned}
&\curl \bfS = \bfPi \\ 
&\bfS = \bfS^{\perp} + \grad \, \bfZ^{s} \\ 
&\divergence\bfS^{\perp} = \textbf{0}  \ \ \ \ 
\mbox{with} \  \bfS^{\perp}\bfn = \textbf{0} \ \ \mbox{on the boundary} \\ 
&\bfalpha = \left( \bfS + \grad\bfW \right) : \bfX,
%%\divergence\bfT &= \textbf{0} ,
\end{aligned}
\end{eqnarray}
where $\bfPi$ and $\bfalpha$ are specified from physical considerations. These equations are solved along with balance of linear and angular momentum involving Cauchy stresses and couple-stresses (with constitutive assumptions) to obtain g.disclination and dislocation stress and couple stress fields. In the companion paper \cite{zhang_acharya_puri} we solve these equations along with
\[
\divergence\bfT = \textbf{0}
\]
with $\bfT$ representing the Cauchy stress as a function of $\bfW$, and we ignore couple stresses for simplicity.

\subsection{Review of Weingarten theorem associated with g.disclinations}\label{sec:weingarten}
\begin{figure}
\centering
\subfigure[The cross-section of a multi-connected body with a toroidal hole. The shaded gray area is the `half-toroid'. The half-toroid is not shown in Figures \ref{fig:weingarten_toroidal_2} and \ref{fig:weingarten_toroidal_3}.]{
\includegraphics[width=0.4\textwidth]{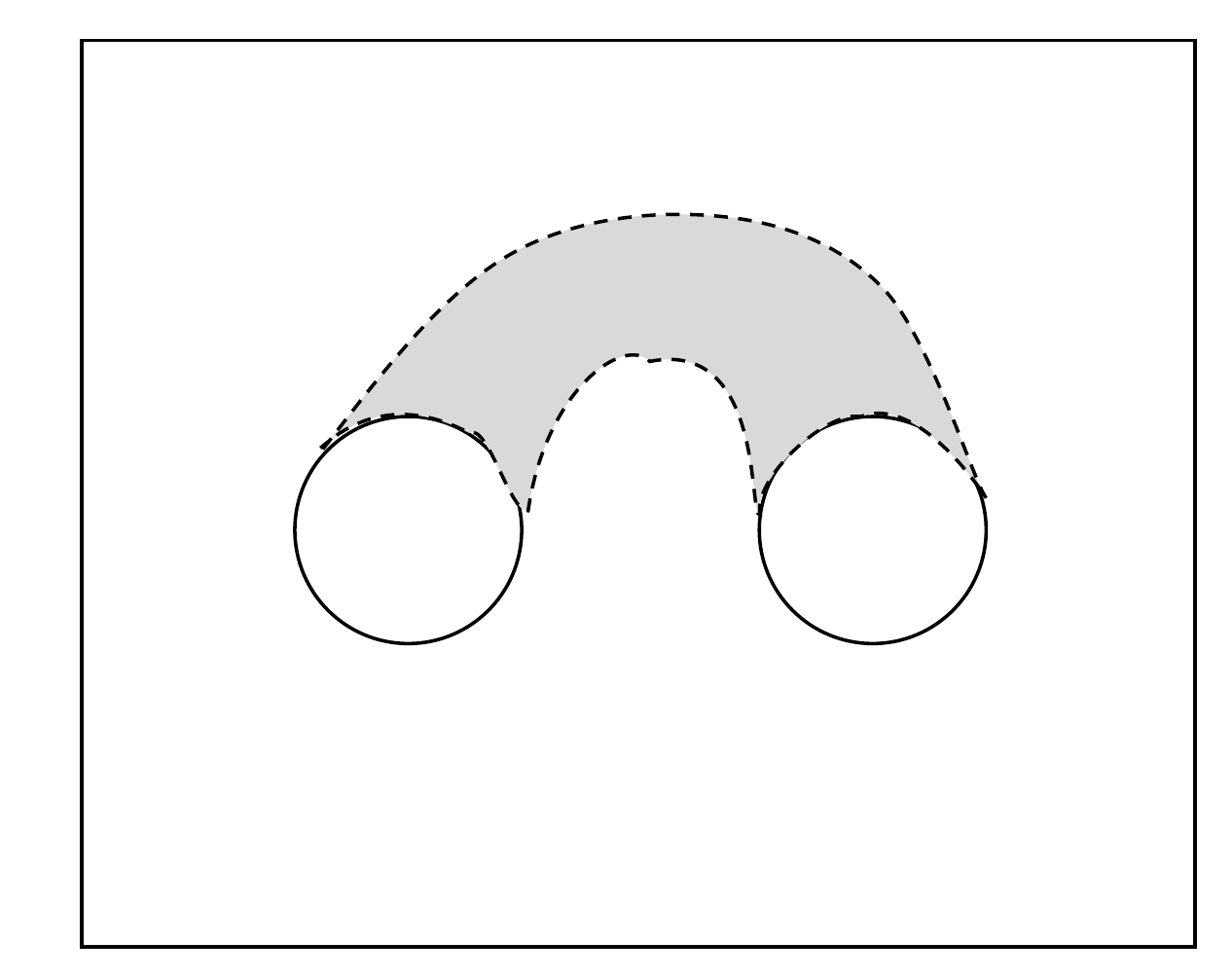}
\label{fig:weingarten_toroidal_1}
}\qquad
\subfigure[The multi-connected body becomes simply-connected after introducing the cut-surface.]{
\includegraphics[width=0.38\textwidth]{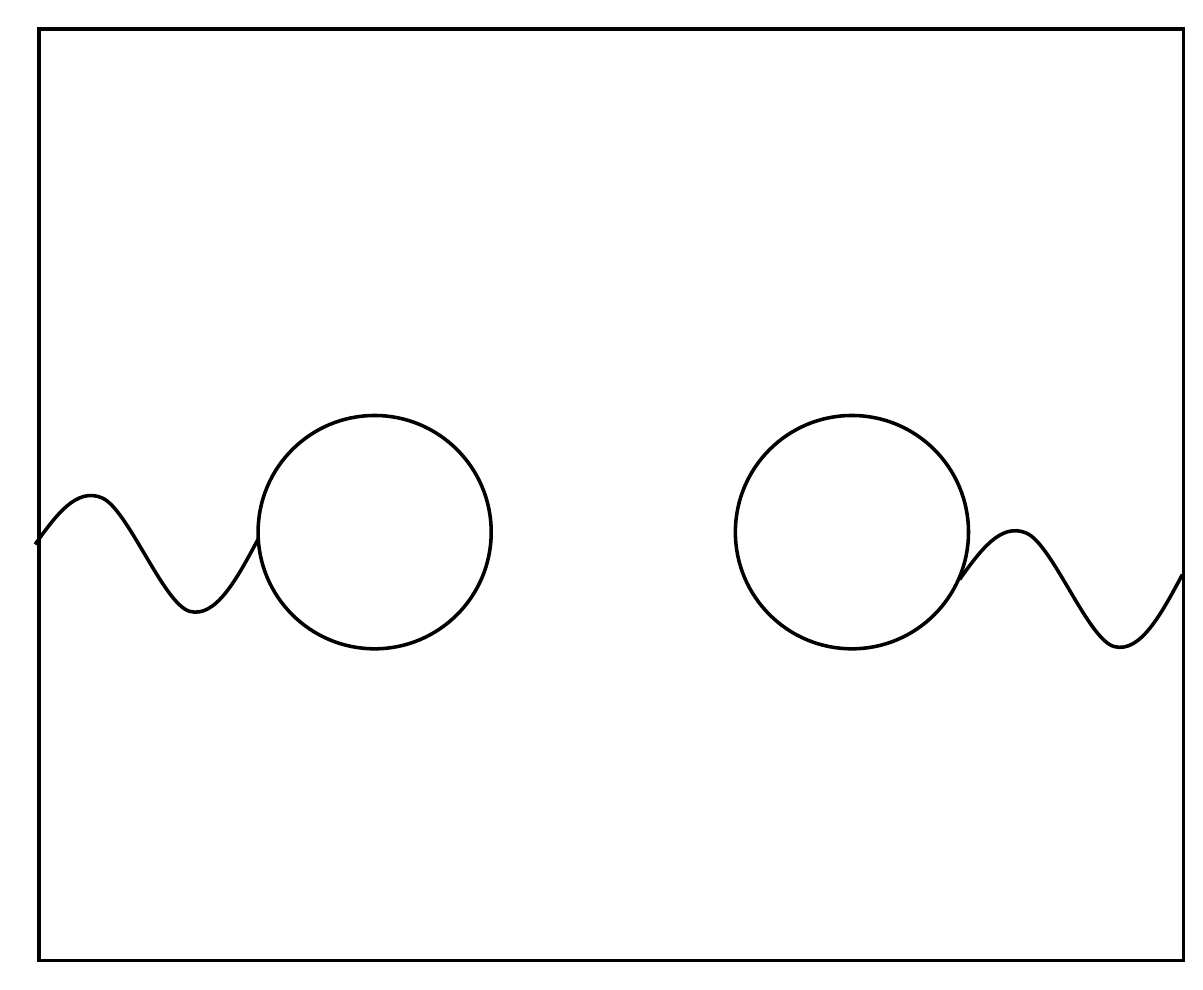}
\label{fig:weingarten_toroidal_2}
}
\subfigure[Another method to introduce the cut surface to make the multi-connected body simply-connected.]{
\includegraphics[width=0.4\textwidth]{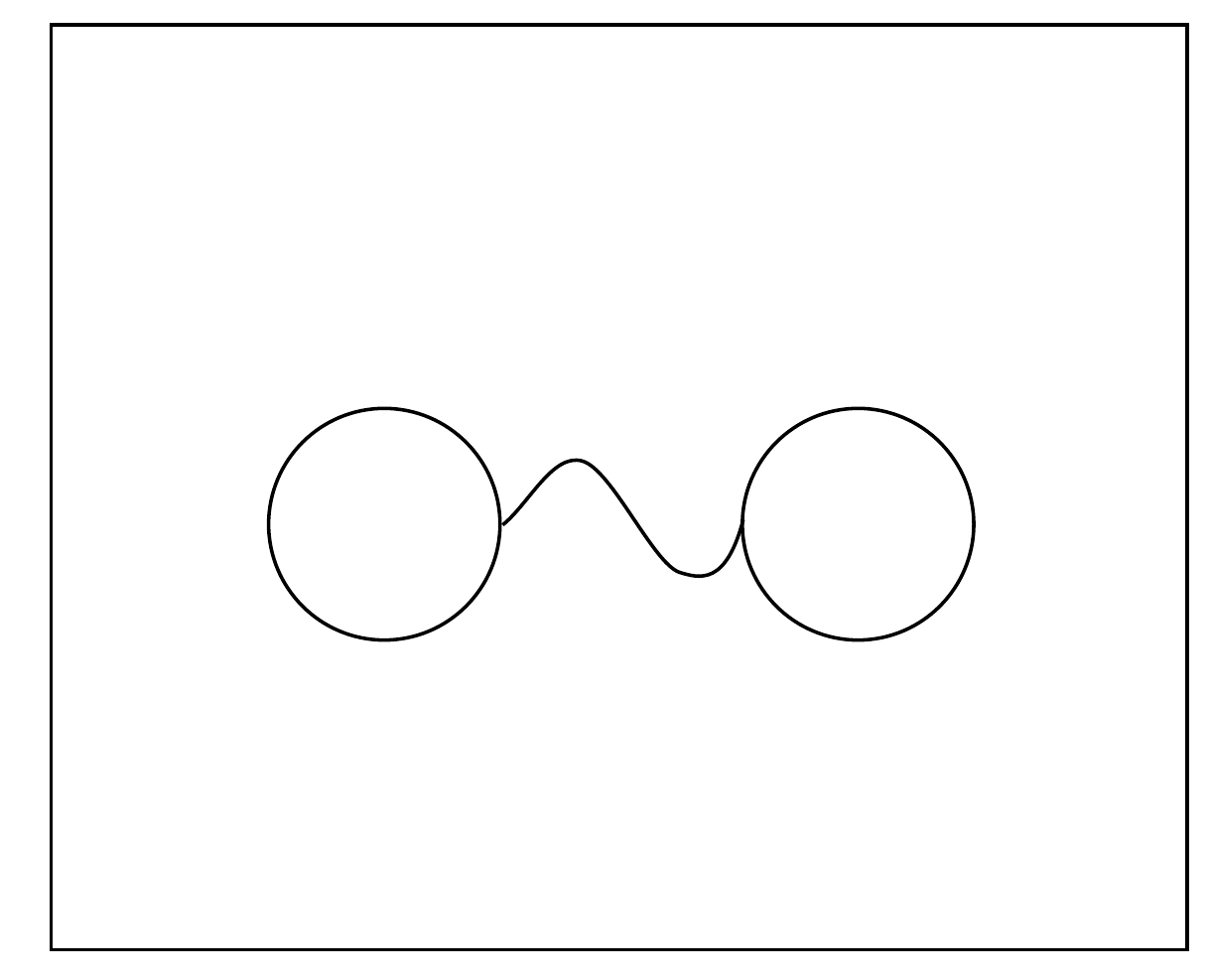}
\label{fig:weingarten_toroidal_3}
}\\
\subfigure[The cross-section of a multi-connected body with a through hole.]{
\includegraphics[width=0.3\textwidth]{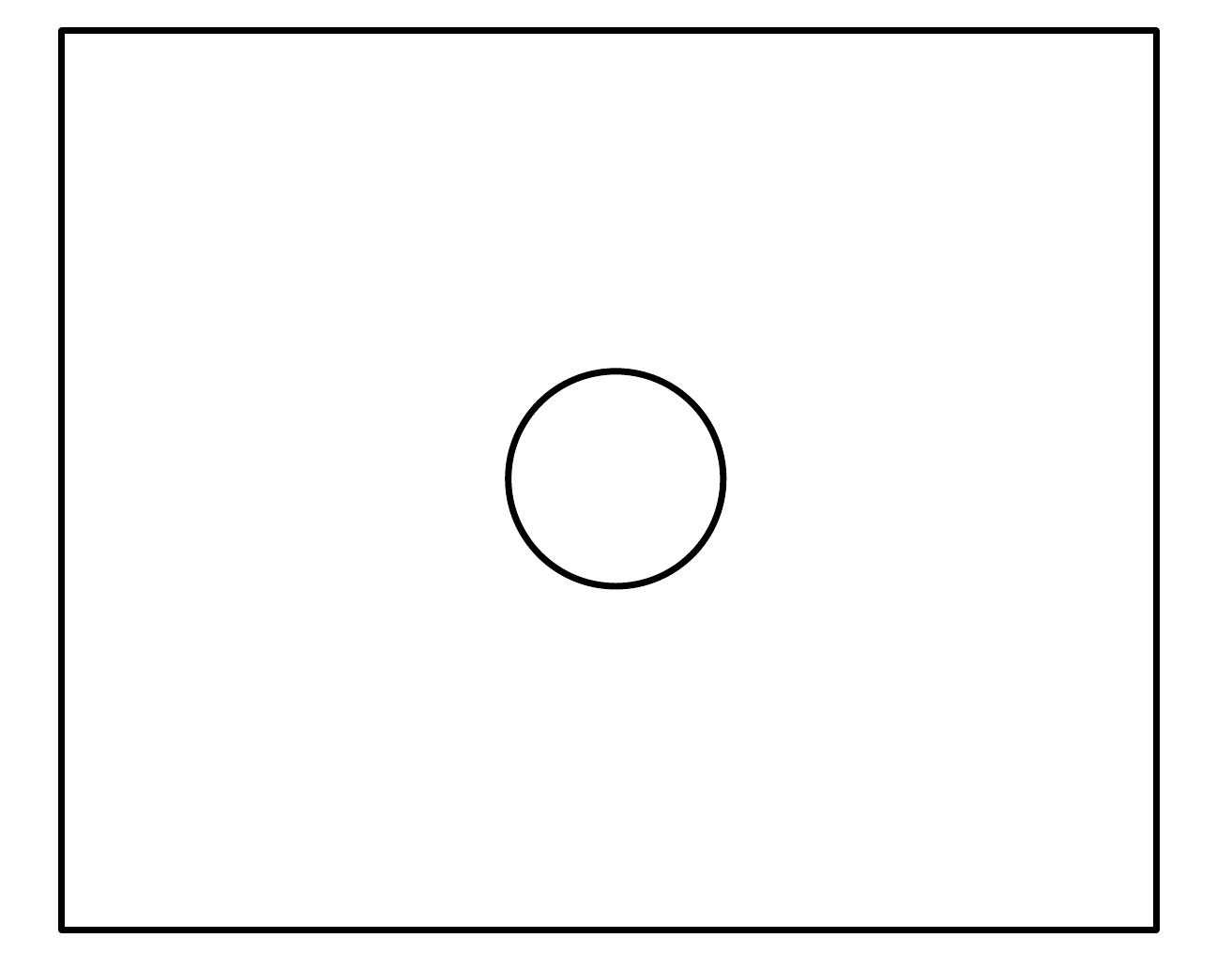}
\label{fig:weingarten_hole_1}
}\qquad
\subfigure[The cross-section of a simply-connected body with a through hole and a cut surface.]{
\includegraphics[width=0.3\textwidth]{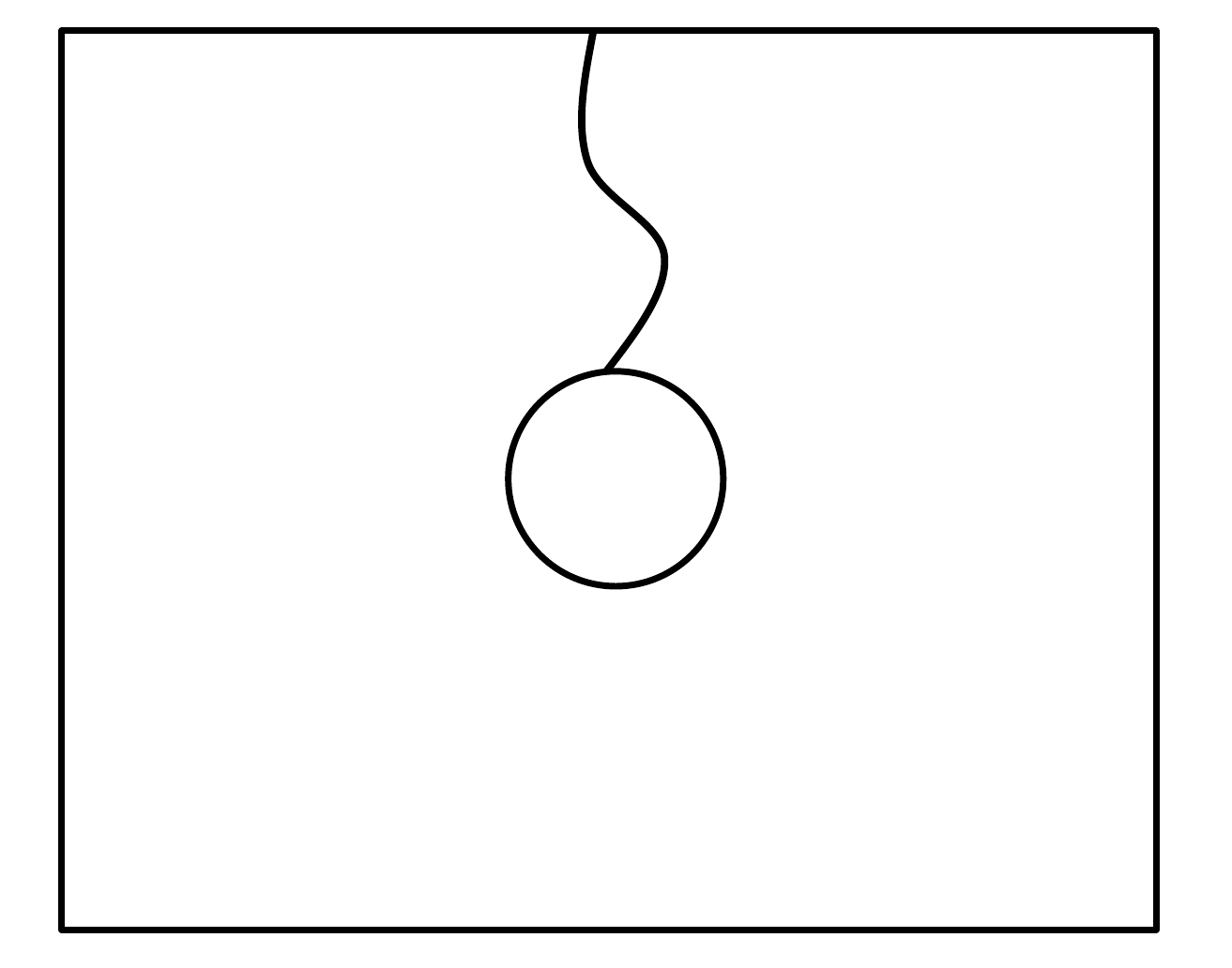}
\label{fig:weingarten_hole_2}
}
\caption{The cross-sections of multi-connected bodies with a toroidal hole or a through hole, and their corresponding simply-connected bodies by introducing cut-surfaces.}
\label{fig:multi_body}
\end{figure}

In this section we provide an overview of the Weingarten-gd theorem for g.disclinations introduced in \cite{acharya2015continuum}. Figure \ref{fig:multi_body} shows cross-sections of three dimensional multi-connected bodies with toroidal (Figure \ref{fig:weingarten_toroidal_1}) and through holes (Figure \ref{fig:weingarten_hole_1}). In both cases, the multi-connected body can be transformed into a simply-connected one by introducing a cut-surface. For the toroidal case, putting the cut-surface either from a curve on the external surface to a curve on the exterior surface of the torus (Figure \ref{fig:weingarten_toroidal_2}) or putting the cut-surface with bounding curve along the interior surface of the torus (Figure \ref{fig:weingarten_toroidal_3}) will make the multi-connected domain into a simply-connected domain. Similarly, the body with the through-hole can be cut by a surface extending from a curve on the external surface to the surface of the hole. Figures \ref{fig:weingarten_toroidal_2} and \ref{fig:weingarten_hole_2} result in topological spheres while Fig. \ref{fig:weingarten_toroidal_3} results in a topological sphere with a contained interior cavity. In terms of g.disclination theory, the holes are associated with the cores of the defect lines.

Given a continuously differentiable 3-order tensor field $\tilde{\bfY}$ on the multi-connected domain such that $\tilde{\bfY}$ is symmetric in the last two indices and $\curl \, \tilde{\bfY} = \bf0$, the Weingarten-gd problem asks if there exists a vector field $\bfy$ on the cut-induced simply-connected domain such that 
\begin{equation*}
\grad  \grad \bfy = \tilde{\bfY},
\end{equation*}
and a formula for the possible jump $[\![ \bfy ]\!]$ of $\bfy$ across the cut-surface. Also, since $\tilde{\bfY}$ is curl-free and continuously differentiable on the multi-connected domain, we can defined a field $\tilde{\bfW}$ such that 
\[
\grad \tilde{\bfW} = \tilde{\bfY}
\]
on the corresponding simply-connected domain. 

In the following, we will assign a unit normal field to any cut-surface. For any point on the cut-surface, say $\bfA$, we will denote by $\bfA^+$ a point arbitrarily close to $\bfA$ from the region into which the normal at $\bfA$ points and as $\bfA^-$ a similar point from the region into which the negative normal points. For any smooth function, say $\bff$, defined on the (multi)-connected domain, we will define 
\begin{equation}\label{+-}
\bff^+(\bfA) := \lim_{\bfA^+ \rightarrow \bfA} \bff(\bfA^+) \ \ \mbox{and} \ \   \bff^-(\bfA) := \lim_{\bfA^- \rightarrow \bfA} \bff(\bfA^-).
\end{equation}

\begin{figure}
\centering
\includegraphics[width=0.6\textwidth]{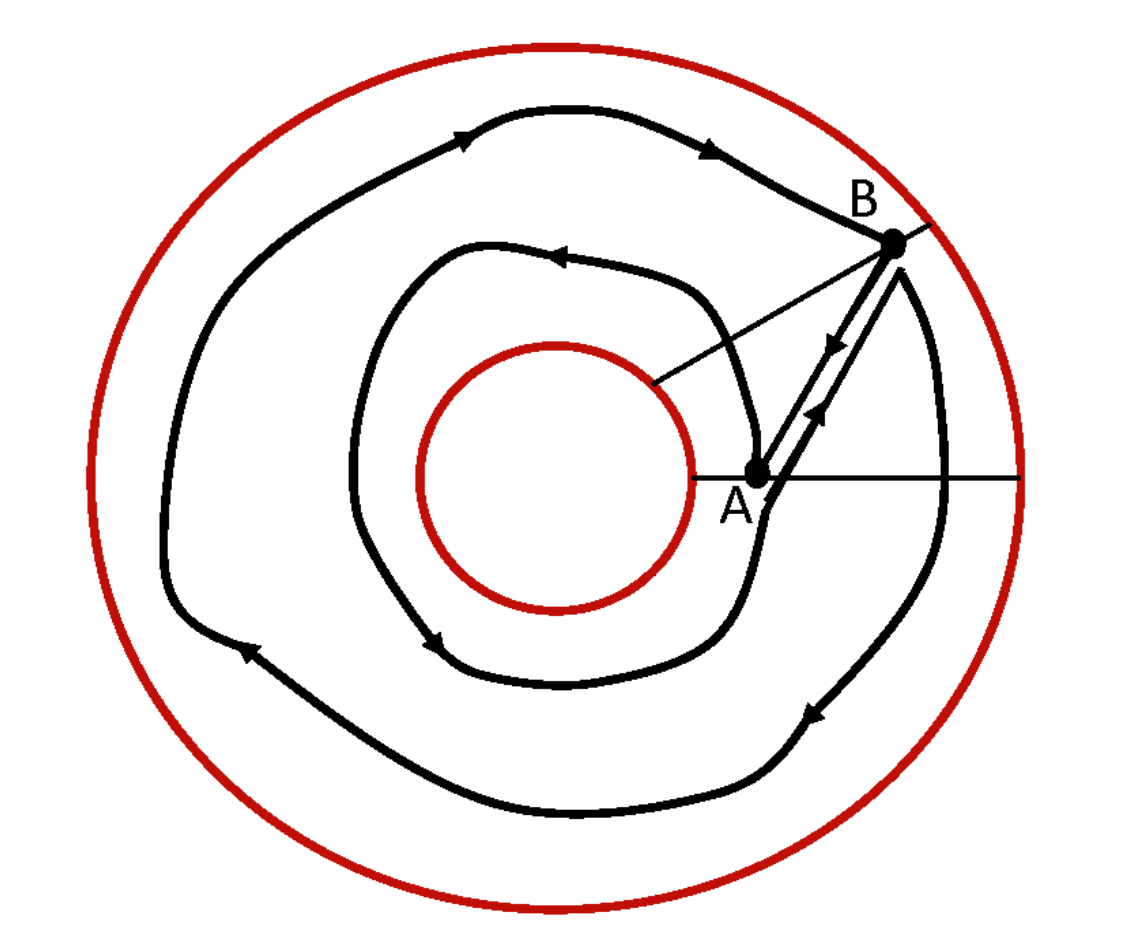}
\caption{A contour enclosing the core on the cross-section of the multi-connected domain. The contour passes through points $\bfA$ and $\bfB$. (Figure reproduced from \cite{acharya2015continuum} with permission from Springer).}
\label{fig:loop}
\end{figure}

Consider a closed contour in Fig. \ref{fig:loop} in the multi-connected domain starting and ending at $\bfA$ and passing through $\bfB$ as shown. In addition, also consider as the `inner' and `outer' closed contours the closed curves that remain by ignoring the overlapping segments, the inner closed contour passing through $\bfA$ and the outer through $\bfB$. Then, because of the continuity of $\tilde{\bfY}$ and its vanishing $curl$, the line integral of $\tilde{\bfY}$ on the inner and outer closed contours must be equal and this statement holds for any closed contour enclosing the hole. The line integral of $\tilde{\bfY}$ on the closed contour is defined as

\begin{equation*}
\int_C \tilde{\bfY} d\bfx =: \bfDelta.
\end{equation*}

Now, considering the cut-surface passing through $\bfA$, if we construct the corresponding $\tilde{\bfW}$, say $\tilde{\bfW}_1 $, then the jump of $\tilde{\bfW}_1$ is given by
\begin{equation*}
[\![ \tilde{\bfW}_1 ]\!] (\bfA) = \int\limits_{C(A^-,A^+)} \tilde{\bfY} d\bfx = \bfDelta,
\end{equation*}
where $C(A^-,A^+)$ is the curve from the inner closed contour with the point $\bfA$ taken out and with start-point $\bfA^-$ and the end-point $\bfA^+$, as shown in Figure \ref{fig:loop}. Similarly, a different cut-surface passing through another point $\bfB$ can be introduced and the corresponding $\tilde{\bfW}_2$ can be constructed with $[\![\tilde{\bfW}_2]\!](\bfB)=\bfDelta$. Since $\bfA$, $\bfB$ and the cut surfaces are chosen arbitrarily, the jump of any of the functions $[\![\tilde{\bfW}]\!]$ across their corresponding cut-surface takes the same value, independent of the invoked cut-surface and the point on the surface. 

In addition, due to the symmetry in the last two indices of $\tilde{\bfY}$, $\curl \, \bf{\tilde\bfW}$ vanishes. Thus, a vector field $\bfy$ can be defined, on the relevant  cut-induced simply-connected domain associated with the construction of $\tilde{\bfW}$, such that 
\begin{equation}\label{y_from_W}
\grad \bfy = \tilde{\bfW}.
 \end{equation} 
 Now choose a point $\bfx_0$ arbitrarily on the cut-surface. Let $\bfx$ be any other point on this cut-surface, as shown in Figure \ref{fig:burgers}. 
 
 \begin{figure}
\centering
\includegraphics[width=0.8\textwidth]{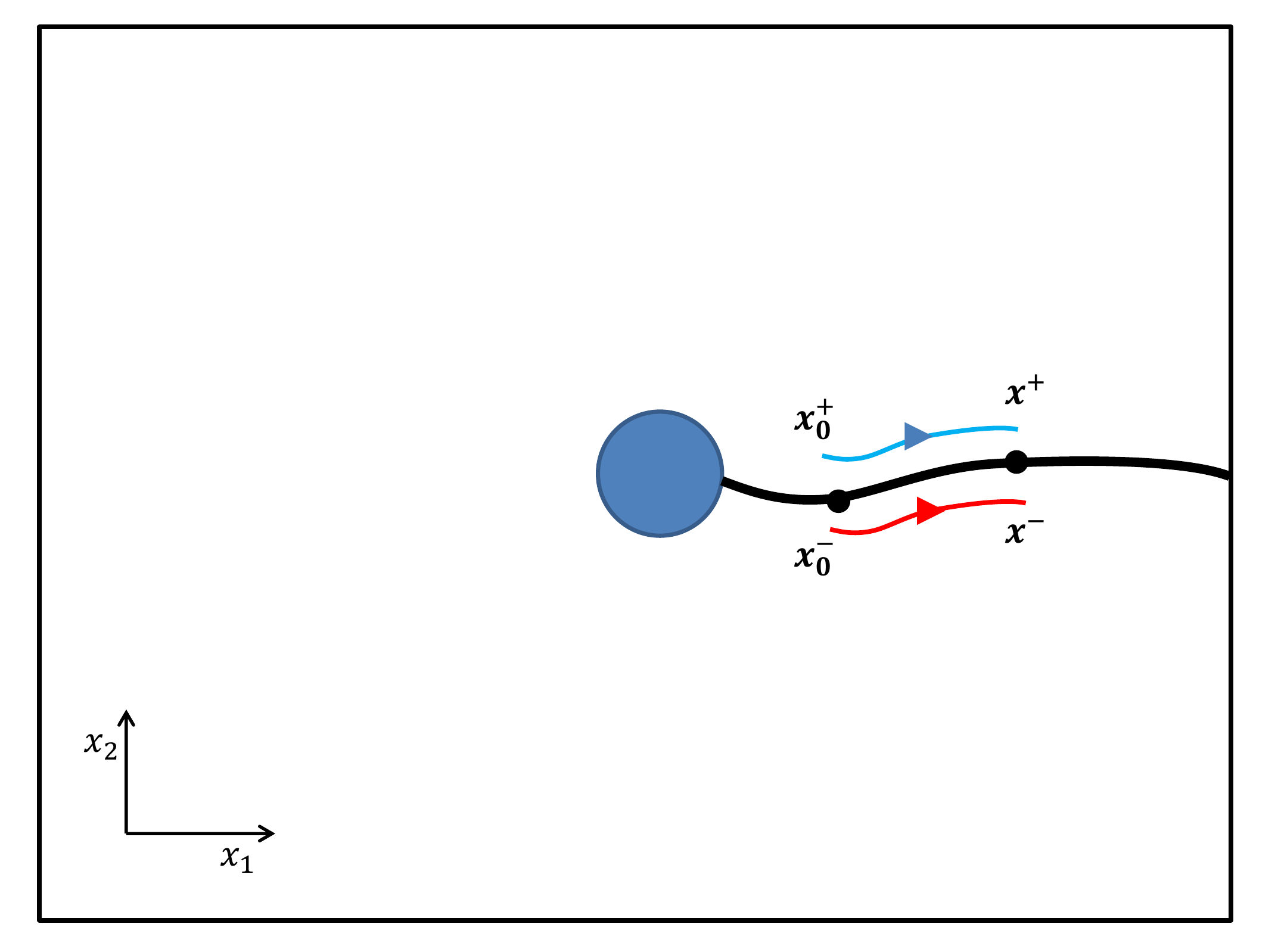}
\caption{Cross-section of a simply-connected domain induced by a cut-surface. The red path is from $\bfx_0^-$ to $\bfx^-$ and the blue path is from $\bfx_0^+$ to $\bfx^+$.}
\label{fig:burgers}
\end{figure}

Since 

\begin{eqnarray*}
\tilde{\bfW}^- = \left(\grad \bfy\right)^- \\
\tilde{\bfW}^+ = \left(\grad \bfy\right)^+,
\end{eqnarray*}

then $\bfy$ at $\bfx$ across the cut-surface is 
\begin{eqnarray*}
\bfy^+(\bfx) = \bfy^+(\bfx_0) + \int_{\bfx_0^+}^{\bfx^+} \tilde{\bfW}^+(\bfx')d\bfx' \\
\bfy^-(\bfx) = \bfy^-(\bfx_0) + \int_{\bfx_0^-}^{\bfx^-} \tilde{\bfW}^-(\bfx')d\bfx'
\end{eqnarray*}
(by working on paths from $\bfx_0^{+/-}$ to $\bfx^{+/-}$ and then taking limits as the paths approach the cut-surface). Then the jump of $\bfy$, $[\![\bfy]\!]$ can be derived as 
\begin{eqnarray*}
[\![\bfy]\!](\bfx) = \bfy^{+}(\bfx)  - \bfy^{-}(\bfx) 
= [\![\bfy]\!](\bfx_0) + \int_{\bfx_0}^{\bfx} [\![\tilde{\bfW}]\!](\bfx')d\bfx'.
\end{eqnarray*} 
Recall that $[\![\tilde{\bfW}]\!](\bfx') = \bfDelta$, which is independent of the cut-surface and the point $\bfx'$ on it.

Therefore,
\begin{equation}\label{eqn:weingarten}
[\![\bfy]\!](\bfx) = [\![\bfy]\!](\bfx_0) + \bfDelta (\bfx-\bfx_0).
\end{equation} 
Furthermore, it can be shown that the jump at any point $\bfx$ on the cut-surface is independent of the choice of the base point $\bfx_0$ (on the cut-surface).

In addition, consider the case where $[\![\tilde{\bfW}]\!] = \bf0$ - i.e. the defect line is a pure dislocation. Then the Burgers vector is defined as 
\begin{equation*}
\bfb(\bfx) := [\![\bfy]\!](\bfx),
\end{equation*}
and from Eqn \ref{eqn:weingarten}, given an arbitrarily fixed cut-surface, we have
\begin{equation}\label{weingarten_jump}
[\![\bfy]\!](\bfx) = [\![\bfy]\!](\bfx_0)
\end{equation}
where $\bfx$ and $\bfx_0$ are arbitrarily chosen points on the cut-surface. Thus, it has been shown that for $\bf\Delta = \bf0$, the  displacement jump is independent of location on a given cut-surface. 

In the next Section \ref{sec:cut_independ}, we furthermore show that the Burgers vector is independent of the choice of the cut-surface as well when $\bfDelta = \bf0$ and $\bfalpha$ is localized in the core.

\subsection{Cut-surface independence in the Weingarten-gd theorem for $\bfDelta = \bf0$} \label{sec:cut_independ}

We now prove that the jump (\ref{weingarten_jump}) across a cut-surface in the Weingarten-gd theorem is independent of the choice of the surface when $\bfDelta = \bf0$. 

By hypothesis, there is a continuous field $\tilde{\bfY}$ in the multi-connected body with $\curl \tilde{\bfY} = \bf0$ and $\tilde{Y}_{ijk} = \tilde{Y}_{ikj}$. Also, after introducing an arbitrary cut-surface, as in Figure \ref{fig:weingarten_hole_2}, we can construct fields $\tilde{\bfW}$ and $\tilde{\bfy}$ such that $\grad (\grad \,\tilde{\bfy}) = \tilde{\bfY}$ and $\grad \,\tilde{\bfW} = \tilde{\bfY}$. Based on the Weingarten-gd theorem, we have on this arbitrary chosen cut-surface
\begin{equation*}
[\![\tilde{\bfy}(\bfx)]\!] = [\![\tilde{\bfy}(\bfx_0)]\!] + \bfDelta (\bfx-\bfx_0),
\end{equation*}
where $\bfDelta = \oint \tilde{\bfY} d\bfx$. Since $\bfDelta = \bf0$, it is clear that
\begin{equation*}
[\![\tilde{\bfy}(\bfx)]\!] = [\![\tilde{\bfy}(\bfx_0)]\!].
\end{equation*}

The goal now is to prove that 
\begin{equation*}
[\![\tilde{\bfy}(\bfx)]\!] =:\bfb
\end{equation*}
where the vector $\bfb$ is independent of the choice of the cut-surface and, hence, independent of $\bfx$ on the cut-surface as well using results of Section \ref{sec:weingarten}.

\begin{figure}
\centering
\includegraphics[width=0.4\textwidth]{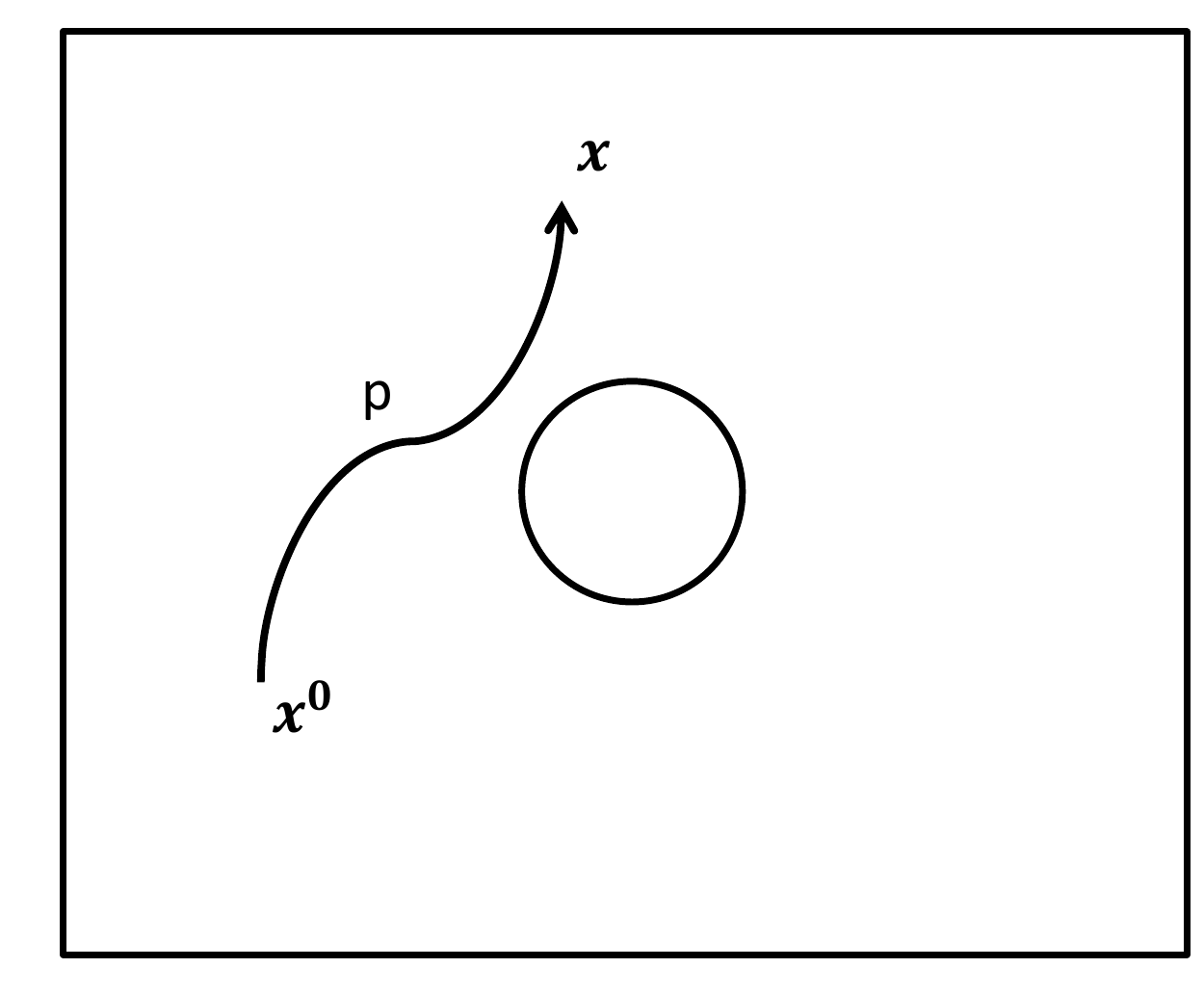}
\caption{Arbitrary path (shown on the cross-section) for construction of a continuous field $\bfW^\tau$ on the simply-connected domain induced by a cut-surface $\tau$  when $\bfDelta = \bf0$.}
\label{fig:cut_proof_3}
\end{figure}

Since the definition of $\tilde{\bfW}$ depends on the cut-surface, given a simply-connected domain induced by a cut-surface $\tau$, we can express any such $\tilde{\bfW}$, say  $\bfW^\tau$, on the simply-connected domain as
\begin{equation}\label{eqn:w_definition}
W^\tau_{ij}(\bfx,\bfx^0,p) := \underset{p}{\int_{\bfx^0}^{\bfx}} E_{ijk} \, dx_k + W^\tau_{ij}(\bfx^0)
\end{equation}
where $p$ is a curve from $\bfx^0$ to $\bfx$ as shown in Figure \ref{fig:cut_proof_3}, and $\bfE := \tilde{\bfY}$ with the field $\tilde{\bfY}$ satisfying the constraint $\oint \tilde{\bfY} \, d\bf{x} = \bfDelta = \bf0$. Since the line integral of $\bfE$ on any closed loop is zero, $\bfW^\tau$ as defined is independent of path on the original multi-connected domain and hence thinking of the constructed $\bfW^\tau$ as a continuous function on it makes sense.

Now with the constructed $\bfW^\tau$, we can define the line integral 
\begin{equation*}
\bfb^{\tau,p}:= \underset{p}{\oint} \bfW^\tau d\bfx 
\end{equation*}
on a closed loop $p$ enclosing the core.

We now show first that $\bfb^{\tau,p}$ is independent of the loop used to define it. Since $\grad \, \bfW^\tau = \bfE$ from the definition (\ref{eqn:w_definition}) and $\bfE$ is symmetric in the last two indices by hypothesis, 
\begin{equation*}
\begin{aligned}
&E_{ijk} = W^\tau_{ij,k} \\
&\Rightarrow E_{ijk}-E_{ikj} = W^\tau_{ij,k}-W^\tau_{ik,j} = 0 \\
&\Rightarrow e_{mkj}( W^\tau_{ij,k}-W^\tau_{ik,j}) = 0 \\
&\Rightarrow \curl \bfW^\tau = \bf0
\end{aligned}
\end{equation*}
\begin{figure}
\centering
\includegraphics[width=0.6\textwidth]{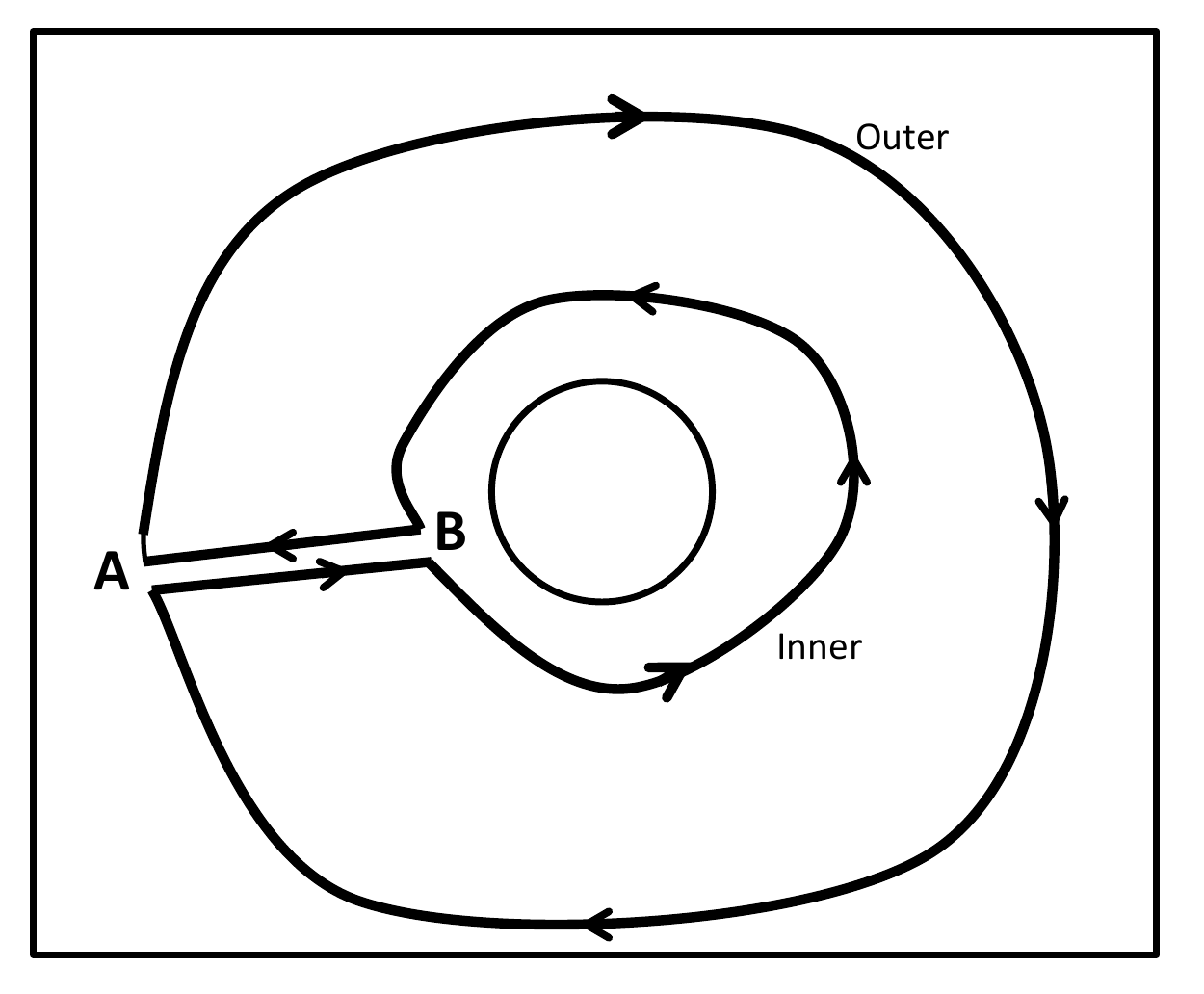}
\caption{A closed loop in the cross-section of a multi-connected domain to justify independence of the Burgers vector from the circuit used to evaluate it for $\bfDelta = \bf0$.}
\label{fig:cut_proof_4}
\end{figure}
Thus, on the multi-connected domain, given any arbitrary closed loop as in Figure \ref{fig:cut_proof_4}, 
\begin{align*}
&\oint \bfW^\tau d\bfx = \bf0 \\
\underset{\text{inner}} {\int} \bfW^\tau d\bfx  -   \underset{\text{outer}}{\int} \bfW^\tau d\bfx &+ \int_B^A \bfW^\tau d\bfx + \int_A^B \bfW^\tau d\bfx  = \bf0
\end{align*}
where $ \!\!\int_{inner} \bfW^\tau d\bfx$ is the integral along the inner loop anti-clockwise and $\!\! \int_{outer} \bfW^\tau d\bfx$ is the integral along the outer loop anti-clockwise. Therefore, since $\int_B^A \bfW^\tau d\bfx + \int_A^B \bfW^\tau d\bfx = \bf0$,
\begin{equation*}
\underset{\text{inner}} {\int} \bfW^\tau d\bfx  = \underset{\text{outer}} {\int} \bfW^\tau d\bfx.
\end{equation*}
Thus, $\bfb^{\tau,p}$ is independent of the loop path $p$ and we will denote it as $\bfb^{\tau}$.

\begin{figure}
\centering
\includegraphics[width=0.6\textwidth]{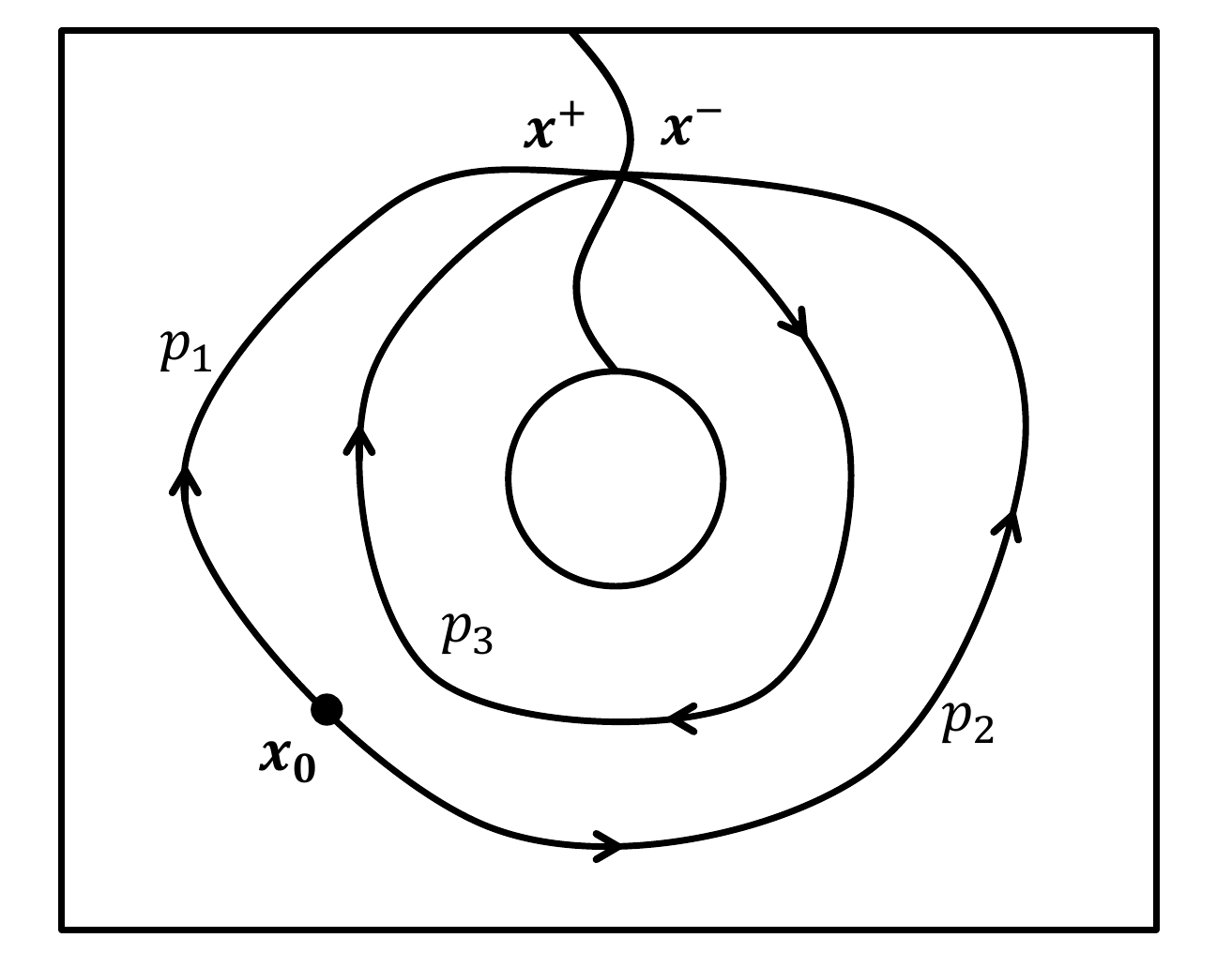}
\caption{The integration path on the cross-section of a simply-connected domain to calculate the jump $\llbracket {\bfy} \rrbracket$ at $\bfx$.}
\label{fig:cut_proof_5}
\end{figure}

Now for a simply-connected domain induced by the cut-surface $\tau$, given a $\bfW^\tau$, there exists (many) $\bfy^\tau$ satisfying $\grad \bfy^\tau = \bfW^\tau$; any such $\bfy^\tau$ may be expressed as 
\begin{equation*}
\bfy^\tau(\bfx;\bfx_0) = \int_{\bfx_0}^{\bfx} \bfW^\tau d\bfx + \bfy^{\tau}(\bfx_0).
\end{equation*}
Then, with reference to Fig. \ref{fig:cut_proof_5}, we have
\begin{align*}
\bfy^\tau(\bfx^+) = \bfy^\tau(\bfx_0) + \underset{p_1} {\int_{\bfx_0}^{\bfx^+}} \bfW^\tau d\bfx \\
\Rightarrow \bfy^\tau(\bfx^+) = {\bfy^\tau}(\bfx_0) + \underset{p_2} {\int_{\bfx_0}^{\bfx^-}}\bfW^\tau d\bfx + \underset{p_3} {\int_{\bfx^-}^{\bfx^+}} \bfW^\tau d\bfx
\end{align*}
and 
\begin{equation*}
\bfy^\tau(\bfx^-) = \bfy^\tau(\bfx_0) + \underset{p_2} {\int_{\bfx_0}^{\bfx^-}} \bfW^\tau d\bfx.
\end{equation*}
Thus,
\begin{equation}\label{eqn:b_tau}
[\![ {\bfy^\tau}]\!] =\underset{p_3}{\int_{\bfx^-}^{\bfx^+}}\bfW^\tau d\bfx = \bfb^\tau.
\end{equation}

We note next that if $\tau$ and $\tau'$ are two cut-surfaces, \eqref{eqn:w_definition} and the continuity of $\bfW^{\tau}, \bfW^{\tau'}$ imply that $\bfW^{\tau} -\bfW^{\tau'}$ is a constant tensor on the original multi-connected domain, and therefore \eqref{eqn:b_tau} implies that $\bfb^\tau = \bfb^{\tau'} =:\bfb$, a constant vector independent of the cut-surface.

Therefore, we have shown that when $\bfDelta=\bf0$, the Burgers vector is cut-surface independent.

 \section{Interpretation of the Weingarten theorem in terms of g.disclination kinematics}\label{sec:rel_gdisclin_wein}
 
The Weingarten-gd theorem for generalized disclinations (\ref{eqn:weingarten}) was reviewed in Section \ref{sec:weingarten}. We recall that the jump of the inverse-deformation $\bfy$ across a cut-surface is characterized, in general, by the jump at an arbitrarily chosen point on the surface, $\llbracket \bfy(\bfx_0) \rrbracket$, and $\bfDelta = \llbracket \tilde{\bfW} \rrbracket$, and all of these quantities are defined from the knowledge of the field $\tilde{\bfY}$. However, given a g.disclination and a dislocation distribution $\bfPi$ and $\bfalpha$, respectively, on the body, it is natural to ask as to what ingredients of g.disclination theory correspond to a candidate $\tilde{\bfY}$ field. A consistency condition we impose is that in the absence of a g.disclination density, the jump of the inverse deformation field should be characterized by the Burgers vector of the given dislocation density field.
 
 \subsection{Derivation of $\tilde{\bfY}$ in g.disclination theory}
 
\begin{figure}
\centering
\includegraphics[width=0.5\textwidth]{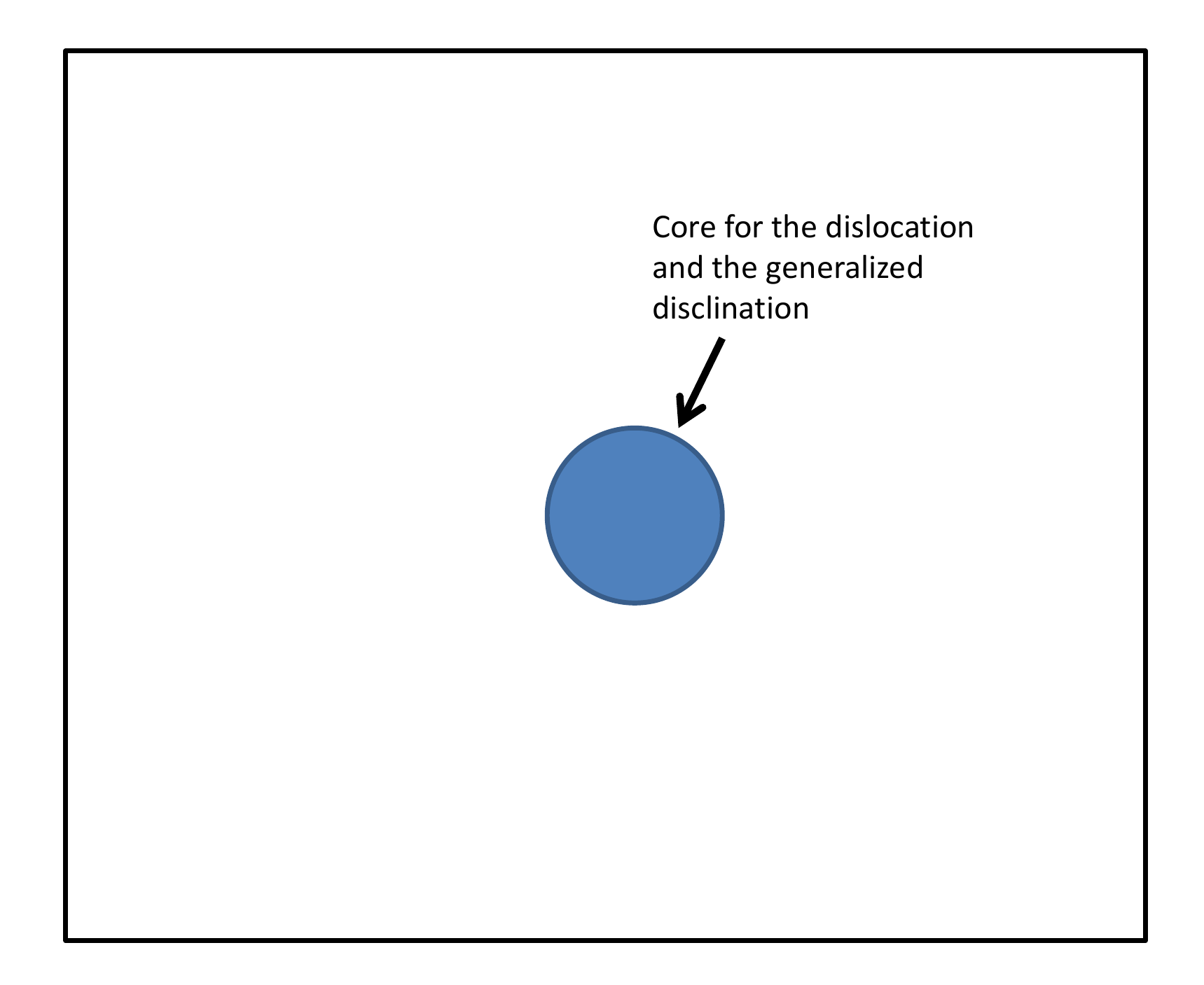}
\caption{The dislocation and the generalized disclination densities are localized in a defect core.}
\label{fig:burgers_core}
\end{figure}

  Let $\bfPi$ and $\bfalpha$ be localized in a core as shown in the Figure \ref{fig:burgers_core}. In g.disclination theory,
\begin{equation*}
\begin{aligned}
&\bfY  = \bfS + \grad \bfW \\
&\bfalpha := \bfS :\bfX + \grad \bfW : \bfX
\end{aligned}
\end{equation*}
with 
\begin{equation*}
\curl \, \bfY = \curl \, \bfS =\bfPi.
\end{equation*}

From the definition of $\bfalpha$, we have 
\begin{eqnarray*}
\begin{aligned}
&\alpha_{ij} = S_{imn}e_{jmn}+W_{im,n}e_{jmn} \\
\Rightarrow \  & \alpha_{ij}e_{jrs} = \left( S_{imn}+W_{im,n} \right) \left(\delta_{mr}\delta_{ns}-\delta_{ms}\delta_{nr}\right) \\
\Rightarrow \  & \left( Y_{irs} - \frac{1}{2} \alpha_{ij}e_{jrs} \right) - \left( Y_{isr} - \frac{1}{2} \alpha_{ij}e_{jsr}\right) = 0.
\end{aligned} 
\end{eqnarray*}

Now, we \emph{define $\tilde{\bfY}$, in terms of ingredients of g.disclination theory,} as 
\begin{equation}\label{eqn:tildey}
\tilde{Y}_{imn} := Y_{imn}-\frac{1}{2}\alpha_{ij}e_{jmn} = S_{imn}+W_{im,n}-\frac{1}{2}\alpha_{ij}e_{jmn}
\end{equation}
and verify that
\begin{equation*}
\tilde{Y}_{imn} = \tilde{Y}_{inm}.
\end{equation*}
Therefore, $\tilde{\bfY}$ is symmetric in the last two indices. In addition, 
\begin{eqnarray*}
\begin{aligned}
&(\curl \, \tilde{\bfY})_{imr} = e_{rpn}\tilde{Y}_{imn,p} = S_{imn,p}e_{rpn} - W_{im,np}e_{rpn}-\frac{1}{2}\alpha_{il,p}e_{lmn}e_{rpn} \\
\Rightarrow \ & (\curl \, \tilde{\bfY})_{imr} = \Pi_{imr}-\frac{1}{2}\alpha_{il,p}e_{lmn}e_{rpn}.
\end{aligned}
\end{eqnarray*}
Since $\bfPi$ and $\bfalpha$ are localized in the core, $\curl \, \tilde{\bfY}$ is localized in the core and $\curl \, \tilde{\bfY} = \bf0$ outside the core.

%%
%%In addition, based on the dislocation density definition,
%%\begin{equation*}
%%\alpha_{ij} = S_{imn}e_{jmn}+W_{im,n}e_{jmn} \\
%%\Rightarrow \alpha_{ij,s}\delta_{js} = S_{imn,s}e_{jmn}\delta_{js} + W_{im,ns}e_{jmn}\delta_{js} \\
%%\Rightarrow \alpha_{ij,j}+ (curl \bfS)_{ijj} = 0 
%%\end{equation*}
%%Recall that $curl \bfS = \bfPi$, then we have
%%\begin{equation}\label{eqn:constraint}
%% \alpha_{ij,j} + \Pi_{ijj} = 0.
%%\end{equation}
%%Eqn[\ref{eqn:constraint}] serves as a necessary condition for the defect density field.

\subsection{Jump of inverse deformation in terms of defect strengths in g.disclination theory} \label{sec:rel_gdisclin_wein_2}

With the definition of $\tilde{\bfY}$ in terms of g.disclination theory (\ref{eqn:tildey}), we will now identify $\bfDelta$  and the jump of the inverse deformation across a cut-surface in a canonical example in terms of prescribed data used to define an isolated defect line (i.e. the strengths of the g.disclination and dislocation contained in it) and location on the surface.

A g.disclination in a thick infinite plate in the $x_1-x_2$ plane is considered, with the g.disclination line in the positive $x_3$ direction. Assume the strength of the g.disclination to be $\bfDelta^{F}$ (cf. Sec. \ref{sec:g_disclination_theory} for definition of the strength). Based on the characterization of $\bfPi$ in (\ref{eqn:pi_characterize}), a candidate for the localized and smooth generalized disclination density $\bfPi$ is assumed to  only have non-zero components $\Pi_{ij3}$, namely $\bfPi = \Pi_{ij3}\bfe_i \otimes \bfe_j \otimes \bfe_3$, with $\Pi_{ij3}$ given as (cf. \cite{acharya2001model})
\begin{equation*}
\Pi_{ij3} = \psi_{ij}(r)=
\begin{cases}
\frac{\Delta^F_{ij}}{\pi r_0} \left( \frac{1}{r} - \frac{1}{r_0} \right) \quad \text{$r < r_0$} \\
0 \quad \text{$r \ge r_0$},
\end{cases}
\end{equation*} 
where $r=\sqrt{x_1^2+x_2^2}$. It is easy to verify that $\bfPi$ is a smooth field and 
\[
\int_{core} \bfPi \bfe_3 da = \bfDelta^F.
\]
Similarly, the localized and smooth dislocation density $\bfalpha$ is assumed as 
\begin{equation*}
\alpha_{i3} =  
\begin{cases}
\frac{b_i}{\pi r_0}\left( \frac{1}{r}-\frac{1}{r_0} \right) & r < r_0 \\
0 & r \ge r_0,
\end{cases}
\end{equation*}
where $\bfb$ is the Burgers vector with
\[
\int_{core} \bfalpha \bfe_3 da = \bfb.
\]

Recall that 
\begin{eqnarray}\label{eqn:def_Y}
\begin{aligned}
&\tilde{Y}_{irs} = Y_{irs} -\frac{1}{2}\alpha_{ij}e_{rsj} \\
&\Rightarrow \tilde{Y}_{irs} = S^{\perp}_{irs} + Z^s_{ir,s} + W_{ir,s} - \frac{1}{2}\alpha_{ij}e_{rsj} \\
&\Rightarrow \tilde{\bfY} = \bfS^{\perp} + \grad\bfZ^s + \grad \bfW - \frac{1}{2}\bfalpha : \bfX,
\end{aligned}
\end{eqnarray}
where $\bfZ^s$ is the compatible part of $\bfS$ and $\bfS^{\perp}$  is incompatible part of $\bfS$ that cannot be represented as the gradient, satisfying the equations
\begin{eqnarray*}
\begin{aligned}
\curl \bfS^{\perp} &= \bfPi \\
\divergence \bfS^{\perp} &= \bf0 \\
\bfS^{\perp}\bfn &= \bf0 \qquad \text{on the boundary}.
\end{aligned}
\end{eqnarray*}
One way to get the solution of $\bfS^{\perp}$ from the above equations is to decompose $\bfS^{\perp}$ as $\bfS^{\perp}= \bfS^*+\grad \bfZ^*$, where $\bfS^*$ satisfies
\begin{eqnarray}\label{eqn:s_star}
\begin{aligned}
\curl \bfS^* &= \bfPi \\
\divergence \bfS^* &= \bf0 
\end{aligned}
\end{eqnarray}
and $\bfZ^*$ satisfies
\begin{eqnarray}\label{eqn:z_star}
\begin{aligned}
&\divergence (\grad \bfZ^*) = \bf0 \\
&(\grad \bfZ^*) \bfn = -\bfS^*\bfn \qquad \text{on the boundary} .
\end{aligned}
\end{eqnarray}
The solution of (\ref{eqn:s_star}) can be acquired from the Riemann-Graves operator as shown in Appendix \ref{sec:app1}. Furthermore, since $\int_{\partial V} \bfS^* \bfn da = \int_V \divergence \bfS^* dv = \bf0$, where $\partial V$ is the boundary of $V$, a unique solution for $\grad \bfZ^*$ from (\ref{eqn:z_star}), which is the (component-wise) Laplace equation for $\bfZ^*$ with Neumann boundary conditions, exists. Substituting $\bfS^{\perp}$ into Eqn (\ref{eqn:def_Y}), we have
\begin{eqnarray}\label{S*}
\begin{aligned}
&\tilde{\bfY} = \bfS^{\perp} + \grad\bfZ^s + \grad \bfW - \frac{1}{2}\bfalpha : \bfX \\
&\Rightarrow  \tilde{\bfY} = \bfS^* + \grad \bfZ^* + \grad\bfZ^s + \grad \bfW - \frac{1}{2}\bfalpha : \bfX \\
& \Rightarrow \tilde{\bfY} := \bfS^* + \grad \bfA - \frac{1}{2}\bfalpha:\bfX,
\end{aligned}
\end{eqnarray}
where $\bfA$ is defined as $\bfA := \bfZ^*+\bfZ^s+\bfW$.

In addition, we have
\begin{eqnarray}\label{eqn:relation_A_S}
&\alpha_{ij} = S_{imn}e_{jmn} + W_{im,n}e_{jmn}  \nonumber \\ 
&\alpha_{ij} = S^*_{imn}e_{jmn} + Z^*_{im,n}e_{jmn}+Z^s_{im,n}e_{jmn}+W_{im,n}e_{jmn} \nonumber \\
&\alpha_{ij} - S^{*}_{imn}e_{jmn} = -(\curl\bfA)_{ij}.
\end{eqnarray}
Denoting $B_{ij} = S^{*}_{imn}e_{jmn} - \alpha_{ij}$, we have
\begin{equation*}
(\curl \bfA)_{ij} = B_{ij}.
\end{equation*}

As given in Appendix \ref{sec:app1},  we obtain $\bfS^{*}$ as 
 \begin{eqnarray*}
&S^{*}_{ij1} =
\begin{cases}
\frac{\Delta^F_{ij}}{2\pi}(-\frac{x_2}{r^2}) & r>r_0 \\
\frac{-x_2 \Delta^F_{ij}}{\pi r^2 r_0} (r-\frac{r^2}{2r_0}) & r \le r_0
\end{cases} \\
&S^{*}_{ij2} =
\begin{cases}
\frac{\Delta^F_{ij}}{2\pi}(\frac{x_1}{r^2}) & r>r_0 \\
\frac{x_1 \Delta^F_{ij}}{\pi r^2 r_0} (r-\frac{r^2}{2r_0})& r \le r_0.
\end{cases}
\end{eqnarray*}
Also, following similar arguments as in  Appendix \ref{sec:app1}, in Appendix \ref{sec:app2} we obtain $\bfA^*$ with $\bfA=\bfA^{*}+\grad \bfz^A$,\footnote{
While not relevant for the essentially topological arguments here, we note that it is in the field $grad\bfz^A$ that the compatible part of $\bfW$ resides which helps in satisfaction of force equilibrium.
} where $\bfA^{*}$ is given by 
\begin{eqnarray*}
&A^{*}_{11}  = 
\begin{cases}
 C_1 \left(-\Delta^F_{12} x_2^2-\Delta^F_{11} x_1 x_2 \right) + \frac{x_2}{r^2}\left[\frac{b_1}{\pi r_0}\left(r-\frac{r^2}{2r_0}\right)\right] &r < r_0 \\
C_2 \left(-\Delta^F_{12} x_2^2 - \Delta^F_{11} x_1 x_2 \right) + \frac{x_2}{r^2}\frac{b_1}{2\pi} & r \ge r_0
\end{cases}
\\
&A^{*}_{12}  = 
\begin{cases}
C_1 \left(\Delta^F_{12} x_2 x_1 + \Delta^F_{11} x_1^2\right)- \frac{x_1}{r^2}\left[\frac{b_1}{\pi r_0}\left(r-\frac{r^2}{2r_0}\right)\right]  &r < r_0 \\
C_2 \left(\Delta^F_{12} x_2 x_1 + \Delta^F_{11} x_1^2\right) - \frac{x_1}{r^2}\frac{b_1}{2\pi} & r \ge r_0
\end{cases}
\\
&A^{*}_{21}  = 
\begin{cases}
 C_1 \left(-\Delta^F_{22} x_2^2-\Delta^F_{21} x_1 x_2\right) + \frac{x_2}{r^2}\left[\frac{b_2}{\pi r_0}\left(r-\frac{r^2}{2r_0}\right)\right] &r < r_0 \\
C_2 \left(-\Delta^F_{22} x_2^2 - \Delta^F_{21} x_1 x_2\right) + \frac{x_2}{r^2}\frac{b_2}{2\pi} & r \ge r_0
\end{cases}
\\
&A^{*}_{22}  = 
\begin{cases}
C_1 \left(\Delta^F_{22} x_2 x_1 + \Delta^F_{21} x_1^2\right) - \frac{x_1}{r^2}\left[\frac{b_2}{\pi r_0}\left(r-\frac{r^2}{2r_0}\right)\right] &r < r_0 \\
C_2 \left(\Delta^F_{22} x_2 x_1 + \Delta^F_{21} x_1^2\right) - \frac{x_1}{r^2}\frac{b_2}{2\pi} & r \ge r_0
\end{cases}
\end{eqnarray*}
and $C_1= \frac{1}{2\pi r_0 r}- \frac{1}{6 \pi r_0^2}$ and $C_2=\frac{1}{3\pi r_0 r}+\frac{r-r_0}{2\pi r^3}$. $\bfA^{*}$ can be decomposed into two parts. The first part is the terms associated with $C_1$ and $C_2$, denoted as $\bfA^o$; the other part is the remaining terms associated with $\bfb$, denoted as $\bfA^{\alpha}$. Thus, $\bfA=\bfA^o+\bfA^{\alpha}+\grad \bfz^A$. $\bfA^o$ and $\bfA^{\alpha}$ are given as 

\begin{eqnarray*}
&A^o_{11}  = 
\begin{cases}
 \left( \frac{1}{2\pi r_0 r}- \frac{1}{6 \pi r_0^2} \right) \left(-\Delta^F_{12} x_2^2-\Delta^F_{11} x_1 x_2 \right) &r < r_0 \\
\left( \frac{1}{3\pi r_0 r}+\frac{r-r_0}{2\pi r^3} \right) \left( -\Delta^F_{12} x_2^2 - \Delta^F_{11} x_1 x_2 \right) & r \ge r_0
\end{cases}
\\
&A^o_{12}  = 
\begin{cases}
 \left( \frac{1}{2\pi r_0 r}- \frac{1}{6 \pi r_0^2} \right) \left( \Delta^F_{12} x_2 x_1 + \Delta^ F_{11} x_1^2 \right) &r < r_0 \\
\left( \frac{1}{3\pi r_0 r}+\frac{r-r_0}{2\pi r^3} \right) \left( \Delta^ F_{12} x_2 x_1 + \Delta^ F_{11} x_1^2 \right) & r \ge r_0
\end{cases}
\\
&A^o_{21}  = 
\begin{cases}
\left( \frac{1}{2\pi r_0 r}- \frac{1}{6 \pi r_0^2} \right)\left( -\Delta ^F_{22} x_2^2-\Delta^ F_{21} x_1 x_2 \right) &r < r_0 \\
\left( \frac{1}{3\pi r_0 r}+\frac{r-r_0}{2\pi r^3} \right) \left(-\Delta^ F_{22} x_2^2 - \Delta^ F_{21} x_1 x_2 \right) & r \ge r_0
\end{cases}
\\
&A^o_{22}  = 
\begin{cases}
 \left( \frac{1}{2\pi r_0 r}- \frac{1}{6 \pi r_0^2} \right) \left( \Delta^ F_{22} x_2 x_1 + \Delta^ F_{21} x_1^2 \right) &r < r_0 \\
\left( \frac{1}{3\pi r_0 r}+\frac{r-r_0}{2\pi r^3} \right) \left(\Delta^ F_{22} x_2 x_1+ \Delta^ F_{21} x_1^2 \right) & r \ge r_0
\end{cases}
\\&
A^{\alpha}_{11}  = 
\begin{cases}
 \frac{x_2}{r^2}\left[\frac{b_1}{\pi r_0}\left(r-\frac{r^2}{2r_0}\right)\right] &r < r_0 \\
\frac{x_2}{r^2}\frac{b_1}{2\pi} & r \ge r_0
\end{cases}
\\
&A^{\alpha}_{12}  = 
\begin{cases}
\frac{x_1}{r^2}\left[\frac{b_1}{\pi r_0}\left(r-\frac{r^2}{2r_0}\right)\right]  &r < r_0 \\
\frac{x_1}{r^2}\frac{b_1}{2\pi} & r \ge r_0
\end{cases}
\\
&A^{\alpha}_{21}  = 
\begin{cases}
 \frac{x_2}{r^2}\left[\frac{b_2}{\pi r_0}\left(r-\frac{r^2}{2r_0}\right)\right] &r < r_0 \\
 \frac{x_2}{r^2}\frac{b_2}{2\pi} & r \ge r_0
\end{cases}
\\
&A^{\alpha}_{22}  = 
\begin{cases}
\frac{x_1}{r^2}\left[\frac{b_2}{\pi r_0}\left(r-\frac{r^2}{2r_0}\right)\right] &r < r_0 \\
 \frac{x_1}{r^2}\frac{b_2}{2\pi} & r \ge r_0.
 \end{cases}
\end{eqnarray*}

We now consider a multiply connected domain by thinking of the cylinder with the core region excluded and introduce a simply-connected domain by a cut-surface. On this simply-connected domain we define a $\tilde{\bfW}$ satisfying $\grad \, \tilde{\bfW} = \tilde{\bfY}$ and recall that $\tilde{\bfY} = \bfS^{*} + \grad \bfA - \frac{1}{2}\bfalpha:\bfX$. Then
\begin{eqnarray}\label{eqn:tildeW}
&\tilde{\bfW}(\bfx) = \int_{\bfx_r}^{\bfx} \tilde{\bfY}(\bfs) d\bfs + \tilde{\bfW}(\bfx_r)\\
&\tilde{\bfW}(\bfx) = \int_{\bfx_r}^{\bfx} [\bfS^{*} + \grad \bfA - \frac{1}{2}\bfalpha:\bfX](\bfs) d\bfs + \tilde{\bfW}(\bfx_r) \\
&\tilde{\bfW}(\bfx) = \int_{\bfx_r}^{\bfx}\bfS^{*}(\bfs) d\bfs -\frac{1}{2} \int_{\bfx_r}^{\bfx}(\bfalpha:\bfX)(\bfs) d\bfs + \bfA(\bfx) + const,
\end{eqnarray}
where $\bfx_r$ is a given point and $const$ is the constant $\tilde{\bfW}(\bfx_r)-\bfA(\bfx_r)$, with $\tilde{\bfW}(\bfx_r)$ being arbitrarily assignable. 

Consider a path $p$ from $\bfz^-$ to $\bfz^+$, both points arbitrarily close to $\bfz \in S$, on opposite sides of $S$ (see (\ref{+-}) for notation) and define 
\[
[\![\bfy (\bfz) ] \!] = \lim_{\substack{\bfz^+ \rightarrow \bfz\\ \bfz^- \rightarrow \bfz}} \int_p \tilde{\bfW}(\bfx) d\bfx
\]
for any $\bfz$ on the cut-surface.

\begin{figure}
\centering
\includegraphics[width=0.5\textwidth]{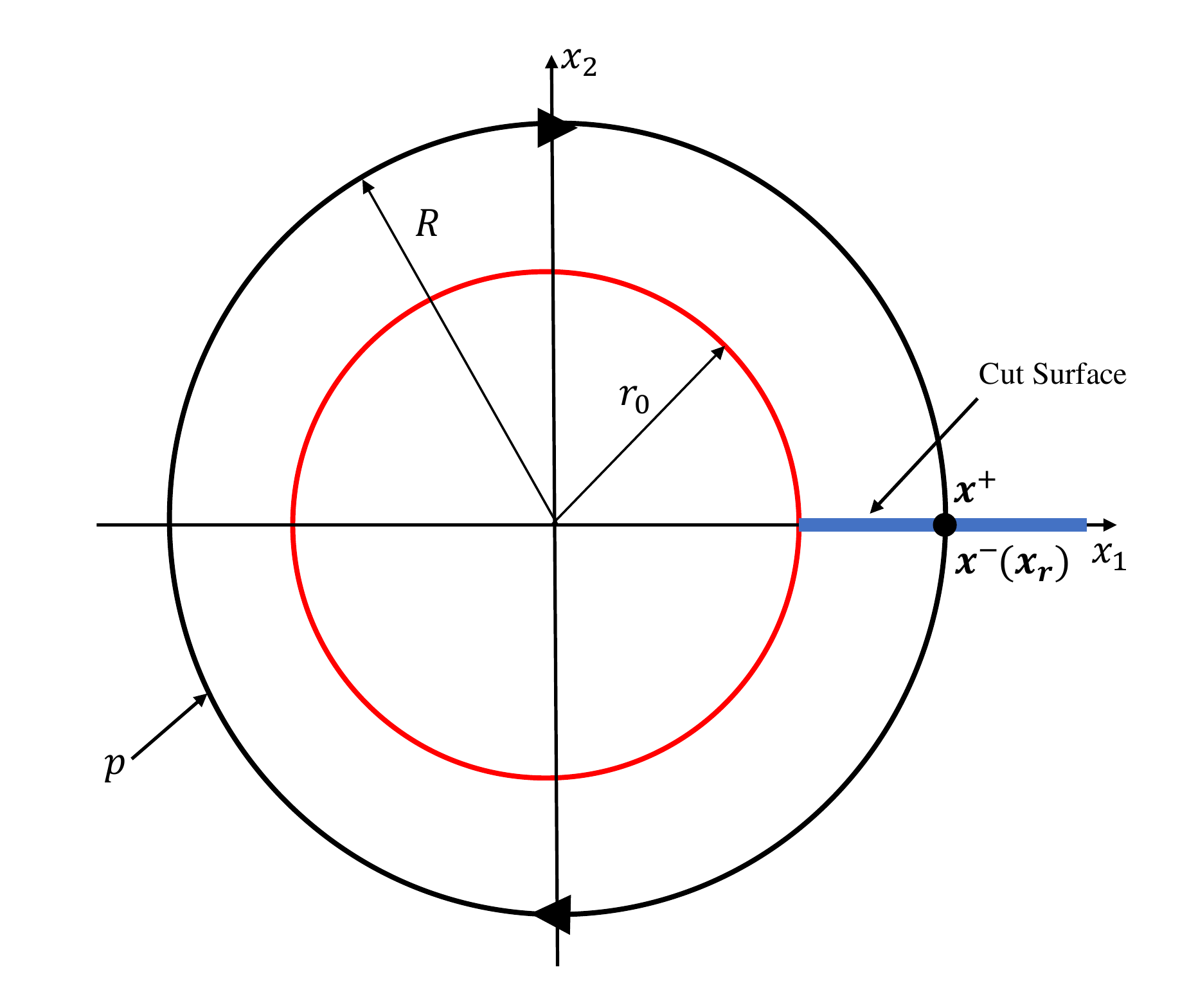}
\caption{Configuration of a closed contour enclosing the disclination and passing through $\bfx$.}
\label{fig:contour}
\end{figure}

After substituting $\tilde{\bfW}$ and noticing $\oint_p const \,d\bfx = \bf0$ and $\oint_p \grad \bfz^A \,d\bfx = \bf0$, the jump at $\bfx$ can be further written as 
\begin{eqnarray} \label{eqn:jump_y}
[\![\bfy]\!] = \int_p \int_{\bfx_r}^{\bfx} \bfS^{*}(\bfs) d\bfs d\bfx + \int_p \bfA(\bfx)d\bfx - \frac{1}{2} \int_p \int_{\bfx_r}^{\bfx} (\bfalpha:\bfX)(\bfs) d\bfs d\bfx \nonumber \\
= \int_p \int_{\bfx_r}^{\bfx} \bfS^{*}(\bfs) d\bfs d\bfx + \int_p \bfA^o(\bfx)d\bfx + \int_p \bfA^{\alpha}(\bfx)d\bfx - \frac{1}{2} \int_p \int_{\bfx_r}^{\bfx} (\bfalpha:\bfX)(\bfs) d\bfs d\bfx.
\end{eqnarray}
 Now suppose $\bfx$ located as $(R,0)$ and $p$ is given as a circle with radius $r=R$, where $R\ge r_0$. Clearly, this circle encloses the whole disclination core. Also, since $A^o_{ij}(x_1,x_2) = A^o_{ij}(-x_1,-x_2)$, then 
\begin{equation*}
\int_p \bfA^o(\bfx)d\bfx = \bf0.
\end{equation*}
Also, for any loop enclosing the core, $\bfalpha = \bf0$ along the loop and thus, 
\begin{equation*}
\int_p \int_{\bfx_r}^{\bfx} (\bfalpha:\bfX)(\bfs) d\bfs d\bfx = \bf0.
\end{equation*}
Therefore, the jump $[\![\bfy]\!] = [\![\bfy]\!]^{s}+ [\![\bfy]\!]^{\alpha}$, where $[\![\bfy]\!]^s = \oint_p \int_{\bfx_r}^{\bfx} \bfS^{*}(\bfs) d\bfs d\bfx$ and $[\![\bfy]\!]^{\alpha} = \oint_p \bfA^{\alpha}(\bfx)d\bfx$. With reference to Figure \ref{fig:contour} and choose $\bfx_r$ as $\bfx^-$, $[\![\bfy]\!]^s$ evaluates to

\begin{eqnarray}\label{eqn:disclination_part}
\begin{aligned}
&[\![\bfy]\!]^s_1= R \Delta^F_{11} \\
&[\![\bfy]\!]^s_2= R \Delta^F_{21}.
\end{aligned}
\end{eqnarray}

$[\![\bfy]\!]^{\alpha} = \oint_p \bfA^{\alpha}(\bfx)d\bfx$ can be obtained as follows:
\begin{eqnarray*}
&[\![\bfy]\!]^{\alpha}_1 = \oint^{0}_{2\pi} \left[-A^{\alpha}_{11}R \sin \beta + A^{\alpha}_{12}R (\cos \beta)\right] d\beta \\
&\Rightarrow [\![\bfy]\!]^{\alpha}_1 = \oint^{0}_{2\pi} - \frac{b_1}{2\pi}d\beta = b_1 \\
&[\![\bfy]\!]^{\alpha}_2 = \oint^{0}_{2\pi} \left[-A^{\alpha}_{21}R \sin \beta + A^{\alpha}_{22}R (\cos \beta)\right] d\beta \\
&\Rightarrow [\![\bfy]\!]^{\alpha}_2 = \oint^{0}_{2\pi} - \frac{b_2}{2\pi}d\beta = b_2 
\end{eqnarray*}

Thus, the jump at point $\bfx$, $(R,0)$, is 
\begin{align*}
[\![\bfy]\!]_1=R \Delta^F_{11} + b_1 \\
[\![\bfy]\!]_2=R \Delta^F_{21} + b_2,
\end{align*}
which can be written in the form
\begin{equation}  \label{eqn:delta}
[\![\bfy]\!](\bfx) = \bfDelta^F \bfx + \bfb = \bfDelta^F(\bfx - \bfx_0) + \llbracket \bfy \rrbracket (\bfx_0),
\end{equation}
where 
\begin{equation*}
\bfDelta^F = 
\begin{bmatrix}
\Delta^F_{11} & \Delta^F_{12} \\
\Delta^F_{21} & \Delta^F_{22}\\
\end{bmatrix}
\end{equation*}
and
\[
\llbracket \bfy \rrbracket (\bfx_0)  = \bfDelta^F \bfx_0 + \bfb.
\]
for an arbitrarily chosen base-point $\bfx_0$ on the cut-surface.

For $\bfPi = \bf0$ (i.e. no generalized disclination in the defect), given a localized dislocation density $\bfalpha$, the jump in the inverse deformation should be the same as the integral of $\bfalpha$ over any arbitrary area threaded by the core, denoted as $\bfb$.
Since $\bfPi = \bf0$, then $\bfDelta^F=\bf0$ and (\ref{eqn:delta}) implies,
\begin{equation*}
[\![\bfy]\!] = \bfb.
\end{equation*}

Thus, in this special, but canonical, example, we have characterized the jump in the inverse deformation due to a defect line in terms of data characterizing the g.disclination and dislocation densities of g.disclination theory and shown that the result is consistent with what is expected in the simpler case when the g.disclination density vanishes.
\subsection{The connection between $\bfW$ and $\bfy$}
\begin{figure}
\centering
\includegraphics[width = 0.5\textwidth]{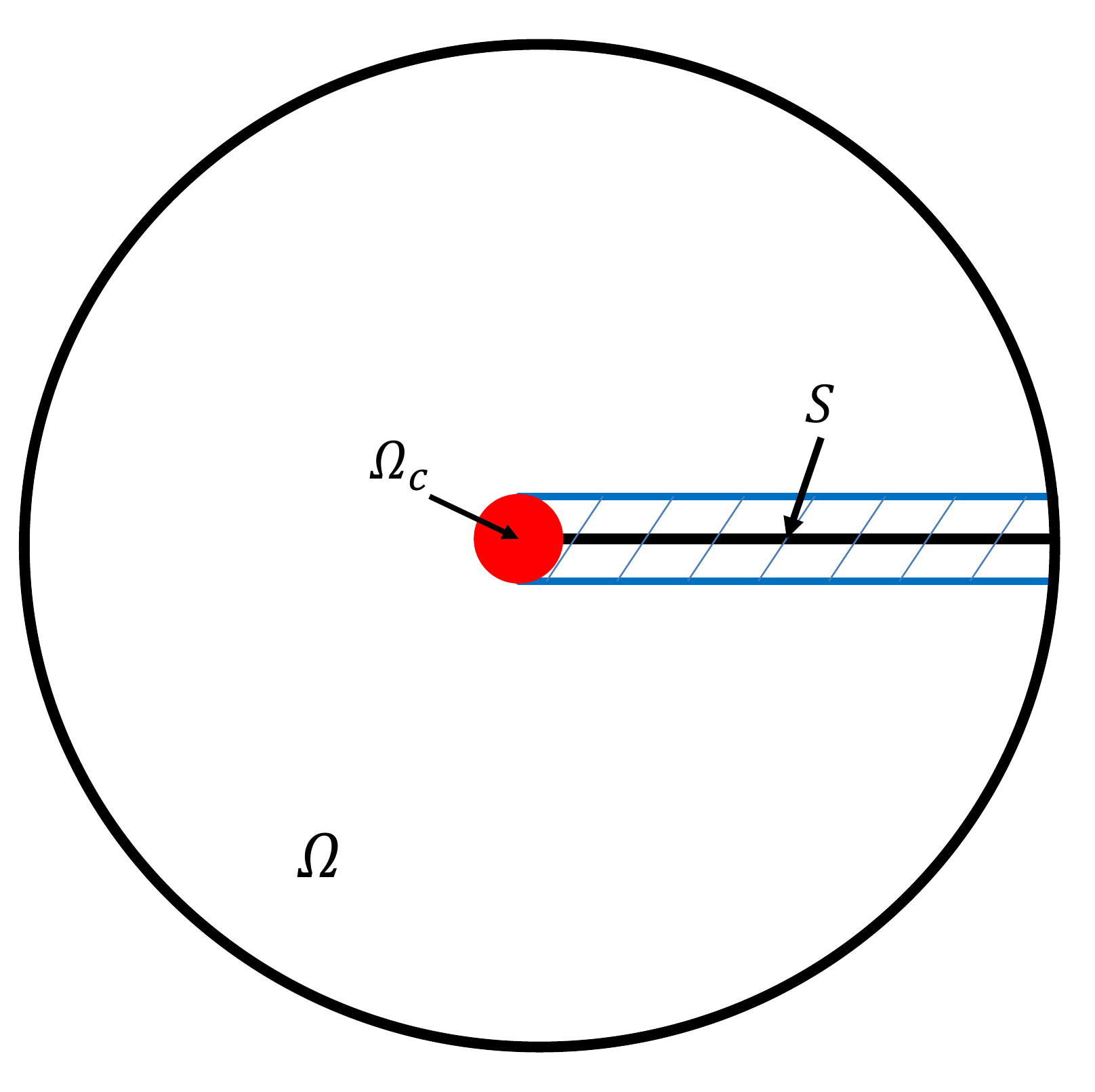}
\caption{The cross-section of a simply-connected domain $\Omega$ with a cut-surface $S$ and a core $\Omega_c$. The blue shaded area is the layer $S_l$ which includes the core.}
\label{fig:layer_core}
\end{figure}

We now deal with the question of how the inverse deformation field $\bfy$ defined on a cut-surface induced simply-connected domain defined from the field $\tilde{\bfY}$ may be related to the i-elastic 1-distortion field $\bfW$ of g.disclination theory. The setting we have in mind is as follows: with reference to Fig. \ref{fig:layer_core}, we consider the domain $\Omega$ with the core comprising the region $\Omega_c \subset \Omega$. Let the cut-surface be $S$, connecting a curve on the boundary of $\Omega_c$ to a curve on the boundary of $\Omega$ so that $(\Omega \backslash \Omega_c)\backslash S$ is simply-connected. Also consider a `layer' region $S_l \subset \Omega$ such that $S \subset S_l$ as well as $\Omega_c \subset S_l$, as shown in Figure \ref{fig:layer_core}. We assume that $\bfS$ has support in $S_l$. We now think that a problem of g.disclination theory has been solved with $\bfS$, $\bfPi$, and $\bfalpha$ as given data on $\Omega$ satisfying the constraint $\curl\, \bfS = \bfPi$.

From (\ref{eqn:tildey}) and (\ref{eqn:tildeW}), we have
\[
\grad \, (\tilde{\bfW} - \bfW) = \bfS - \frac{1}{2} \bfalpha \bfX \ \ \ \ \mbox{on} \ (\Omega \backslash \Omega_c) \backslash S.
\]
We then have
\begin{equation*}
\llbracket \bfW \rrbracket = \llbracket \grad \, \bfy \rrbracket - \bfDelta^F = \bf0 \ \ \ \ \mbox{on} \   S,
\end{equation*}
since the dislocation density $\bfalpha$ is localized in the core, an expected result since $\bfW$ is a continuous field on $\Omega$. Moreover, we have
\[
\bfW = \grad\, \bfy + \bfW^* \ \ \ \ \mbox{on} \ \Omega \backslash S_l,
\]
where $\bfW^*$ is a constant second-order tensor. Thus, when $S_l$ is truly in the shape of a layer around $S$ (including the core as defined), $\bfW$ can indeed be viewed as the gradient of the deformation $\bfy$ constructed from $\tilde{\bfY}$, up to a constant second-order tensor, in most of the domain. On the other hand, when $S_l = \Omega \backslash \Omega_c$ as e.g. when $\curl \, \bfS = \bfPi$ with $\divergence\, \bfS = \bf0$ on $\Omega$ with $\bfPi$ still supported in the core, such an identification is not possible.

 \section{Burgers vector of a g.disclination dipole} \label{sec:burgers}

\begin{figure}
\centering
\includegraphics[width=0.8\textwidth]{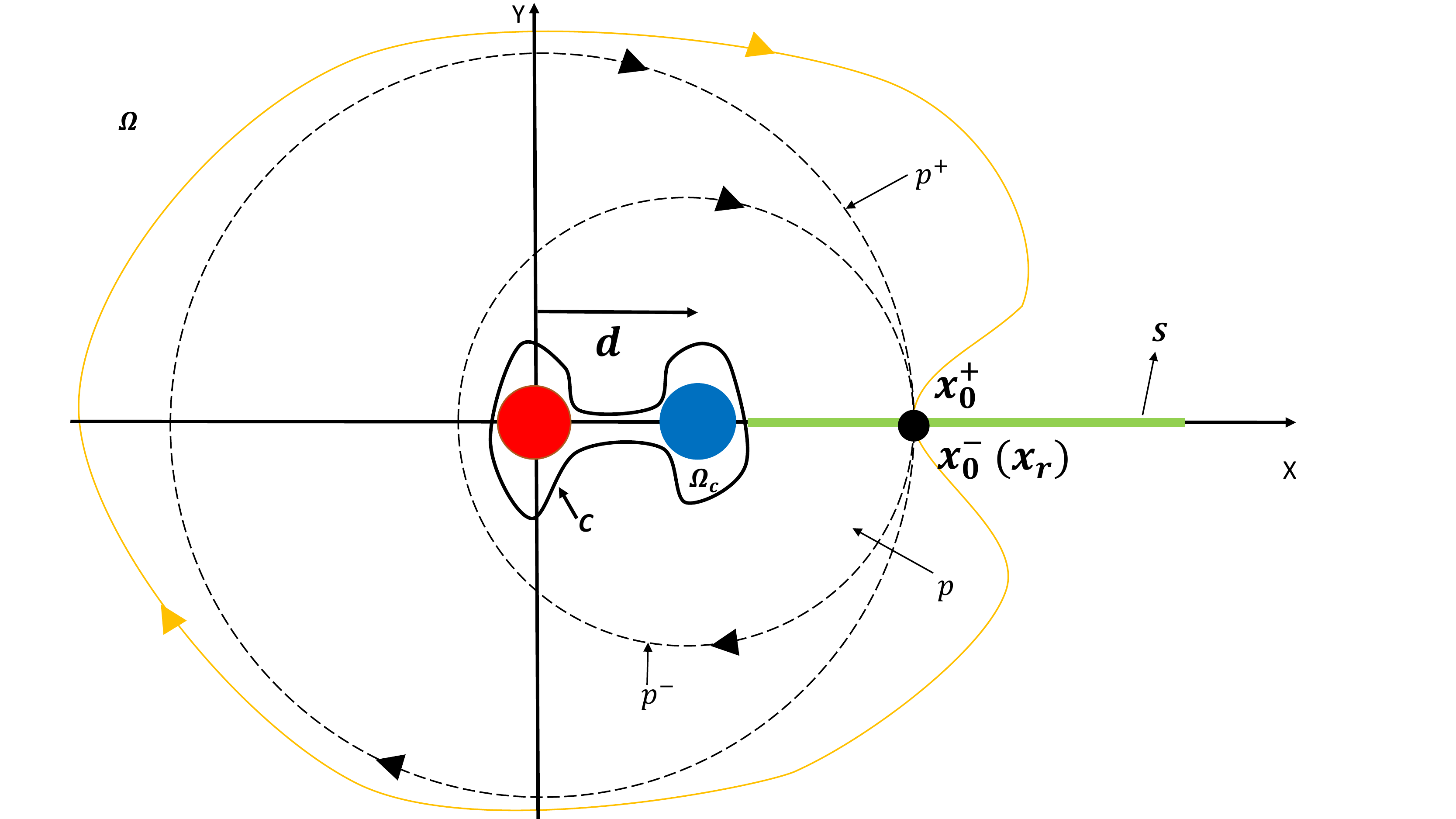}
\caption{Configuration for a disclination dipole with separation vector $\bfd$ and an introduced cut-surface.}
\label{fig:dipole_burgers}
\end{figure}

We will derive the Burgers vector for a given g.disclination dipole with separation vector $\bfd$. In Figure \ref{fig:dipole_burgers}, the red circle is the (cross-section of the) positive g.disclination while the blue circle is that of the  negative g.disclination. The separation vector between these two disclination is $\bfd$.  For the calculations in this section, we assume Cartesian coordinates whose origin is at the positive g.disclination core and the $x$ and $y$ axes are shown as in Figure \ref{fig:dipole_burgers}. The boundary of the positive g.disclination core is the circle with center at $(0,0)$ and radius $r_0$; the boundary of the negative disclination core is the circle with center at $(d,0)$ and radius $r_0$.  Denote the strength of the positive disclination as $\bfDelta^F$ (see Sec. \ref{sec:g_disclination_theory} for the definition of the strength). We denote the whole domain as $\Omega$ and the boundary of the domain as $\partial \Omega$. The positive g.disclination density field $\bfPi^+$ and the negative g.disclination density field $\bfPi^-$ are both localized inside the cores (while being defined in all of $\Omega$).  We define the core of the g.disclination dipole, $\Omega_c$, as a patch including the positive and negative g.disclinations enclosed by the black contour $C$ in Figure \ref{fig:dipole_burgers}. In the following calculation, the core is referred as the g.disclination dipole core. Also, let the cut-surface be $S$, which is along the positive $x$ axis connecting the boundary of $\Omega$ to the curve $C$ at $\bfx_c$ shown as the green line in Figure \ref{fig:dipole_burgers}. Namely, $S=\{(x,y) \in \Omega | y=0, x \ge x_c\}$. In addition, the cut-induced simply-connected domain is denoted as $(\Omega \backslash \Omega_c)\backslash S$. 

We denote the defect density field for the g.disclination dipole as $\bfPi$. Clearly, $\bfPi$ is localized within the g.disclination core, given as $\bfPi = \bfPi^{+}+\bfPi^{-}$ on $\Omega$. Based on the Weingarten-gd theorem, given $\bfS^*$, $\bfalpha$, and $\bfA$, $\tilde{\bfY}$ is defined as in (\ref{eqn:def_Y}) and (\ref{S*}):
\[
\tilde{\bfY} := \bfS^{*} + \grad \bfA - \frac{1}{2}\bfalpha :\bfX \qquad \text{in $\Omega$}.
\]
In this case, $\bfalpha = \bf0$, thus
\begin{equation} \label{eqn:Y_dipole}
\tilde{\bfY} = \bfS^{*} + \grad \bfA.
\end{equation}

Recall that given the g.disclination density $\bfPi$, $\bfS^{*}$ is calculated from 
\begin{eqnarray*}
\begin{rcases}
\curl \bfS^{*} = \bfPi = \bfPi^{+}+\bfPi{-} \\
\divergence \bfS^{*} = \bf0
\end{rcases}
&&\qquad \text{ in $\Omega$}.
\end{eqnarray*}
Since $\bfS^{*}$ is a solution to linear equations, $\bfS^{*}$ can be written as  
\[
\bfS^{*} = \bfS^{*+} +\bfS^{*-},
\]
with $\bfS^{*+}$ and $\bfS^{*-}$ calculated from
\begin{eqnarray*}
\begin{rcases}
\curl \bfS^{*+} = \bfPi^+ \\
\divergence \bfS^{*+} = \bf0 
\end{rcases}
&&\qquad \text{ in $\Omega$} \\
\begin{rcases}
\curl \bfS^{*-} = \bfPi^- \\
\divergence \bfS^{*-} = \bf0 
\end{rcases}
&&\qquad \text{ in $\Omega$}.
\end{eqnarray*}

Similarly, given $\bfS^{*}$, $\bfA = \bfA^{*}+\grad \bfz^A$, where $\bfA^{*}$ satisfies (\ref{eqn:relation_A_S})
\begin{eqnarray*}
\begin{rcases}
\curl \bfA^{*} = \bfS^{*}:\bfX \\
\divergence \bfA^{*} = \bf0 
\end{rcases}
&&\qquad \text{ in $\Omega$}.
\end{eqnarray*}
Therefore, $\bfA$ can be written as $\bfA = \bfA^{*+} + \bfA^{*-} + \grad \bfz^A$, where $\bfA^{*+}$ and $\bfA^{*-}$ are calculated from 
\begin{eqnarray*}
\begin{rcases}
\curl \bfA^{*+} = \bfS^{*+}:\bfX \\
\divergence \bfA^{*+} = \bf0 
\end{rcases}
&&\qquad \text{ in $\Omega$} \\
\begin{rcases}
\curl \bfA^{*-} = \bfS^{*-}:\bfX \\
\divergence \bfA^{*-} = \bf0 
\end{rcases}
&&\qquad \text{ in $\Omega$}.
\end{eqnarray*}

After substituting $\bfS^{*}$ and $\bfA$, $\tilde{\bfY}$ can be written as 
\[
\tilde{\bfY} = \bfS^{*+}+\bfS^{*-}+ \grad \bfA^{*+} + \grad \bfA^{*-} + \grad \, \grad \bfz^A \ \ \ \mbox{on} \ \ \ \Omega.
\]
In the cut-induced simply-connected domain $(\Omega \backslash \Omega_c)\backslash S$, given the core $\Omega_c$ and cut-surface $S$  shown in Figure \ref{fig:dipole_burgers}, $\tilde{\bfW}$ is defined as (\ref{eqn:tildeW})
\[
\tilde{\bfW}(\bfx) := \int_{\bfx_r}^{\bfx} \tilde{\bfY}(\bfs) d\bfs+ \tilde{\bfW}(\bfx_r)
\]
where $\bfx_r$ is a fixed point and $\tilde{\bfW}(\bfx_r)$ is an arbitrary constant. After substituting $\tilde{\bfY}$, we have
\[
\tilde{\bfW}(\bfx) = \int_{\bfx_r}^{\bfx}\bfS^{*+}(\bfs) d\bfs + \int_{\bfx_r}^{\bfx}\bfS^{*-}(\bfs) d\bfs+ \bfA^{*+}(\bfx) + \bfA^{*-}(\bfx) + \grad\bfz^A(\bfx)+ const,
\]
where $const$ is the constant equal to $\tilde{\bfW}(\bfx_r)-\bfA^{*+}(\bfx_r) - \bfA^{*-}(\bfx_r) - \grad\bfz^A(\bfx_r)$. Write $\tilde{\bfW}(\bfx)$ as 
\[
\tilde{\bfW}(\bfx) = \tilde{\bfT}^{+}(\bfx; \bfx_r) + \tilde{\bfT}^{-}(\bfx; \bfx_r) + \grad \bfz^A(\bfx) + const,
\]
where 
\begin{eqnarray*}
\tilde{\bfT}^{+}(\bfx; \bfx_r)  :=  \int_{\bfx_r}^{\bfx}\bfS^{*+}(\bfs) d\bfs  + \bfA^{*+}(\bfx)  \\
\tilde{\bfT}^{-}(\bfx; \bfx_r)  :=  \int_{\bfx_r}^{\bfx}\bfS^{*-}(\bfs) d\bfs  + \bfA^{*-}(\bfx)
\end{eqnarray*}

With reference to Fig. \ref{fig:dipole_burgers}, it follows from (\ref{y_from_W}) that the jump of $\bfy$ at $\bfx_0$ is
\[
\llbracket\bfy\rrbracket  (\bfx_0)= \int_p \tilde{\bfW}(\bfx) d\bfx,
\]
where $p$ is a path shown in Fig. \ref{fig:dipole_burgers} from $\bfx_0^-$ to $\bfx_0^+$. Since $\bfz$ is continuous on $\Omega$, $\int_p \grad \bfz (\bfx) d\bfx=\bf0$. Also, with $\int_p const d\bfx=\bf0$, we have
\[
\llbracket \bfy \rrbracket (\bfx_0)= \int_p \tilde{\bfT}^+(\bfx;\bfx_r) d\bfx + \int_p \tilde{\bfT}^-(\bfx;\bfx_r) d\bfx,
\]

Let $\bfx_0$ be the point located at $(x_0,0)$ and denote $p^+$ as a clockwise circle centered at $(0,0)$ with radius $x_0$ and  $p^-$ as a clockwise circle centered at $(d,0)$ with radius $x_0-d$. Also, we choose $\bfx_0$  big enough so that $p^+$ and $p^-$ enclose the g.disclination core - this induces no loss in generality for our final result, as we show in the discussion surrounding (\ref{eqn:point_indep}) and (\ref{eqn:burgers_general}). Also, let $\bfx_r$ be $\bfx_0^-$. Based on the argument in Appendix \ref{sec:append_loop_independ}, 
\begin{eqnarray*}
 \oint_p \tilde{\bfT}^+(\bfx; \bfx_r) d\bfx = \oint_{p^+} \tilde{\bfT}^+(\bfx; \bfx_0^-) d\bfx =: \bfI^+\\
 \oint_p \tilde{\bfT}^-(\bfx; \bfx_r) d\bfx = \oint_{p^-} \tilde{\bfT}^-(\bfx; \bfx_0^-) d\bfx =: \bfI^-
\end{eqnarray*}
(This step is utilized to facilitate the computation of the line integrals, which are most conveniently calculated on circular paths).

Given the $\bfPi^+$ and $\bfPi^-$ fields, based on the calculation in Appendix \ref{append:calculation_dipole}, we have
\begin{eqnarray*}
I^+_1 =  x_0 \Delta^F_{11} \\
I^+_2 = x_0 \Delta^F_{21} \\
I^-_1 = -(x_0-d)\Delta^F_{11}\\
I^-_2 =  -(x_0-d) \Delta^F_{21}.
\end{eqnarray*}

Thus, the jump $\llbracket \bfy \rrbracket$ at $\bfx_0$ is
\begin{eqnarray*}
&\llbracket y \rrbracket_1 (\bfx_0) = x_0 \Delta^F_{11} -(x_0-d)\Delta^F_{11} = d\Delta^F_{11} \\
 &\llbracket y \rrbracket_2 (\bfx_0) = x_0 \Delta^F_{21} -(x_0-d) \Delta^F_{21} = d\Delta^F_{21}
\end{eqnarray*}

Therefore, the jump at $\bfx_0$ is 
\begin{eqnarray*}
\llbracket \bfy \rrbracket (\bfx_0) = \bfDelta^F \bfd
\end{eqnarray*}

Recall from (\ref{eqn:Y_dipole}) that
\[
\tilde{\bfY} = \bfS^{*} + \grad \bfA,
\]
so that
\[
\curl \tilde{\bfY} = \curl \bfS^{*} = \bfPi.
\]
Therefore, 
\[
\bfDelta =  \int_C \tilde{\bfY} d\bfx = \int_{\Omega_c} \curl \tilde{\bfY} \bfn da = \int_{\Omega_c} (\bfPi^+ + \bfPi^-) \bfn da
\]
where $C$ is the boundary of the g.disclination core and $\Omega_c$ is the area of the core enclosed by $C$. Because $\bfPi^+$ and $\bfPi^-$ have opposite signs but are otherwise identical (shifted) fields, 
\[
\bfDelta = \int_{\Omega_c} (\bfPi^+ + \bfPi^-) \bfn da = \bf0.
\]
Then, invoking (\ref{weingarten_jump}), the jump of $\bfy$ satisfies
\begin{equation}\label{eqn:point_indep}
\llbracket \bfy \rrbracket (\bfx) = \llbracket \bfy \rrbracket(\bfx_0) = \bfDelta^F \bfd\ \ \ \mbox{for} \ \ \ \bfx \in S.
\end{equation}

In addition, as proved in Section \ref{sec:cut_independ}, the jump $\llbracket \bfy \rrbracket$ is independent of the cut-surface when $\bfDelta =\bf0$. Therefore, defining the Burgers vector $\bfb$ of the g.disclination dipole as the jump $\llbracket \bfy \rrbracket$ across \emph{any} arbitrary cut-surface rendering $\Omega \backslash \Omega_c$ simply-connected, we have
\begin{equation}\label{eqn:burgers_general}
\bfb = \bfDelta^F \bfd.
\end{equation}

Now, considering a disclination dipole with separation vector $\bfd = \delta \bfr$, where the Frank vector for the positive disclination is given  as $\bfOmega = \Omega_3 \bfe_3$, $\bfDelta^F$ is given by the difference of two inverse rotation tensors, written as 
\begin{equation*}
\left\lbrack \bfDelta^F \right\rbrack = 
\begin{bmatrix}
\cos \Omega_3-1& -\sin \Omega_3 \\
\sin \Omega_3 & \cos \Omega_3-1 \\
\end{bmatrix}.
\end{equation*}
When the strength of the disclination is low, $|\Omega_3| \ll 1$, $\cos \Omega_3 - 1 \approx 0$ and $\sin \Omega_3 \approx \Omega_3$, and we have
\begin{equation*}
\bfDelta^F = 
\begin{bmatrix}
0& - \Omega_3 \\
 \Omega_3 & 0 \\
\end{bmatrix},
\end{equation*}
a skew symmetric tensor.

From (\ref{eqn:burgers_general}), the components of the Burgers vector $\bfb$ are given as 
\begin{eqnarray*}
b_1 =  -\Omega_3 \delta r_2, \ \ \  b_2 = \Omega_3 \delta r_1 
\end{eqnarray*}
which matches the result (\ref{eqn:dipole_burgers_elasticity}) from linear elasticity. 
\section*{Acknowledgments}
CZ and AA acknowledge support from grant NSF-DMS-1434734. AA also acknowledges support from grants NSF-CMMI-1435624 and ARO W911NF-15-1-0239. It is a pleasure to acknowledge a discussion with Jianfeng Lu.

\appendix
\section{Analytical solution for $\bfS^{*}$ in the generalized disclination model} \label{sec:app1}

We recall the governing equations of $\bfS^{*}$ in g.disclination theory given as 
\begin{eqnarray*}
&\curl \bfS^{*} = \bfPi \\
&\divergence \bfS^{*} = \bf0.
\end{eqnarray*}

To solve $\bfS^{*}$ from these two equations, the Riemann-Graves integral operator \cite{edelen1985applied, edelen1988dispersion, acharya2001model} is applied. Suppose $\bfPi$ satisfies $\Pi_{ijk,k}=0$, and define 

\begin{equation*}
\hat{\Pi}_{injk} = e_{jkm} \Pi_{inm}.
\end{equation*}

In the 2-D case, the only non-zero components of $\bfPi$ are $\Pi_{ij3}$ with $i,j=1,2$. Thus $\hat{\bfPi}$ has no non-zero component with index $3$. Then the relation between $\bfPi$ and $\hat{\bfPi}$ is 
\begin{eqnarray*}
&\hat{\Pi}_{1112} = \Pi_{113} \qquad \hat{\Pi}_{2212} = \Pi_{223} \qquad \hat{\Pi}_{1212} = \Pi_{123} \qquad \hat{\Pi}_{2112} = \Pi_{213} \\
&\hat{\Pi}_{1121} = -\Pi_{113} \qquad \hat{\Pi}_{2221} = -\Pi_{223} \qquad \hat{\Pi}_{1221} = -\Pi_{123} \qquad \hat{\Pi}_{2121} = -\Pi_{213} 
\end{eqnarray*}

The integral is now introduced as 
\begin{equation*}
H_{ink} = (x_j-x^0_j) \int_0^1 \hat{\Pi}_{injk} (x^0 + \lambda(x-x^0))\lambda d\lambda,
\end{equation*}
where $\bfx^0$ is any fixed point in the body (and we assume a star-shaped domain).

Suppose $\Pi_{ij3}$ takes the form 
\begin{equation*}
\Pi_{ij3} = \psi_{ij}(\sqrt{x_1^2+x_2^2}),
\end{equation*}
and also assume $\bfx^0$ be the origin of the coordinates. Then for a positive disclination, the $\bfH$ at $\bfx$ is given as 
\begin{eqnarray*}
&H_{ij1} = 
\begin{cases}
\frac{\Delta F_{ij}}{2\pi}(-\frac{x_2}{r^2}) & r>r_0 \\
\frac{-x_2}{r^2}\int_0^r \psi_{ij}(s)sds & r \le r_0
\end{cases} \\
&H_{ij2} = 
\begin{cases}
\frac{\Delta F_{ij}}{2\pi}(\frac{x_1}{r^2}) & r>r_0 \\
\frac{x_1}{r^1}\int_0^r \psi_{ij}(s)sds & r \le r_0
\end{cases}
\end{eqnarray*}

Then it can be shown that $\bfH$ is the solution of $\bfS^{*}$ which satisfies $\curl \bfS^{*} = \bfPi$ and $\divergence \bfS^{*} = \bf0$.

In this work, we assume $\psi$ takes the form
\begin{equation*}
\psi_{ij}(r) = 
\begin{cases}
\frac{\Delta F_{ij}}{\pi r_0} (\frac{1}{r} - \frac{1}{r_0}) \quad \text{$r < r_0$} \\
0 \quad \text{$r \ge r_0$}.
\end{cases}
\end{equation*}
Thus,

\begin{eqnarray*}
&H_{ij1} = 
\begin{cases}
\frac{\Delta F_{ij}}{2\pi}(-\frac{x_2}{r^2}) & r>r_0 \\
\frac{-x_2 \Delta F_{ij}}{\pi r^2 r_0} (r-\frac{r^2}{2r_0}) & r \le r_0
\end{cases} \\
&H_{ij2} = 
\begin{cases}
\frac{\Delta F_{ij}}{2\pi}(\frac{x_1}{r^2}) & r>r_0 \\
\frac{x_1 \Delta F_{ij}}{\pi r^2 r_0} (r-\frac{r^2}{2r_0})& r \le r_0.
\end{cases}
\end{eqnarray*}

\section{Calculation of $\bfA^*$ for $\tilde{\bfY}$} \label{sec:app2}

The governing equations for $\bfA^*$ are given as 
\begin{equation*}
\begin{aligned}
\curl \bfA^* & = \bfS^{*} :\bfX \\
\divergence \bfA^* & = \bf0
\end{aligned}
\end{equation*}

Since in the 2-D case the last index of all non-zero components of $\bfPi$ is 3 and the first two are 1 or 2, $\left(div \, (\bfS^{*}:\bfX)\right)_{ipp} = - \Pi_{ipp} = 0$, indicating solutions to $\bfA^*$ exist and we again utilize the Riemann-Graves integral.

Denote $\bfB = \bfS^{*} :\bfX$, then $B_{13} = S^{*}_{112}-S^{*}_{121}$ and $B_{23} =S^{*}_{212}-S^{*}_{221}$. After substituting the $\bfS^{*}$ from Appendix \ref{sec:app1}, 
\begin{eqnarray*}
&B_{13} =  
\begin{cases}
\frac{1}{2\pi r^2} (\Delta F_{12}x_2 + \Delta F_{11}x_1) & r \ge r_0 \\
\frac{1}{r^2} (x_2 \int_0^r \psi_{12}sds + x_1 \int_0^r \psi_{11}sds) & r<r_0
\end{cases} \\
&B_{23} = 
\begin{cases}
\frac{1}{2\pi r^2} (\Delta F_{22} x_2 + \Delta F_{21} x_1) & r \ge r_0\\
\frac{1}{r^2} (x_2 \int_0^r \psi_{22}s ds + x_1 \int_0^r \psi_{21} s ds ) & r < r_0
\end{cases}
\end{eqnarray*}

In addition, we can assume $\psi_{ij}$ also takes the form (as in Appendix \ref{sec:app1})
\begin{equation*}
\psi_{ij}(r) = 
\begin{cases}
\frac{\Delta F_{ij}}{\pi r_0} (\frac{1}{r} - \frac{1}{r_0}) & r < r_0 \\
0  & r \ge r_0.
\end{cases}
\end{equation*}

Thus, 
\begin{eqnarray*}
&B_{13} = 
\begin{cases}
\frac{1}{2\pi r^2} (\Delta F_{12} x_2 + \Delta F_{11} x_1) & r \ge r_0 \\
\frac{1}{\pi r^2r_0}(r - \frac{r^2}{2 r_0}) ( \Delta F_{12}x_2 +  \Delta F_{11}x_1) & r <r_0
\end{cases}
\\
&B_{23} = 
\begin{cases}
\frac{1}{2\pi r^2} (\Delta F_{22} x_2 + \Delta F_{21} x_1) & r \ge r_0 \\
\frac{1}{\pi r^2r_0}(r - \frac{r^2}{2 r_0}) (\Delta F_{22}x_2 + \Delta F_{21}x_1) & r <r_0
\end{cases}
\end{eqnarray*}

The Riemann-Grave integral operator for the equation $\curl \bfA^* = \bfB$ is 
\begin{equation*}
A^*_{ij} = (x_m-x^0_m) \int_0^1 \hat{B}_{imj} (x^0 + \lambda(x-x^0))\lambda d\lambda,
\end{equation*}
where $\hat{B}_{imj} = e_{mjr}B_{ir}$. Assuming $\bfx^0$ to be the origin, $\bfA^*$ can be written as 
\begin{eqnarray*}
&A^*_{11} = -x_2 \int_0^1 B_{13}(\lambda \bfx) \lambda d\lambda \\
&A^*_{12} = x_1 \int_0^1 B_{13}(\lambda \bfx) \lambda d\lambda \\
&A^*_{21} = -x_2 \int_0^1 B_{23}(\lambda \bfx) \lambda d\lambda \\
&A^*_{22} = x_1 \int_0^1 B_{23}(\lambda \bfx) \lambda d\lambda.
\end{eqnarray*}

Also, with the choice of $\bfx^0$ as the origin, and if $r < r_0$, then $\lambda r < r_0$, and we have
\begin{eqnarray*}
&\int_0^1 B_{13}(\lambda \bfx) \lambda d\lambda  = (1-\frac{r}{3r_0}) (\frac{\Delta F_{12} \sin \theta }{2 \pi r_0}+\frac{\Delta F_{11} \cos \theta}{2 \pi r_0} ) \\
&\int_0^1 B_{23}(\lambda \bfx) \lambda d\lambda  = (1-\frac{r}{3r_0}) (\frac{\Delta F_{22} \sin \theta }{2 \pi r_0}+\frac{\Delta F_{21} \cos \theta}{2 \pi r_0} )
\end{eqnarray*}

If $r \ge r_0$, then
\begin{eqnarray*}
&\int_0^1 B_{13}(\lambda \bfx)\lambda d\lambda  = \int_0^{r_0/r} B_{13}(\lambda \bfx)\lambda d\lambda + \int_{r_0/r}^1 B_{13}(\lambda \bfx)\lambda d\lambda \\
&\int_0^1 B_{23}(\lambda \bfx)\lambda d\lambda  = \int_0^{r_0/r} B_{23}(\lambda \bfx)\lambda d\lambda + \int_{r_0/r}^1 B_{23}(\lambda \bfx)\lambda d\lambda.
\end{eqnarray*}

Since,
\begin{eqnarray*}
&\int_{r_0/r}^r B_{13}(\lambda \bfx)\lambda d\lambda  = \frac{r-r_0}{2\pi r^2}(\Delta F_{12} \sin \theta + \Delta F_{11} \cos \theta) \\
&\int_{r_0/r}^r B_{23}(\lambda \bfx)\lambda d\lambda  = \frac{r-r_0}{2\pi r^2}(\Delta F_{22} \sin \theta + \Delta F_{21} \cos \theta),
\end{eqnarray*}
therefore
\begin{eqnarray*}
&\int_0^1 B_{13}(\lambda \bfx) \lambda d\lambda  = (\frac{1}{3\pi r_0}+\frac{r-r_0}{2\pi r^2})(\Delta F_{12} \sin \theta + \Delta F_{11} \cos \theta) \\
&\int_0^1 B_{23}(\lambda \bfx) \lambda d\lambda  = (\frac{1}{3 \pi r_0}+ \frac{r-r_0}{2\pi r^2})(\Delta F_{22} \sin \theta + \Delta F_{21} \cos \theta).
\end{eqnarray*}

Converting from polar coordinates to Cartesian coordinates, 
\begin{eqnarray*}
&\int_0^1 B_{13}(\lambda \bfx) \lambda d\lambda  = (\frac{1}{3\pi r_0 r}+\frac{r-r_0}{2\pi r^3})(\Delta F_{12} x_2 + \Delta F_{11} x_1) \\
&\int_0^1 B_{23}(\lambda \bfx) \lambda d\lambda  = (\frac{1}{3 \pi r_0 r}+ \frac{r-r_0}{2\pi r^3})(\Delta F_{22} x_2 + \Delta F_{21} x_1),
\end{eqnarray*}
  
and the solution for $\bfA^*$ is 
\begin{eqnarray*}
&A^*_{11}  = 
\begin{cases}
 (\frac{1}{2\pi r_0 r}- \frac{1}{6 \pi r_0^2}) (-\Delta F_{12} x_2^2-\Delta F_{11} x_1 x_2) &r < r_0 \\
(\frac{1}{3\pi r_0 r}+\frac{r-r_0}{2\pi r^3})(-\Delta F_{12} x_2^2 - \Delta F_{11} x_1 x_2) & r \ge r_0
\end{cases}
\\
&A^*_{12}  = 
\begin{cases}
 (\frac{1}{2\pi r_0 r}- \frac{1}{6 \pi r_0^2}) (\Delta F_{12} x_2 x_1 + \Delta F_{11} x_1^2) &r < r_0 \\
(\frac{1}{3\pi r_0 r}+\frac{r-r_0}{2\pi r^3})(\Delta F_{12} x_2 x_1 + \Delta F_{11} x_1^2) & r \ge r_0
\end{cases}
\\
&A^*_{21}  = 
\begin{cases}
 (\frac{1}{2\pi r_0 r}- \frac{1}{6 \pi r_0^2}) (-\Delta F_{22} x_2^2-\Delta F_{21} x_1 x_2) &r < r_0 \\
(\frac{1}{3\pi r_0 r}+\frac{r-r_0}{2\pi r^3})(-\Delta F_{22} x_2^2 - \Delta F_{21} x_1 x_2) & r \ge r_0
\end{cases}
\\
&A^*_{22}  = 
\begin{cases}
 (\frac{1}{2\pi r_0 r}- \frac{1}{6 \pi r_0^2}) (\Delta F_{22} x_2 x_1 + \Delta F_{21} x_1^2) &r < r_0 \\
(\frac{1}{3\pi r_0 r}+\frac{r-r_0}{2\pi r^3})(\Delta F_{22} x_2 x_1 + \Delta F_{21} x_1^2) & r \ge r_0.
\end{cases}
\end{eqnarray*}

\section{An auxiliary path-independence result in the dislocation-free case}\label{sec:append_loop_independ}

\begin{figure}
\centering
\includegraphics[width=0.5\textwidth]{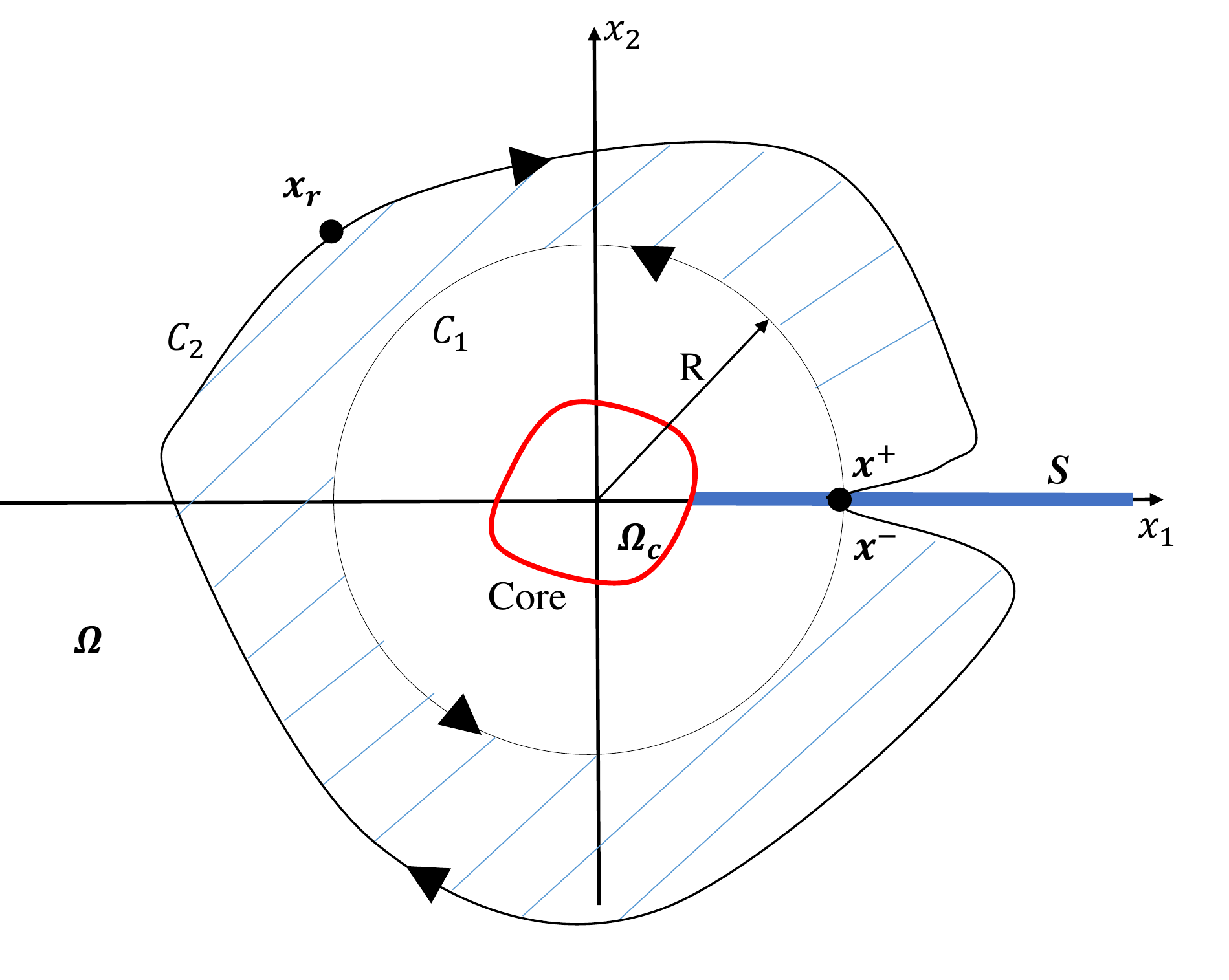}
\caption{A single g.disclination with a circular and an arbitrary path connecting $\bfx^-$ and $\bfx^+$ in the simply-connected domain $(\Omega\backslash\Omega_c)\backslash S$.}
\label{fig:appen_loop_indep}
\end{figure}

Consider the configuration shown in Figure \ref{fig:appen_loop_indep} with the whole domain denoted as $\Omega$, the core as $\Omega_c$, the cut-surface as $S$, and the cut-induced simply-connected domain as $(\Omega\backslash\Omega_c)\backslash S$. Given any g.disclination density localized in the core $\Omega_c$ (the patch enclosed by the red line in Figure \ref{fig:appen_loop_indep}) and the dislocation density $\bfalpha = \bf0$, we can calculate $\bfS^{*}$ and $\bfA^{*}$ from the following
\begin{eqnarray*}
\begin{rcases}
\curl \bfS^{*} = \bfPi \\
\divergence \bfS^{*} = \bf0 
\end{rcases}
&&\qquad \text{in $\Omega$} \\
\begin{rcases}
\curl \bfA^{*} = \bfS^{*}:\bfX \\
\divergence \bfA^{*} = \bf0 
\end{rcases}
&&\qquad \text{in $\Omega$} 
\end{eqnarray*} 
 
Given a fixed point $\bfx_r$, define $\tilde{\bfT}^p$ as
\[
\tilde{\bfT}^p(\bfx; \bfx_r)  :=   \underset{p}{\int_{\bfx_r}^{\bfx}}\bfS^{*}(\bfs) d\bfs  + \bfA^{*}(\bfx),
 \]
 where $p$ is a path from $\bfx_r$ to $\bfx$. Since $curl \bfS^{*} = \bf0$ outside the core $\Omega_c$, $\tilde{\bfT}^p$ is path-independent in the simply-connected domain $(\Omega\backslash \Omega_c ) \backslash S$; hence, we denote $\tilde{\bfT}^p(\bfx; \bfx_r)$ as $\tilde{\bfT}(\bfx; \bfx_r)$. Also, since $\bfA^{*}$ is calculated from $curl \bfA^{*} = \bfS^{*}:\bfX$, we have
\begin{eqnarray*}
\begin{aligned}
\curl \tilde{\bfT}(\bfx; \bfx_r) &= \curl \int_{\bfx_r}^{\bfx}\bfS^{*}(\bfs) d\bfs + curl \bfA^{*} \\
\Rightarrow \curl \tilde{\bfT}(\bfx; \bfx_r) &= - \grad \left(\int_{\bfx_r}^{\bfx}\bfS^{*}(\bfs) d\bfs\right) : \bfX + \curl \bfA^{*}\\
\Rightarrow \curl \tilde{\bfT}(\bfx; \bfx_r) &= -\bfS^{*} :\bfX + \curl \bfA^{*} \\
\Rightarrow \curl \tilde{\bfT}(\bfx; \bfx_r) &= \bf0.
\end{aligned}
\end{eqnarray*}
 
Let $C_1$ be the anti-clockwise circular path from $\bfx^+$ to $\bfx^-$ with radius $R$ and $C_2$ be the clockwise outer path from $\bfx^-$ to $\bfx^+$ shown in the Figure \ref{fig:appen_loop_indep}. Then $C = C_1+C_2$ is the closed contour enclosing the blue shaded area. Take the integral of $\tilde{\bfT}$ along the closed contour $C$ shown as in Figure \ref{fig:appen_loop_indep},
\begin{eqnarray*}
 \int_{C} \tilde{\bfT}(\bfx; \bfx_r) d\bfx = \int_A \curl \tilde{\bfT}(\bfx; \bfx_r) \bfn d a,
\end{eqnarray*}
where $A$ is the patch enclosed by the loop $C$ and $\bfn$ is the normal unit vector of the patch $A$. Therefore, 
\begin{eqnarray*}
 \int_{C} \tilde{\bfT}(\bfx; \bfx_r) d\bfx = \bf0 \\
 \Rightarrow \int_{C_1} \tilde{\bfT}(\bfx; \bfx_r) d\bfx + \int_{C_2} \tilde{\bfT}(\bfx; \bfx_r) d\bfx = \bf0 \\
 \Rightarrow \int_{-C_1}\tilde {\bfT}(\bfx; \bfx_r) d\bfx = \int_{C_2} \tilde{\bfT} (\bfx; \bfx_r)d\bfx.
\end{eqnarray*}
Namely, the integration of $\tilde{\bfT}(\bfx;\bfx_r)$ from $\bfx^-$ to $\bfx^+$ along any path can be calculated as the integration of $\tilde{\bfT}(\bfx;\bfx_r)$ along a circular path from $\bfx^-$ to $\bfx^+$.

\section{Calculations for Burgers vector of a g.disclination dipole}\label{append:calculation_dipole}

Given the configuration in Section \ref{sec:burgers}, and for in-plane variations of fields, $\bfPi^+$ and $\bfPi^-$ only have non-zero component $\Pi^+_{ij3}$ and $\Pi^-_{ij3}$, which are given as follows:
\begin{equation*}
\Pi^+_{ij3} = 
\begin{cases}
\frac{\Delta^F_{ij}}{\pi r_0} \left( \frac{1}{r_+} - \frac{1}{r_0} \right) \quad \text{$r_+ < r_0$} \\
0 \quad \text{$r_+ \ge r_0$},
\end{cases}
\end{equation*} 
and
\begin{equation*}
\Pi^-_{ij3} = 
\begin{cases}
-\frac{\Delta^F_{ij}}{\pi r_0} \left( \frac{1}{r_-} - \frac{1}{r_0} \right) \quad \text{$r_- < r_0$} \\
0 \quad \text{$r_- \ge r_0$},
\end{cases}
\end{equation*} 
where $r_+=\sqrt{x_1^2+x_2^2}$ and $r_-=\sqrt{(x_1-d)^2+x_2^2}$. As given in Appendix \ref{sec:app1},  we obtain $\bfS^{*+}$ as 
 \begin{eqnarray*}
&S^{*+}_{ij1} =
\begin{cases}
\frac{\Delta^F_{ij}}{2\pi}(-\frac{x_2}{r_+^2}) & r_+>r_0 \\
\frac{-x_2 \Delta^F_{ij}}{\pi r_+^2 r_0} (r_+-\frac{r_+^2}{2r_0}) & r_+ \le r_0
\end{cases} \\
&S^{*+}_{ij2} =
\begin{cases}
\frac{\Delta^F_{ij}}{2\pi}(\frac{x_1}{r_+^2}) & r_+>r_0 \\
\frac{x_1 \Delta^F_{ij}}{\pi r_+^2 r_0} (r_+-\frac{r_+^2}{2r_0})& r_+ \le r_0,
\end{cases}
\end{eqnarray*}
and $\bfS^{*-}$ as 
 \begin{eqnarray*}
&S^{*-}_{ij1} =
\begin{cases}
\frac{\Delta^F_{ij}}{2\pi}(\frac{x_2}{r_-^2}) & r_->r_0 \\
\frac{x_2 \Delta^F_{ij}}{\pi r_-^2 r_0} (r_--\frac{r_-^2}{2r_0}) & r_- \le r_0
\end{cases} \\
&S^{*-}_{ij2} =
\begin{cases}
\frac{\Delta^F_{ij}}{2\pi}(-\frac{x_1-d}{r_-^2}) & r_->r_0 \\
\frac{-(x_1-d) \Delta^F_{ij}}{\pi r_-^2 r_0} (r_--\frac{r_-^2}{2r_0})& r_- \le r_0.
\end{cases}
\end{eqnarray*}

Also, following the same arguments as in Appendix \ref{sec:app2}, we obtain for $\bfA^{*+}$ and $\bfA^{*-}$:
\begin{eqnarray*}
&A^{*+}_{11}  = 
\begin{cases}
 C_1^+ \left(-\Delta^F_{12} x_2^2-\Delta^F_{11} x_1 x_2 \right)  &r_+ < r_0 \\
C_2^+ \left(-\Delta^F_{12} x_2^2 - \Delta^F_{11} x_1 x_2 \right)  & r_+ \ge r_0
\end{cases}
\\
&A^{*+}_{12}  = 
\begin{cases}
C_1^+ \left(\Delta^F_{12} x_2 x_1 + \Delta^F_{11} x_1^2\right)  &r_+ < r_0 \\
C_2^+ \left(\Delta^F_{12} x_2 x_1 + \Delta^F_{11} x_1^2\right)& r_+ \ge r_0
\end{cases}
\\
&A^{*+}_{21}  = 
\begin{cases}
 C_1^+ \left(-\Delta^F_{22} x_2^2-\Delta^F_{21} x_1 x_2\right)  &r_+ < r_0 \\
C_2^+ \left(-\Delta^F_{22} x_2^2 - \Delta^F_{21} x_1 x_2\right) & r_+ \ge r_0
\end{cases}
\\
&A^{*+}_{22}  = 
\begin{cases}
C_1^+ \left(\Delta^F_{22} x_2 x_1 + \Delta^F_{21} x_1^2\right) &r_+ < r_0 \\
C_2^+ \left(\Delta^F_{22} x_2 x_1 + \Delta^F_{21} x_1^2\right) & r_+ \ge r_0
\end{cases}
\\&A^{*-}_{11}  = 
\begin{cases}
 C_1^- \left(\Delta^F_{12} x_2^2+\Delta^F_{11} (x_1-d) x_2 \right)  &r_- < r_0 \\
C_2^- \left(\Delta^F_{12} x_2^2 + \Delta^F_{11} (x_1-d) x_2 \right)  & r_- \ge r_0
\end{cases}
\\
&A^{*-}_{12}  = 
\begin{cases}
C_1^- \left(-\Delta^F_{12} x_2 (x_1-d) - \Delta^F_{11} (x_1-d)^2\right)  &r_- < r_0 \\
C_2^- \left(-\Delta^F_{12} x_2 (x_1-d) - \Delta^F_{11} (x_1-d)^2\right)& r_- \ge r_0
\end{cases}
\\
&A^{*-}_{21}  = 
\begin{cases}
 C_1^- \left(\Delta^F_{22} x_2^2 + \Delta^F_{21} (x_1-d) x_2\right)  &r_- < r_0 \\
C_2^- \left(\Delta^F_{22} x_2^2 + \Delta^F_{21} (x_1-d) x_2\right) & r_- \ge r_0
\end{cases}
\\
&A^{*-}_{22}  = 
\begin{cases}
C_1^- \left(-\Delta^F_{22} x_2 (x_1-d) - \Delta^F_{21} (x_1-d)^2\right) &r_- < r_0 \\
C_2^- \left(-\Delta^F_{22} x_2 (x_1-d) - \Delta^F_{21} (x_1-d)^2\right) & r_- \ge r_0
\end{cases}
\end{eqnarray*}
where $C_1^+= \frac{1}{2\pi r_0 r_+}- \frac{1}{6 \pi r_0^2}$, $C_2^+=\frac{1}{3\pi r_0 r_+}+\frac{r_+-r_0}{2\pi r_+^3}$, $C_1^-= \frac{1}{2\pi r_0 r_-}- \frac{1}{6 \pi r_0^2}$ and $C_2^-=\frac{1}{3\pi r_0 r_-}+\frac{r_--r_0}{2\pi r_-^3}$. 
Given $\bfx_+$ as $(-x_0,0)$, $\bfx_-$ as $(d-x_0,0)$ and two paths $p^+$ and $p^-$ as in Figure \ref{fig:dipole_burgers}, following the same notation as in Section \ref{sec:burgers}, we have
\begin{eqnarray*}
\oint_{p^+} \left \{\int_{\bfx_-}^{\bfx}\bfS^{*+}(\bfs) d\bfs  + \bfA^{*+}(\bfx)\right \} d\bfx =: \bfI^+\\
\oint_{p^-} \left \{\int_{\bfx_-}^{\bfx}\bfS^{*-}(\bfs) d\bfs  + \bfA^{*-}(\bfx)\right \} d\bfx =: \bfI^-.
\end{eqnarray*}
After substituting $\bfS^{*+}$, $\bfS^{*-}$, $\bfA^{*+}$, and $\bfA^{*-}$, applying polar coordinates and some algebraic calculation, we obtain
\begin{eqnarray*}
I^+_1 =  x_0 \Delta^F_{11} \\
I^+_2 = x_0 \Delta^F_{21} \\
I^-_1 = -(x_0-d)\Delta^F_{11}\\
I^-_2 =  -(x_0-d) \Delta^F_{21}.
\end{eqnarray*}

\newpage
\bibliography{bibtex}
\bibliographystyle{amsalpha}

\end{document}